\documentclass[twocolumn, times, resetfootnote]{aastex7}

\usepackage{graphicx}
\usepackage{natbib}
\usepackage{color}
\usepackage{wrapfig}
\usepackage{xspace}
\usepackage{amsmath}

\newcommand{\rj}{\ensuremath{R_{\rm Jup}}}
\newcommand{\mearth}{$M_\oplus$}

\newcommand{\mum}{$\mu$m}
\newcommand{\degree}{$^\circ$}
\newcommand{\acenA}{$\alpha$~Cen~A\xspace}
\newcommand{\acen}{$\alpha$~Cen\xspace}

\newcommand{\acenB}{$\alpha$~Cen~B\xspace}
\newcommand{\acenAB}{$\alpha$~Cen~AB\xspace}
\newcommand{\acenABC}{$\alpha$~Cen~ABC\xspace}
\newcommand{\emus}{$\epsilon$~Mus\xspace}
\newcommand{\sone}{$S1$\xspace}

\begin{document}
\shorttitle{JWST/MIRI Observations of \acenA: Paper I}
\shortauthors{Beichman \& Sanghi et al.}

\title{Worlds Next Door: A Candidate Giant Planet Imaged in the Habitable Zone of \acenA.\\ I. Observations, Orbital and Physical Properties, and Exozodi Upper Limits}

\correspondingauthor{Charles Beichman \& Aniket Sanghi}

\author[0000-0002-5627-5471]{Charles Beichman}
\altaffiliation{Shared first authorship.}
\affiliation{NASA Exoplanet Science Institute, Caltech-IPAC, Pasadena, CA 91125, USA}
\affiliation{Jet Propulsion Laboratory, California Institute of Technology, Pasadena, CA 91109, USA}
\email[show]{chas@ipac.caltech.edu}

\author[0000-0002-1838-4757]{Aniket Sanghi}
\altaffiliation{Shared first authorship.}
\affiliation{Cahill Center for Astronomy and Astrophysics, California Institute of Technology, 1200 E. California Boulevard, MC 249-17, Pasadena, CA 91125, USA}
\affiliation{NSF Graduate Research Fellow}
\email[show]{asanghi@caltech.edu}

\author[0000-0002-8895-4735]{Dimitri Mawet}
\affiliation{Cahill Center for Astronomy and Astrophysics, California Institute of Technology, 1200 E. California Boulevard, MC 249-17, Pasadena, CA 91125, USA}
\affiliation{Jet Propulsion Laboratory, California Institute of Technology, Pasadena, CA 91109, USA}
\email{dmawet@astro.caltech.edu}

\author[0000-0003-0626-1749]{Pierre Kervella}
\affiliation{LIRA, Observatoire de Paris, Universit\'e PSL, Sorbonne Universit\'e, Universit\'e Paris Cit\'e, CY Cergy Paris Universit\'e, CNRS, 5 place Jules Janssen, 92195 Meudon, France}
\affiliation{French-Chilean Laboratory for Astronomy, IRL 3386, CNRS and U. de Chile, Casilla 36-D, Santiago, Chile}
\email{pierre.kervella@obspm.fr}

\author[0000-0002-4309-6343]{Kevin Wagner}
\affiliation{Department of Astronomy and Steward Observatory, University of Arizona, USA}
\email{kevinwagner@arizona.edu}

\author[0000-0002-9644-8330]{Billy Quarles}
\affiliation{Department of Physics and Astronomy, East Texas A\&M University Commerce, TX 75428, USA}
\email{billylquarles@gmail.com}

\author[0000-0001-6513-1659]{Jack J. Lissauer}
\affiliation{Space Science \& Astrobiology Division, Building 245, NASA Ames Research Center, Moffett Field, CA 94035, USA}
\email{jack.lissauer@nasa.gov}

\author[0000-0003-4761-5785]{Max Sommer}
\affiliation{Institute of Astronomy, University of Cambridge, Madingley Road, Cambridge CB3 0HA, UK}
\email{ms3078@cam.ac.uk}

\author[0000-0001-9064-5598]{Mark Wyatt}
\affiliation{Institute of Astronomy, University of Cambridge, Madingley Road, Cambridge CB3 0HA, UK}
\email{wyatt@ast.cam.ac.uk}

\author[0000-0003-0958-2150]{Nicolas Godoy}
\affiliation{Aix Marseille Univ., CNRS, CNES, LAM, Marseille, France}
\email{nicolas.godoy@lam.fr}

\author[0000-0001-6396-8439]{William O. Balmer}
\affiliation{Department of Physics \& Astronomy, Johns Hopkins University, 3400 N. Charles Street, Baltimore, MD 21218, USA}
\affiliation{Space Telescope Science Institute, 3700 San Martin Drive, Baltimore, MD 21218, USA}
\email{wbalmer1@jhu.edu}

\author[0000-0003-3818-408X]{Laurent Pueyo}
\affiliation{Space Telescope Science Institute, 3700 San Martin Drive, Baltimore, MD 21218, USA}
\email{pueyo@stsci.edu}

\author[0000-0002-3414-784X]{Jorge Llop-Sayson}
\affiliation{Jet Propulsion Laboratory, California Institute of Technology, Pasadena, CA 91109, USA}
\email{jorge.llop.sayson@jpl.nasa.gov}

\author[0000-0003-3184-0873]{Jonathan Aguilar}
\affiliation{Space Telescope Science Institute, 3700 San Martin Drive, Baltimore, MD 21218, USA}
\email{jaguilar@stsci.edu}

\author[0000-0001-9674-1564]{Rachel Akeson}
\affiliation{NASA Exoplanet Science Institute, Caltech-IPAC, Pasadena, CA 91125, USA}
\email{rla@ipac.caltech.edu}

\author[0000-0002-4951-8025]{Ruslan Belikov}
\affiliation{Space Science \& Astrobiology Division, Building 245, NASA Ames Research Center, Moffett Field, CA 94035, USA}
\email{ruslan.belikov-1@nasa.gov}

\author[0000-0001-9353-2724]{Anthony Boccaletti}
\affiliation{LIRA, Observatoire de Paris, Universit\'e PSL, Sorbonne Universit\'e, Universit\'e Paris Cit\'e, CY Cergy Paris Universit\'e, CNRS, 5 place Jules Janssen, 92195 Meudon, France}
\email{anthony.boccaletti@obspm.fr}

\author[0000-0002-9173-0740]{Elodie Choquet}
\affiliation{Aix Marseille Univ., CNRS, CNES, LAM, Marseille, France}
\email{elodie.choquet@lam.fr}

\author[0000-0002-9036-2747]{Edward Fomalont}
\affiliation{National Radio Astronomy Observatory, 520 Edgemont Rd., Charlottesville, VA 22903, USA}
\affiliation{ALMA, Vitacura, Santiago, Chile}
\email{efomalon@nrao.edu}

\author[0000-0002-1493-300X]{Thomas Henning}
\affiliation{Max-Planck-Institut f\"ur Astronomie (MPIA), K\"onigstuhl 17, 69117 Heidelberg, Germany}
\email{henning@mpia.de}

\author[0000-0003-4653-6161]{Dean Hines}
\affiliation{Space Telescope Science Institute, 3700 San Martin Drive, Baltimore, MD 21218, USA}
\email{hines@stsci.edu}

\author[0000-0003-2215-8485]{Renyu Hu}
\affiliation{Jet Propulsion Laboratory, California Institute of Technology, Pasadena, CA 91109, USA}
\email{renyu.hu@jpl.nasa.gov}

\author{Pierre-Olivier Lagage}
\affiliation{Universit\'e Paris-Saclay, Universit\'e Paris Cit\'e, CEA, CNRS, AIM, 91191 Gif-sur-Yvette, France}
\email{pierre-olivier.lagage@cea.fr}

\author[0000-0002-0834-6140]{Jarron Leisenring}
\affiliation{Steward Observatory, University of Arizona, Tucson, AZ 85721, USA}
\email{jarronl@arizona.edu}

\author[0000-0001-5864-9599]{James Mang}
\affiliation{Department of Astronomy, University of Texas at Austin, Austin, TX 78712, USA}
\affiliation{NSF Graduate Research Fellow}
\email{j_mang@utexas.edu}

\author[0000-0001-5644-8830]{Michael Ressler}
\affiliation{Jet Propulsion Laboratory, California Institute of Technology, Pasadena, CA 91109, USA}
\email{michael.e.ressler@jpl.nasa.gov}

\author[0009-0009-2491-0694]{Eugene Serabyn}
\affiliation{Jet Propulsion Laboratory, California Institute of Technology, Pasadena, CA 91109, USA}
\email{eugene.serabyn@jpl.nasa.gov}

\author[0000-0001-6172-3403]{Pascal Tremblin}
\affiliation{Universit\'e Paris-Saclay, UVSQ, CNRS, CEA, Maison de la Simulation, 91191 Gif-sur-Yvette, France}
\email{pascal.tremblin@cea.fr}

\author[0000-0001-7591-2731]{Marie Ygouf}
\affiliation{Jet Propulsion Laboratory, California Institute of Technology, Pasadena, CA 91109, USA}
\email{marie.ygouf@jpl.nasa.gov}

\author[0000-0002-8749-823X]{Mantas Zilinskas}
\affiliation{Jet Propulsion Laboratory, California Institute of Technology, Pasadena, CA 91109, USA}
\email{mantas.zilinskas@jpl.nasa.gov}

\begin{abstract}
We report on coronagraphic observations of the nearest solar-type star, \acenA, using the MIRI instrument on the \emph{James Webb} Space Telescope. The proximity of \acen (1.33 pc) means that the star's habitable zone is spatially resolved at mid-infrared wavelengths, so sufficiently large planets or quantities of exozodiacal dust would be detectable via direct imaging. With three epochs of observation (August 2024, February 2025, and April 2025), we achieve a sensitivity sufficient to detect $T_{\rm eff}\approx$ 225--250~K (1--1.2~$R_{\rm Jup}$) planets between 1\arcsec--2\arcsec\ and exozodiacal dust emission at the level of $>$5--8$\times$ the brightness of our own zodiacal cloud. The lack of exozodiacal dust emission sets an unprecedented limit of a few times the brightness of our own zodiacal cloud---a factor of $\gtrsim$~5--10 more sensitive than measured toward any other stellar system to date. In August 2024, we detected a F$_\nu$(15.5 \mum) = 3.5~mJy point source, called \sone, at a separation of 1.5\arcsec\ from \acenA at a contrast level of $5.5\times10^{-5}$. Because the August 2024 epoch had only one successful observation at a single roll angle, it is not possible to unambiguously confirm \sone as a bona fide planet. Our analysis confirms that \sone is neither a background nor a foreground object. \sone is not recovered in the February and April 2025 epochs. However, if \sone is the counterpart of the object, $C1$, seen by the VLT/NEAR program in 2019, we find that there is a 52\% chance that the $S1+C1$ candidate was missed in both follow-up JWST/MIRI observations due to orbital motion. Incorporating constraints from the non-detections, we obtain families of dynamically stable orbits for $S1+C1$ with periods between 2--3 years. These suggest that the planet candidate is on an eccentric ($e \approx 0.4$) orbit significantly inclined with respect to the \acenAB orbital plane ($i_{\rm mutual} \approx 50^\circ$, prograde, or $\approx 130^\circ$, retrograde). Based on the photometry and inferred orbital properties, the planet candidate could have a temperature of 225~K, a radius of $\approx$1--1.1~\rj\ and a mass between 90--150~\mearth, consistent with RV limits. This paper is first in a series of two papers: Paper II (Sanghi \& Beichman et al. 2025, in press) discusses the data reduction strategy and finds that \sone is robust as a planet candidate, as opposed to an image or detector artifact. 
\end{abstract}

\section{Introduction}

$\alpha$ Centauri A is the closest solar-type star to the Sun and offers a unique opportunity for direct imaging with the \emph{James Webb} Space Telescope (JWST) to detect an exoplanet within its habitable zone and to achieve an unprecedented level of sensitivity for the detection of an exozodiacal dust cloud \citep{Beichman2020, Sanghi2025}. Among the nearby stars, \acenA, \textit{primus inter pares}, offers a nearly 3-fold improvement in the angular scale of its Habitable Zone and a 7.5-fold boost in the absolute brightness of any planet compared to the next nearest solar type star, $\tau$ Ceti. Specifically, the F1550C coronagraph onboard the Mid-InfraRed Instrument (MIRI) can be used to probe the 1--3 au ($<$4\arcsec) region around \acenA which is predicted to be stable within the \acenAB system for exoplanets and/or an exozodiacal dust cloud \citep{Quarles2018a, Cuello2024}. The detection of a planet or exozodiacal emission, or more stringent limits on either, would advance our understanding of the formation of planetary systems in binary stellar systems and yield an important target for future observations with both JWST and the extremely large ground-based telescopes. Of particular interest is the ability of JWST/MIRI to confirm the detection of a candidate ($C1$) identified using the VISIR mid-infrared camera (10--12.5 $\mu$m) on ESO's Very Large Telescope (VLT) as part of the NEAR (New Earths in Alpha Centauri Region) Breakthrough Watch Project \citep{Wagner2021}.

In this paper, we present the results of  a deep search for planets and zodiacal dust emission obtained with  three epochs of JWST/MIRI coronagraphic imaging observations of \acenA. This paper is the first in a series and is followed by \citet[][also referred to as Paper II]{Aniket2025}. It is organized as follows. Section~\ref{sec:obs} describes the observational strategy and program execution. Section~\ref{sec:results} summarizes key aspects of the data processing strategy, the detection of a candidate exoplanet in the August 2024 data, the planet temperature sensitivity of our observations, and upper limits on the presence of exozodiacal emission. Section~\ref{sec:orbits} analyzes possible orbital configurations for the candidate planet. The planet's physical properties, as constrained by its observed brightness and orbit, as well as by radial velocity measurements \citep{Wittenmyer2016, Zhao2018}, are considered in Section~\ref{sec:phot}. Section~\ref{sec:discussion} discusses the importance of the presence of the candidate planet and the upper limits on exozodiacal emission in the context of theories of planet and disk formation in binary systems, as well as prospects for recovering the candidate in future observations. Finally, Section~\ref{sec:conclusions} presents our conclusions. Appendix~\ref{app:obs-details} provides the complete details of observation preparation and Appendix~\ref{sec:system} includes new ALMA astrometry and an updated ephemeris for the \acenAB system.
 
\section{Observations\label{sec:obs}}

\begin{deluxetable*}{lcccc}[!thb]
    \scriptsize
    \tablecaption{Stellar Properties\label{tab:stars}}
    \tablehead{\colhead{Property} & \colhead{\acenA} & \colhead{\acenB} & \colhead{\emus} & \colhead{References}}
    \startdata
    Spectral Type &G2V&K1V&M4III&1, 2\\
    Mass ($M_\odot$) &1.0788 $\pm$ 0.0029&0.9092 $\pm$ 0.0025&\nodata&3\\
    Luminosity ($L_\odot$) &1.5059 $\pm$ 0.0019&0.4981 $\pm$ 0.0007&\nodata&3\\
    $K$ (mag) &$-$1.48 $\pm$ 0.05&$-$0.60 $\pm$ 0.05&$-$1.42 $\pm$ 0.05&2, 4\\
    F1065C& $-1.51\pm 0.05$ (160 Jy)&$-0.59\pm 0.05$ (51 Jy)&$-1.9\pm 0.1$ (194 Jy)&5, 6, 7\\
    F1140C& $-1.51\pm 0.05$ (120 Jy)&$-0.59\pm 0.05$ (59 Jy)&$-1.9\pm 0.1$ (180 Jy)&5, 6, 7\\
    F1550C& $-1.51\pm 0.05$ (63 Jy)&$-0.59\pm 0.05$ (28 Jy)&$-2.0\pm 0.1$ (100 Jy)&5, 6, 7\\
    Parallax (mas) &\multicolumn{2}{c}{$750.81\pm0.38$}& $9.99\pm 0.20$&3, 8\\
    Distance (pc) &\multicolumn{2}{c}{1.33}&100&3, 8\\
    Proper Motion ($\mu_\alpha, \mu_\delta$, mas yr$^{-1}$) &\multicolumn{2}{c}{($-3639.95\pm 0.42$, $+700.40\pm 0.17$ )}&($-231.04 \pm 0.19$, $-26.39\pm  0.26$)&3, 8\\
    R.A. &14:39:26.155&14:39:25.9421&12:17:33.620 &3, 8, 9 \\
    Decl. &$-$60:49:56.287&$-$60:49:51.334&$-$67:57:39.072&3, 8, 9 \\
    \enddata
    \tablerefs{(1) \citet{Valenti2005}; (2) \citet{Ducati2002}; (3) \citet{Akeson2021}; (4) \citet{Engels1981};  (5) from angular size of \acenA, $\Theta=8.502$ mas combined with a Kurucz-Castelli model with $T_{\rm eff}=5795$ K and log $g=4.312$ dex \citep[cgs units;][]{Kervella2017};
    (6) From fit to Kurucz model atmosphere using VOSA SED utility \citep{Engels1981,Ducati2002,Castelli2003,Bayo2008}; (7) \citet{LRS}; (8) \citet{Gaia2016}; (9) for \emus Epoch 2016.0 (Gaia DR3) and for \acen Epoch 2019.5 \citep{Akeson2021}.}
\end{deluxetable*}

\subsection{Observational Strategy}

We elected to observe with MIRI and its Four Quadrant Phase Mask Coronagraph \citep[4QPM;][]{Rieke2015, Wright2015, Boccaletti2022} centered at 15.5 \mum\ (the F1550C filter) for a number of reasons: (1)~favorable star-planet contrast ratio for the 200--350~K temperatures expected for a planet heated by \acenA at 1--3 au; (2)~low susceptibility to the effects of wavefront drift at this long wavelength; and (3)~the reduced brightness of background objects with typical stellar photospheres. However, despite these advantages, the \acenAB system presents numerous challenges in planning and executing coronagraphic measurements with JWST at any wavelength.

\begin{itemize}

\item The presence of \acenB only 7\arcsec--9\arcsec\ away from \acenA puts the full intensity of this bright, [F1550C] $\sim -0.59$ mag star in the focal plane at a position that cannot be attenuated. We developed a strategy ($\S$\ref{sec:AcenB}) to place \emus at the position \acenB would occupy (unocculted) during the observation of \acenA (occulted). This observation would provide a PSF reference to mitigate the effects of \acenB.

\item The selection of a reference star is complicated by the requirement that it be both comparably bright to \acenA and have similar photospheric properties in the F1550C waveband. 

\item The moment-by-moment position of \acenA is the result of a complex interplay of its high proper motion and parallax (as calculated for the location of JWST at the epoch of observation), and of the orbital motion of \acenA and \acenB about their common center of mass \cite[see][and Table~\ref{tab:stars}]{Akeson2021}.

\item With [F1550C] $\sim -1.5$ mag, \acenA is too bright for direct target acquisition (TA) with MIRI, necessitating a blind offset from a nearby star with an accuracy of $<$10 mas to avoid degradation of the coronagraphic contrast \citep{Boccaletti2015}. The chosen reference star \emus ($\S$\ref{sec:refstar}) is similarly too bright for direct TA, also necessitating a blind offset. 

\item Offset stars must be of sufficient astrometric accuracy, be as close as possible to \acenA or \emus, but not affected by diffraction or other artifacts from the target stars, and be of sufficient brightness to be readily detectable in a short TA observation at F1000W (Figure~\ref{fig:TAobs}).

\item The time of observation should minimize the change in solar aspect angle between target and reference star observations and thus minimize the change in the telescope's thermal environment.

\item Finally, all MIRI coronagraphic observations using the 4QPM are affected by excess background radiation appearing around the quadrant boundaries referred to as the ``Glow Sticks" \citep{Boccaletti2022}. 
\end{itemize}

\subsection{Planned Observation Sequences \label{sec:sequences}}

\begin{figure*}[!htb]
\centering
\includegraphics[width=0.5\textwidth]{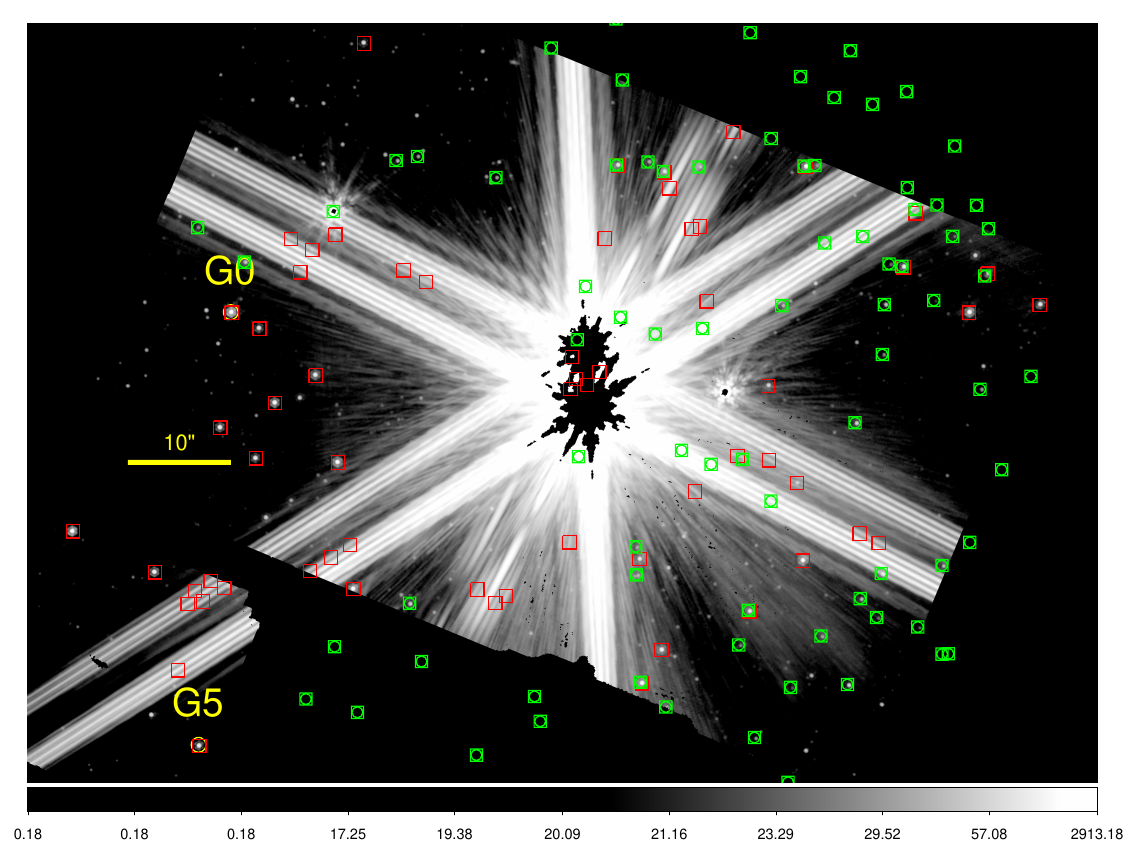}\includegraphics[width=0.5\textwidth]{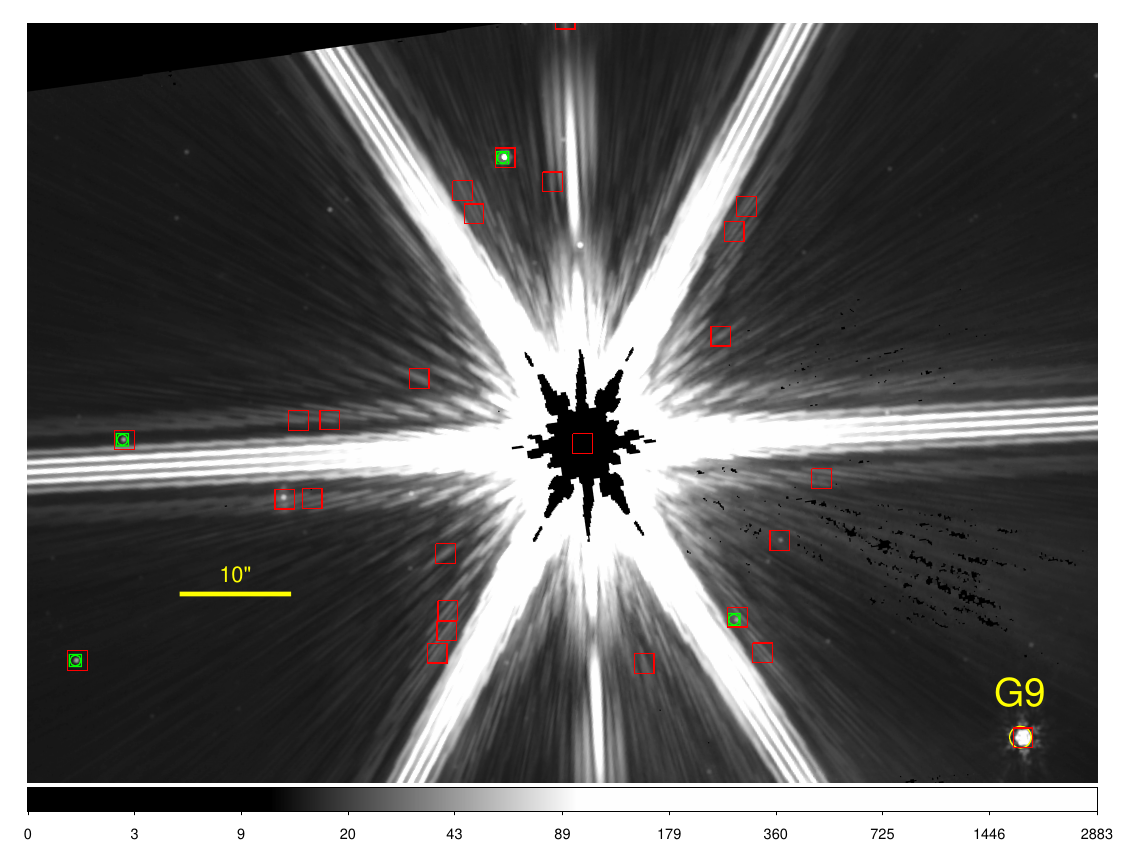}
\caption{\emph{Left:} F1000W image of \acenAB showing Gaia stars (green boxes) and MIRI detections (red boxes). The stars labeled $G0$ and $G5$ were used for target acquisition of \acenA. \emph{Right:} similar F1000W image for \emus. The star labeled $G9$ was used for target acquisition of \emus. \label{fig:TAobs}}
\end{figure*}

The above considerations led to the observational sequences described below and detailed in Appendix~\ref{app:obs-details}. Based on in-flight performance, JWST can place both the target and reference star with an accuracy of $\sim$5--7~mas (1$\sigma$, each axis) behind the MIRI/4QPM. To provide diversity in determining the PSF for post-processing, we selected a 9-point dither pattern for observing the reference star. The multiple reference PSF observations improve the ability of post-processing algorithms to remove residual stellar speckles and help to mitigate wavefront error (WFE) drifts over the 32~hr duration of the entire sequence. The measurement strategy was as follows:
\begin{enumerate}
\item Offset from a Gaia star ($G9$ in Figure~\ref{fig:TAobs}) to place \emus at the center of the F1550C coronagraphic mask and make a 9-point dithered set of image observations \added{of the reference star behind the MIRI/4QPM.} This is followed by observations of a background field to subtract the Glow Stick.
\item Place \emus at the detector location that \acenB\ would occupy in the Roll \#1 observation to help mitigate speckles from the unocculted star at the position of \acenA.
\item Offset from a Gaia star ($G0$ or $G5$ in Figure~\ref{fig:TAobs}) to place \acenA\ at the center of the F1550C coronagraphic mask at the Roll \#1 V3PA\footnote{V3PA is the position angle (PA) of the V3 reference axis eastward relative to north when projected onto the sky.} angle for a sequence of 1250 images, followed by observations of a background field to subtract the Glow Stick.
\item Repeat the \acenA\ sequence (\#3) at a second V3PA angle (Roll \#2).
\item Repeat the \emus 9-point dither sequence (\#1) at the mask center. 
\item Repeat the off-axis \emus observation sequence (\#2) but at the detector position of \acenB in the Roll \#2 observation.
\end{enumerate}

\begin{figure*}[!htb]
\centering
\includegraphics[width=\textwidth]{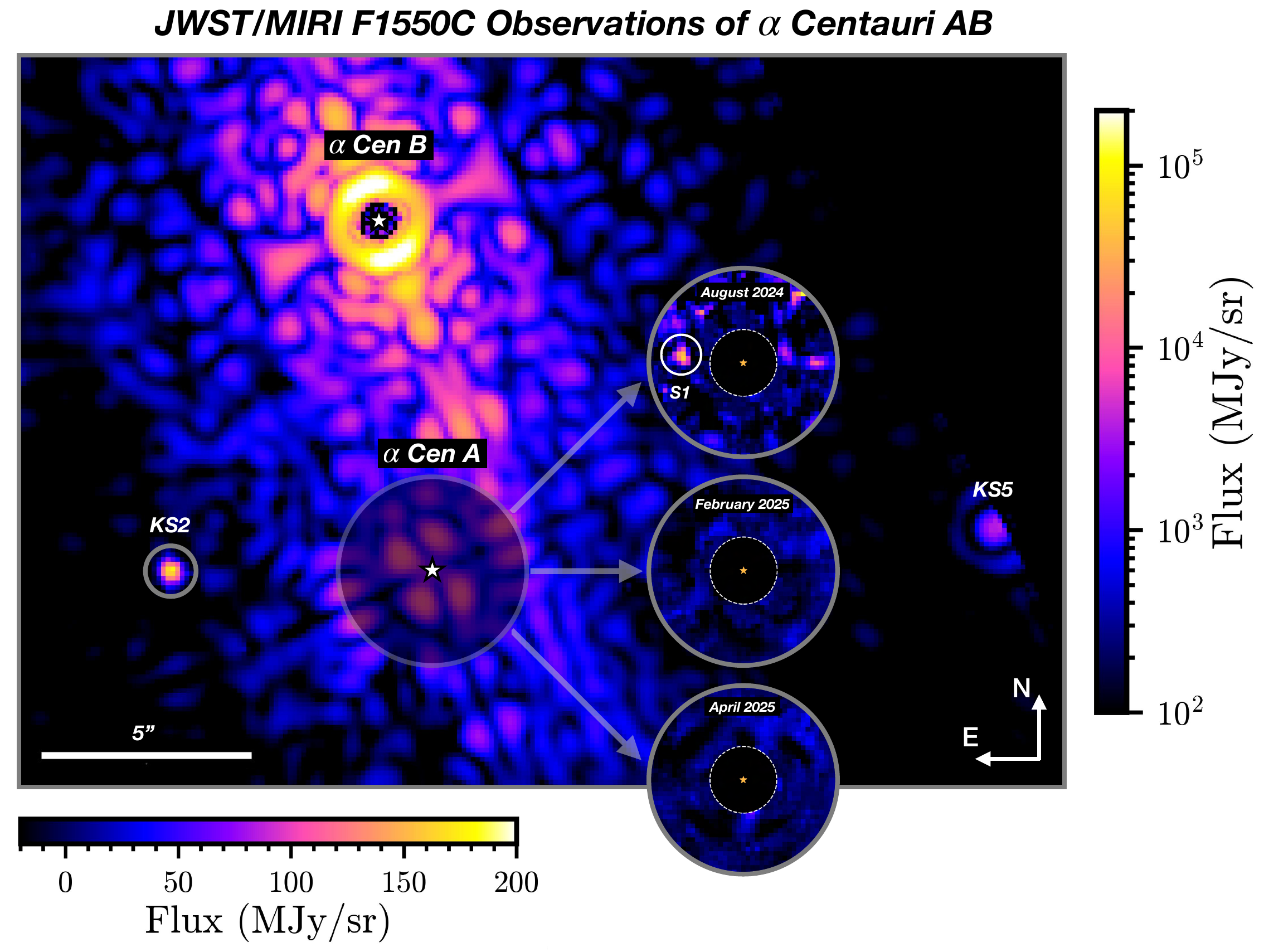}
\caption{JWST's view of the \acenAB system. Shown above is a background-subtracted Stage 2b F1550C image of the \acenAB\ system from August 2024. The image is oriented North up and East left. The white stars denote the approximate positions of \acenB, saturated near the top of the image, and \acenA\ in the lower part of the image hidden behind the F1550C mask. The right colorbar (logarithmically scaled) is associated with this image. At the edge of the detector, to the West of \acenA, is a known background source KS5 \citep{Kervella2016}. To the East of \acenA is known background source KS2 \citep{Kervella2016}, shown as an inset, as it is only detected after performing PSF subtraction (no colorbar shown for the inset, scaled linearly between $-$5 and 50 MJy/sr). An $\approx$2\farcs75 radius region around \acenA is shaded and mapped to three PSF-subtracted images, one for each observation epoch. Candidate \sone is seen only in August 2024. The bottom colorbar (linearly scaled) is associated with the three PSF-subtracted images. For reference, 1~MJy/sr $\approx 4.6 \times 10^{-6}$ Jy/AiryCore, where AiryCore is defined as the area of a circular aperture of diameter 1~FWHM ($\approx 0\farcs5$).\label{fig:AcenTotal}}
\end{figure*}

\subsection{Executed Observation Sequences}
The observations of \acenA (Cycle 1 GO, PID \#1618; PI: Beichman, Co-PI: Mawet) were initiated in August~2023, but were unsuccessful due to target acquisition and offset failures. A sequence of short test images was obtained in June and July~2024 to validate the target acquisition strategy. Specifically, in July~2024, we executed \#1 without dithering and \#3 with fewer integrations.\footnote{No \emus reference star observation was acquired at the detector position of \acenB in the July 2024 test observations. This severely compromised the quality of PSF subtraction. Hence, these observations are not presented.} Following successful execution of the test program, we conducted our full-set of science observations in August 2024. We successfully executed steps \#1, \#4, and \#6 (\#2 was not part of the sequence planned for this observation). However, the first roll on \acenA\ (\#3) and the second \emus observation (\#5) were unsuccessful due to guide star failures.

Based on results from the August 2024 data, the STScI Director's Office approved a follow-up Director's Discretionary Time (DDT) program (PID \#6797; PI: Beichman, Co-PI: Sanghi). The complete two roll sequence with associated reference star observations (\#1--\#6) was attempted in February 2025 as part of this DDT program, but due to a telescope pointing anomaly, the first \acenA\ roll (\#3) was not executed. All other observations were successful.

The STScI Director's Office approved a second follow-up DDT program (PID \#9252; PI: Beichman, Co-PI: Sanghi), which resulted in the successful execution of a full two roll sequence in April 2025. A summary log of all successful observations is provided in Table~\ref{tab:MIRIObs}. In all cases, the accuracy of the offsets from Gaia stars was consistent with the expected initial pointing accuracy ($1\sigma$, 5--7 mas), the offset accuracy ($1\sigma$, 1.5 mas) and the line-of-sight jitter ($1\sigma$, 1.5 mas) during the observing sequence at each position. The February 2025 Roll 2 and April 2025 Roll 1 observations showed offsets of $>$~10 mas from the 4QPM center or from the Eps Mus dither pattern and were thus of lower quality \citep[see dither map in][]{Aniket2025}. 

\begin{deluxetable*}{cc|cc|cccc}
    \tabletypesize{\scriptsize}
    \tablecaption{Observations of a Candidate Planet Orbiting \acenA \label{tab:S1props}}
    \tablehead{
    ID	& Epoch& ($\Delta \alpha$, $\Delta \delta$)& ($\rho, \theta)$&Wavelength&Flux&Contrast &S/N\\
    & & (\arcsec)&(\arcsec, $^\circ$)&(\mum)&(mJy)& to \acenA&}
    \startdata
    $C1$ &June 1, 2019&($-0.64, -0.56)\;\pm$ 0.05&(0.85 $\pm$ 0.05, 228.9 $\pm$ 3.3)&11.25&1.2 $\pm$ 0.4&$0.8\times10^{-5}$&3\\ \hline
    $S1$&August 10, 2024&(1.50, 0.17) $\pm$ 0.13&(1.51 $\pm$ 0.13, 83.5 $\pm$ 4.9)&15.5&3.5 $\pm$ 1.0&$5.5\times10^{-5}$&4--6\\
    \enddata
    \tablecomments{Observations of $C1$ were obtained in 2019 by the VLT/NEAR experiment \citet{Wagner2021}. Position angle ($\theta$) is measured East of North.}
\end{deluxetable*}

\section{Results\label{sec:results}}
\subsection{Summary of Data Reduction \label{sec:data}}

Paper II \citep{Aniket2025} describes in detail the initial pipeline processing, PSF subtraction techniques for both \acenA and \acenB, source identification steps, the photometry and astrometry estimation procedures, and detection sensitivity analysis. Here, we provide a short summary. Level 0 data products were downloaded from MAST, processed for best up-the-ramp calculation of source brightness and bad pixel rejection \citep{Brandt2024, Carter2025}, and post-processed to remove the residual stellar diffraction from \acenA and \acenB. We assembled distinct reference PSF libraries for each epoch consisting of the individual 400 frames (per dither position) of each 9-pt SGD observation of \emus behind the 4QPM and the individual 1250 frames of \emus at the unocculted position of \acenB (for a given roll) obtained at the corresponding epoch. We employed reference star differential imaging (RDI) and jointly subtracted \acenAB from the 1250 \acen integrations using the principal component analysis-based Karhunen-Lo\`eve Image Processing algorithm \citep{Soummer2012}. Signal-to-noise ratio maps were generated to search for point sources \citep{Mawet2014} and extended emission (custom method), and assess detection significance.

\subsection{Detection of a Point Source Around \acenA}

A comprehensive search of the $\sim$3\arcsec\ region around \acenA revealed a single point-like source in the August~2024 data, \sone (Figure~\ref{fig:AcenTotal}). The source was detected $\approx$1\farcs5 East of \acenA at a S/N between 4--6 (corresponding to a 3.3--4.3$\sigma$ Gaussian significance for the equivalent false positive probability, see Paper II) with a flux density of $\approx 3.5$ mJy (Table~\ref{tab:S1props}). The contrast of \sone with respect to \acenA\ in the F1550C bandpass is $\approx 5.5\times10^{-5}$. \sone is not recovered in the February and April 2025 observations (Figure~\ref{fig:AcenTotal}). At wider separations, in all three epochs, we identified two objects denoted KS2 and KS5 that are known from deep 2 \mum\ VLT/NACO imaging to be background stars \citep{Kervella2016}. In the August 2024 data, KS2 is seen $\approx6$\arcsec\ East of \acenA, after post-processing, exactly in the position expected for a distant, low proper motion star (Figure~\ref{fig:AcenTotal}). The bright object KS5 ($K_s\sim 7$ mag) is detected just off the edge of the coronagraphic field and will eventually pass within a few mas of \acenA \citep[mid-2028;][]{Kervella2016}. 
 
Paper II \citep{Aniket2025} discusses the robustness of the detection of \sone and with the help of several tests, presents reasonable evidence that \sone is a celestial signal, as opposed to an image artifact. Three primary artifact scenarios are shown to be unlikely:

\begin{itemize}
 \item \textit{\sone is not likely a short-lived detector artifact in the \acenAB integrations.} \sone was independently detected in multiple subsets of the full 1250 frame integration sequence \added{(Section 4.2.3, Paper II)}. Additionally, there was no evidence for transient ``hot pixels" in the data, centered on \sone.

 \item \textit{\sone is not likely a PSF-subtraction artifact from the \emus coronagraphic reference images.} \sone was detected in post-processing analyses performed by iteratively excluding each one of the nine dither positions (``leave-one-out" analysis, \added{Section 4.2.4, Paper II)}.

 \item \textit{\sone is not likely a PSF-subtraction artifact from imperfect subtraction of \acenB.} \sone is well matched to the expected PSF profile and behaves differently with respect to changes in subtraction parameters from another point-like object ($A1$) identified as an artifact from \acenB. \added{$A1$'s signal disappears both when the number of azimuthal subsections and number of principal components increases. \sone's signal persists in both cases (Section 4.2.2, Paper II).} 
\end{itemize}

To assess whether \sone is physically associated with \acenA, we address whether it could be either a background or foreground (Solar System) object. Multiple arguments rule out these scenarios:

\begin{itemize}

\item First and most conclusively, the JWST data themselves provide definitive evidence against the hypothesis that \sone is a background object. No point source counterparts to \sone are detected at the expected location for a background source in the February~2025 and April~2025 observations (Figure~\ref{fig:bkgnd}). See Paper II \citep{Aniket2025} for further details.

\item Archival images taken by Spitzer/IRAC \citep{Rieke2004}, 2MASS, and VLT/NACO \citep{Kervella2016} when \acenA was up to one arcminute away from its current position do not show any sources at the \sone position. We also considered the effects of interstellar extinction on background source detectability in archival imaging. Extinction maps from Planck and stellar data along the line-of-sight toward \acen\ provide a range $20<A_V {\rm (mag)} <40$ \citep{Planck2016, Zhang2022}, making more extreme $A_V$ values unlikely.\footnote{Planck Extinction maps:\\ \url{https://irsa.ipac.caltech.edu/data/Planck/release\_2/all-sky-maps/maps/component-maps/foregrounds/COM\_CompMap\_Dust-DL07-AvMaps\_2048\_R2.00.fits}} As shown in Figure~\ref{fig:AV}, if \sone were a reddened star \citep[an M0III Kurucz model with $T_{\rm eff}=3800$ K is shown;][]{Buser1992} or a normal star-dominated galaxy, its emission would be 4 to 25 times brighter at IRAC wavelengths than at F1550C and would have been detectable by Spitzer, or in the deep NACO $K$-band image. This argument applies to any stellar temperature, since at these wavelengths the emission is approximately Rayleigh-Jeans.

\item Figure~\ref{fig:AV} also shows the spectral energy distribution for a non-photosphere dominated galaxy, the prototypical starburst galaxy or ULIRG, Arp~220, at zero redshift \citep{Polletta2007}. Such an object could have escaped detection in the archival datasets, but the probability of chance alignment with an extragalactic background object is extremely low based on source-counting studies in the MIRI broadband filters. \citet{Stone2024} find that the background density of $F_\nu(F1500W)\gtrsim$ 1~mJy sources is $<$0.05~arcmin$^{-2}$, corresponding to a chance alignment likelihood $<4\times10^{-4}$ within a 3\arcsec\ field-of-view. 

\begin{figure}[!t]%
 \centering
 \includegraphics[width=0.45\textwidth]{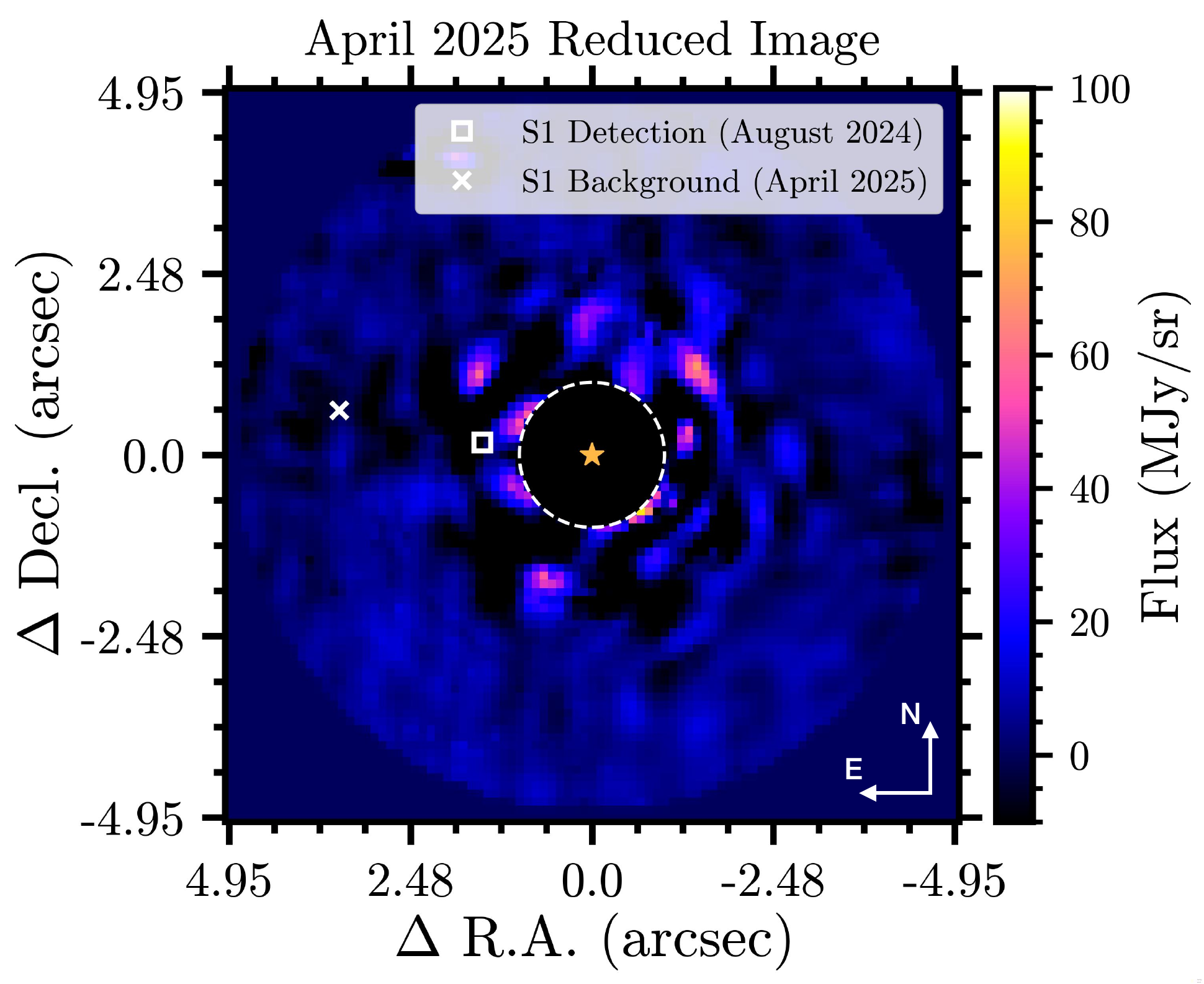}
 \caption{PSF-subtracted image centered on \acenA for the April 2025 observations, which provides the longest time-baseline to test the stationary background source hypothesis. A square marks the location where \sone was detected in the August 2024 epoch. A cross marks the expected location of \sone if it were a fixed background object and showed apparent motion with respect to \acenA due to the star's parallactic and proper motion. No source is detected at this location.
 \label{fig:bkgnd}}
\end{figure}
 
\begin{figure}[t]%
 \centering
 \includegraphics[width=0.5\textwidth]{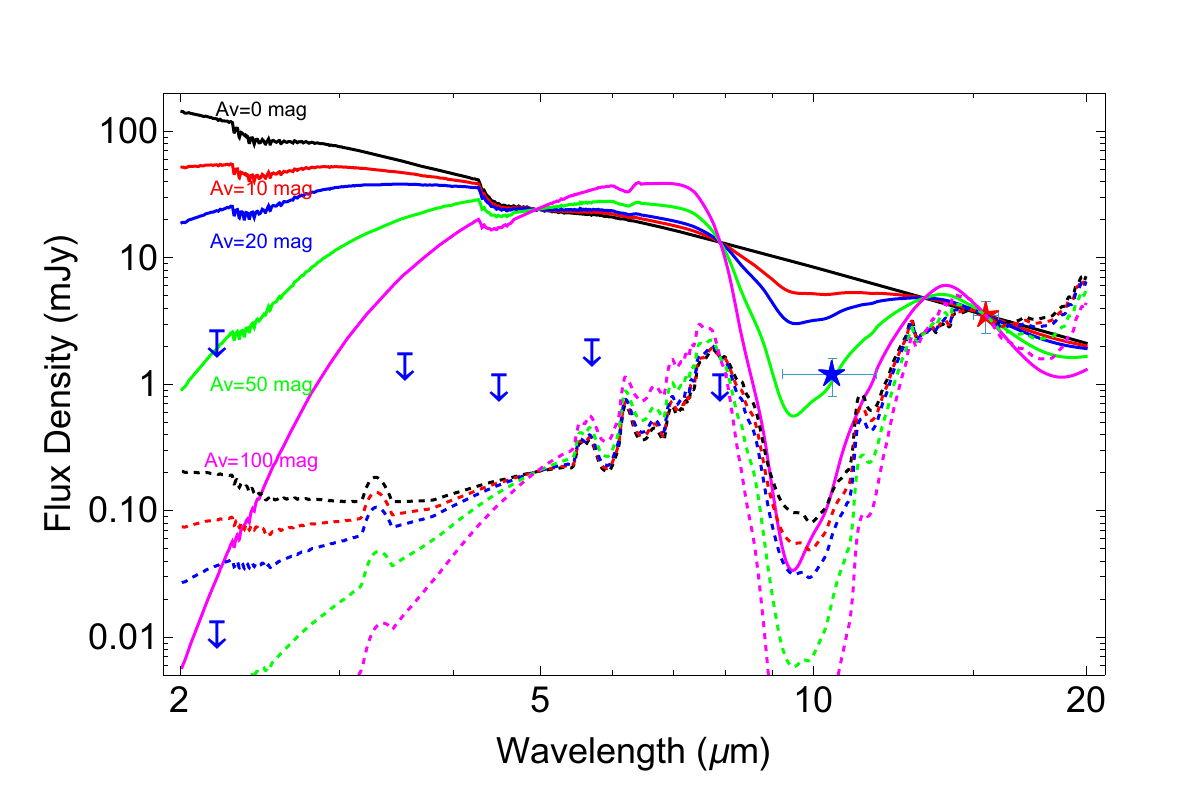} 
 \caption{Limits from archival imaging at \sone's position. The solid, color-coded lines show a photospheric model for an M0III star ($T_{\rm eff} = 3800$~K) reddened by increasing levels of extinction, all normalized to 3.5 mJy at 15.5 \mum\ (red star). The blue star denotes the flux density of the object denoted $C1$ detected by the VLT/NEAR experiment \citep{Wagner2021}. The dashed lines show the spectral energy distribution of a typical star-burst galaxy or ULIRG (Arp~220) similarly reddened. 
 Upper limits at the position of \sone come from observations at earlier epochs with Spitzer/IRAC (3--8~\mum), 2MASS, and NACO \citep{Kervella2016}. 
 \label{fig:AV}}
\end{figure}

\begin{figure*}[!htb]
\centering
\includegraphics[width=\linewidth]{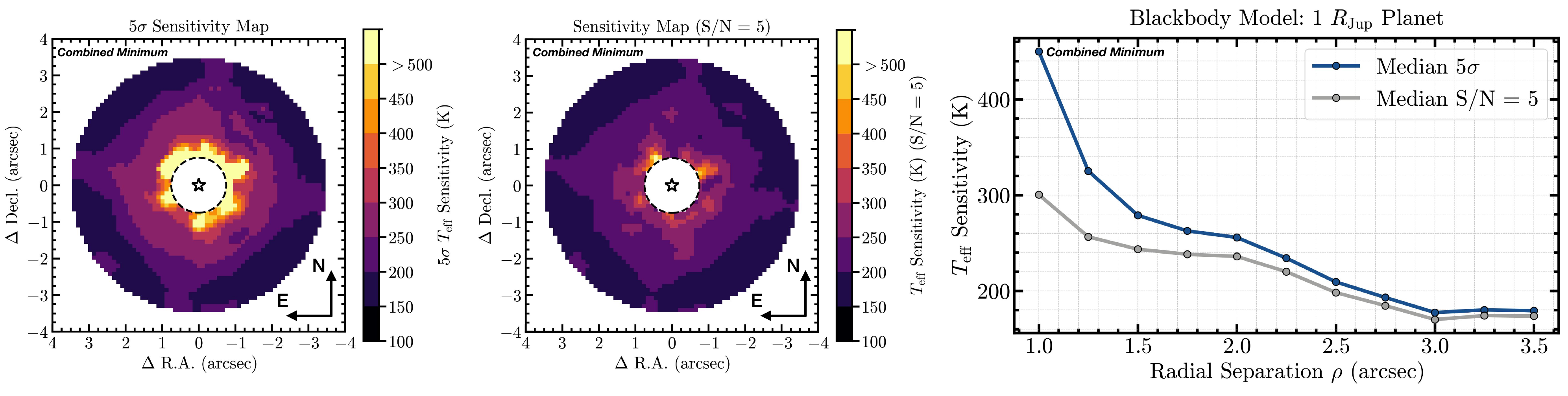}
\caption{\emph{Left:} two-dimensional 5$\sigma$ planet effective blackbody temperature sensitivity map, combined across all epochs by selecting the best sensitivity (``combined minimum", see Paper II), in sky coordinates (North up, East left). The central region ($<$ 0\farcs75, or $<$ 1.5 FWHM, radial separations) is masked (poor detectability). A discrete colormap is chosen to highlight the different sensitivity zones across the image. \emph{Center:} same as the left panel for a S/N = 5 detection threshold. \emph{Right:} 5$\sigma$ and S/N = 5 median planet effective  blackbody temperature curves combined across all epochs.}
\label{fig:teff-sensitivity}
\end{figure*}

\item We eliminate the possibility that \sone is a foreground Solar System object in a number of ways. An inner main belt asteroid (MBA) at 2.2 AU with a typical temperature of 200 K would have to have a diameter of $>$ 2 km to emit $\sim$ 3 mJy at 15.5 \mum. Such objects are extremely rare, $<10^{-4}$ brighter than 3 mJy at 12 \mum\ in a 5\arcmin$\times$5\arcmin\ field at \acenA's ecliptic latitude of $\beta= -42^\circ$ \citep{Brooke2003}. Furthermore, the completeness for such large MBAs is over 90\% and the Minor Planet Catalog shows no known objects at the position of \acenA at the August epoch\footnote{\url{https://minorplanetcenter.net/cgi-bin/checkmp.cgi}}. Finally, as described in Paper II, there is no angular motion seen between the beginning and the end of $\sim$2.5 hour MIRI observation compared to the expected $>$10~arcsec/hr motion for an MBA at the solar elongation of our observations, $\approx 100^\circ$ \citep{Brooke2003}. 
\end{itemize}

Based on all of the above considerations, we pursue the hypothesis that \sone is a planet physically associated with and in orbit around \acenA, as opposed to an artifact or an astrophysical contaminant, and investigate its properties. Given that \sone is only detected in a single roll observation in August~2024, we emphasize that it is, at the moment, a \emph{planet candidate}. Additional sightings of \sone are required with JWST, or other upcoming facilities, to confirm what would be ``$\alpha$~Cen~Ab".

\subsection{Planet Detection Limits with JWST/MIRI}
We assess our sensitivity to planets around \acenA across all three epochs of MIRI F1550C observations. Paper II \citep{Aniket2025} presented the calculation of 2D flux/contrast sensitivity maps. Here, we use the ``combined minimum" map from Paper II, which corresponds to the 2D sensitivity map with the best flux sensitivity across all three epochs at each location where the PSF injection-recovery test was performed. We convert the 5$\sigma$ and S/N = 5 flux sensitivities (see Paper II) to effective temperatures ($T_{\rm eff}$) assuming a blackbody model and a typical planet radius of 1~$R_{\rm Jup}$ (for smaller planets, the minimum detectable planet $T_{\rm eff}$ increases). The results are shown in Figure~\ref{fig:teff-sensitivity}. The MIRI F1550C observations are sensitive to $T_{\rm eff}\approx$ 250 K (1~$R_{\rm Jup}$) planets between 1\arcsec--2\arcsec\ for a S/N = 5 detection threshold. Planets colder than 200~K can be detected at wider separations ($>$~2\farcs5). We note here that more realistic planet atmospheric models may have a higher brightness temperature (and thus flux) in the F1550C bandpass relative to the effective blackbody temperature assumed here (see \S\ref{sec:AtmModels}, for example). This would improve the detectability of colder planets at smaller separations than presented here. 

\subsection{Limits on Extended Emission around \acenA \label{sec:zodi}}

\begin{figure*}[!htb]
\centering
\includegraphics[width=\linewidth]{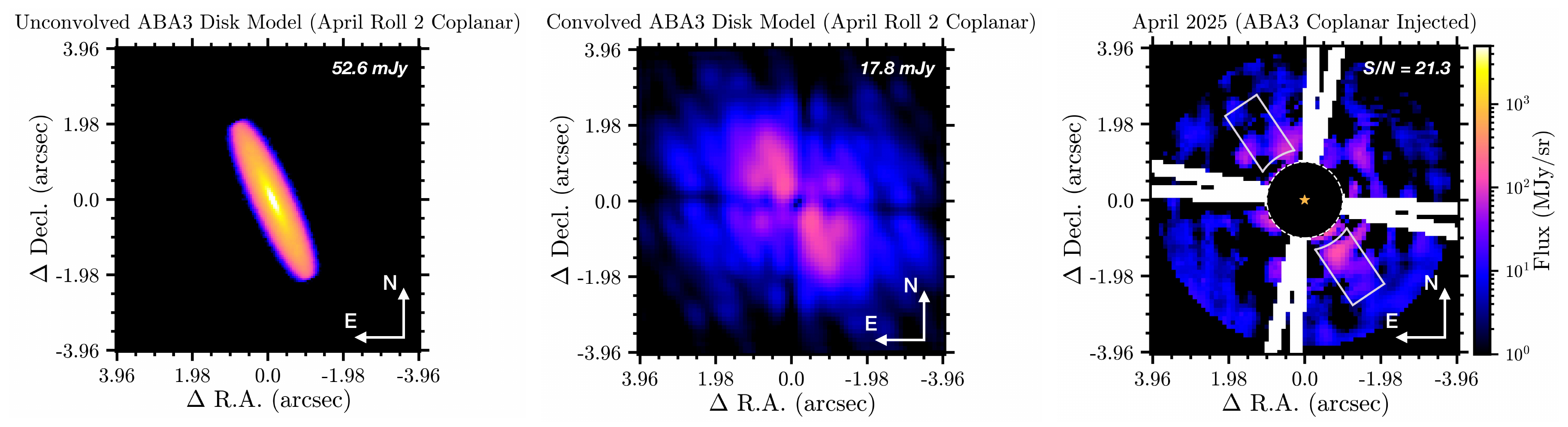}
\caption{\emph{Left:} unconvolved asteroid belt analogue-3 (ABA-3) exozodi model coplanar with the \acenAB binary. \emph{Center:} the exozodi model in the left panel after PSF convolution (for the April 2025 observation orientation). \emph{Right}: PSF-subtracted image showing the recovery of the ABA 3 exozodi model injected in the raw \acenAB dataset from the April 2025 observations. The regions of the image affected by the MIRI 4QPM transition boundaries at each roll are masked. The aperture that yielded the highest S/N ($=$ 21.3) detection for the injected disk is shown (see Paper II for details). All images are plotted on a common logarithmic color scale.}
\label{fig:zodi_convolve}
\end{figure*}

\citet{Beichman2020} predicted that JWST's ability to resolve the habitable zone around \acenA would result in unprecedented sensitivity to warm dust---the analog of the thermal emission from dust generated by collisions the asteroid belt in our solar system. We show below that the current observations have not only met but exceeded those expectations with limits as low as a few times the solar system brightness levels at 15.5~\mum.

\subsubsection{Exozodi Model Description}
As described in Paper II, we injected a number exozodiacal cloud models into the processed \acen datacubes to set limits on extended ``exozodiacal" emission around \acenA. In the case of \acenA, stable orbits—and thus significant dust buildup—are limited to within the stable zone, approximately $<$3~au from the star. The Solar System zodiacal cloud is therefore a poor proxy for a potential exozodi around \acenA. For a more realistic representation, we consider the scenario of an asteroid belt analogue (ABA) located between 2--3 au, where dust is produced in collisions, which is then transported inward under Poynting-Robertson (PR) drag. This scenario is captured by the semi-analytical model of \citet{Rigley2020}, which combines approaches to determine (1)~the size distribution arising in a planetesimal belt under collisions and PR drag loss \citep{Wyatt2011}, and (2)~how it evolves interior to the belt under further collisional and drag-induced evolution \citep{Wyatt2005}. The resulting optical depth distribution (across grain sizes and disk radii) is then combined with the particles' thermal emission properties, determined using Mie theory, to compute the disk's surface brightness distribution. 

Following the approach of \citet{Sommer2025}, we generate astrophysical scenes of the inclined, edge-on disks from the respective surface brightness profiles, and convolve them with the spatially varying PSF of the F1550C coronagraphic filter \citep[modeled using \texttt{STPSF},][]{perrin14}, before injecting them into the MIRI datacubes. An example of an exozodi scene, before and after PSF convolution, alongside a PSF subtracted image obtained after model injection, is shown in Figure~\ref{fig:zodi_convolve}. Note that all exozodiacal disks considered here are assumed to be coplanar with the \acenAB\ plane, which is a reasonable assumption for potential circumstellar debris disks in binaries (see Section~\ref{sec:exozodi_in_binaries}), although the invariable plane about which the orbit precesses could also be affected by the gravity of massive planets in the system.

\begin{deluxetable*}{lcccccccccccc}
    \centering
    \tabletypesize{\small}
    \tablecaption{Exozodiacal Disk Models \label{tab:ExoZodi}}
    \vspace*{2mm}
    \tablehead{
    &  \colhead{$M_\text{dust}$} 
    & \colhead{$M_\text{belt}$}
    & \colhead{$T_\text{coll,1000km}$}
    & \colhead{$F_\text{d,15}$} 
    & \colhead{$\displaystyle F_\mathrm{d,15}/F_{\star,15}$}
    &\colhead{} 
    &\colhead{} 
    &\colhead{}  
    & \colhead{$\displaystyle L_\mathrm{d}/L_\star$}
    & 
    &
    & \\[-1mm]
    \colhead{Model}
    &\colhead{($10^{-8}\,$\mearth)}
    &\colhead{($M_\text{MAB}$)}
    &\colhead{(Gyr)}
    &\colhead{(mJy)}
    &\colhead{ $\times 10^{-4}$}
    & \colhead{$\frac{F_\text{d,24}}{\sigma_\text{24}}$}
    & \colhead{$\frac{F_\text{d,70}}{\sigma_\text{70}}$}
    & \colhead{$\frac{F_\text{d,100}}{\sigma_\text{100}}$} 
    &\colhead{ $\times 10^{-7}$}
    & \colhead{ $Z_L$ }
    &\colhead{$Z_\Sigma$}
    & \colhead{ S/N }} \vspace*{2mm}
    
    \startdata
    ABA 1  & 20  & 28  & 0.88 & 596 & 69  & 1.04 & 0.29 & 0.76 & 94   & 58   & 84    &78.7  \\
    ABA 2  & 4   & 5.6 & 4.40 & 156 & 18  & 0.27 & 0.07 & 0.17 & 27   & 17   & 29   &52.6  \\
    ABA 3  & 2   & 2.8 & 8.80 & 53  & 6.1 & 0.09 & 0.04 & 0.09 & 8.3  & 5.1  & 8.4   &21.3  \\
    ABA 4  & 1.2 & 1.7 & 14.5 & 20  & 2.3 & 0.03 & 0.02 & 0.06 & 3.0  & 1.8  & 3.0   &5.8  \\
    1-zodi & \nodata   & \nodata   & \nodata    & 8.8 & 1.0 & 0.02 & 0.02 & 0.06 & 1.5  & .94  & 1.0  & $-$0.7 \\ \hline
    \enddata
    \tablecomments{\small
    Exozodi models used for injection and derived quantities. Columns: $M_\text{dust}$, ABA model dust mass parameter (belt mass up to $1\,$cm grain size);
    $M_\text{belt}$, ABA model belt mass up to largest planetesimal ($1000\,\text{km}$) in units of Solar System main asteroid belts ($M_\text{MAB}=4\times10^{-4}\,$\mearth); 
    $T_\text{coll,1000km}$, collisional lifetime of largest planetesimal;
    $F_\text{d,15}$, total disk flux at 15.5~$\mu$m;
    $F_\mathrm{d,15}/F_{\star,15}$, fractional disk flux at 15.5~$\mu$m;
    $\frac{F_\text{d,24}}{\sigma_\text{24}}$, photometric significance (phot.\@ sig.\@) for MIPS24;
    $\frac{F_\text{d,70}}{\sigma_\text{70}}$, phot.\@ sig.\@ for PACS70;
    $\frac{F_\text{d,100}}{\sigma_\text{100}}$, phot.\@ sig.\@ for PACS100 \citep[all uncertainties from][]{Wiegert2014};
    $L_\mathrm{d}/L_\star$, fractional disk luminosity;
    $Z_L$, zodi level by fractional luminosity;
    $Z_\Sigma$, zodi level by Earth Equivalent Insolation Distance (EEID) surface density;
    and S/N ratios of injection recovery tests with the April 2025 dataset for the case of binary-coplanar disks.}
\end{deluxetable*}

\begin{figure}[t!]
\centering
\includegraphics[width=\columnwidth]{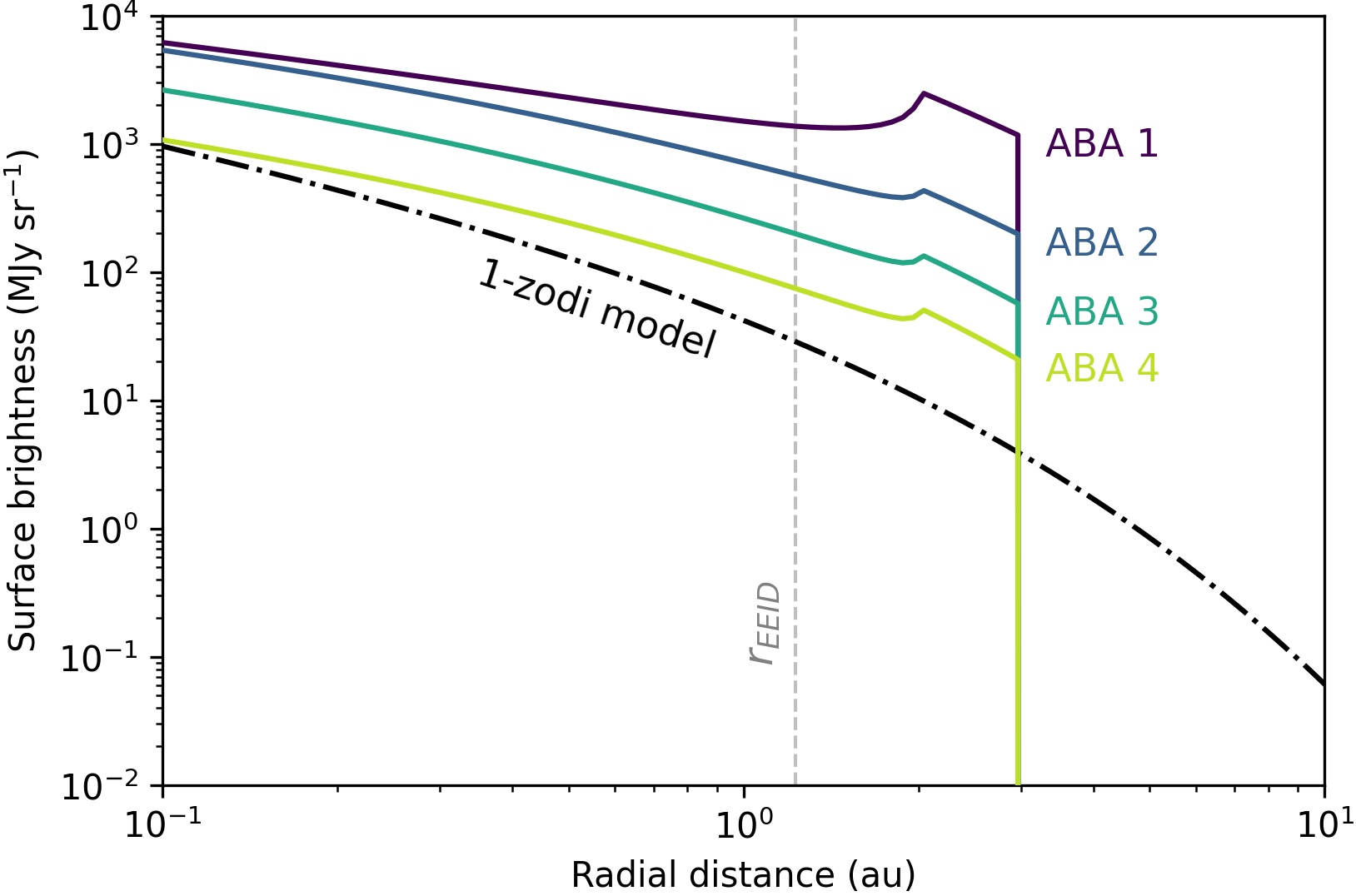}
\caption{Surface brightness distribution of injected zodi models. ABA-scenario zodis for different belt masses are represented by solid lines. Our fiducial 1-zodi model based on the \citet{Kelsall1998} model is represented by the dash-dot line.}
\label{fig:zodi_surfbright}
\end{figure}

For the model parameters, we further assume a belt opening angle of 5\degree, a size-independent catastrophic disruption threshold of $10^7$ erg g$^{-1}$ (representing the grains' collisional strength), as well as a grain composition of 1/3 amorphous silicates and 2/3 organic refractories by volume, which determines their Mie-theory-derived optical properties. Four models of different belt dust masses, ABA~1--4, are considered, for which derived quantities are summarized in Table~\ref{tab:ExoZodi}. The resulting surface brightness profiles of the different models are compared in Figure~\ref{fig:zodi_surfbright}. 
Here, the ABA-1 model differs from the other models in having a sharper decline of surface brightness at the inner belt edge. This is because, at that mass, even small dust is effectively ground down to blowout sizes before it can migrate inward past the belt. As a result, further increases in belt mass only enhance the local brightness within the belt, while the interior regions reach saturation \citep{Wyatt2005}. 

To give an indication of the plausibility of the ABA models, we also conduct a simplified analysis of total belt mass and collisional lifetimes within the belt, assuming a canonical collisional cascade with a size distribution following a power-law slope of $-3.5$ \citep{Dohnanyi1969} extending up to a maximum planetesimal size of $1000\,\text{km}$. Comparing the collisional lifetime of the largest planetesimal to the system age of \acen ($\sim$~5~Gyr) shows that the ABA-1 model is likely not viable, since even with planetesimals as large as $1000\,\text{km}$, collisions would have inevitably eroded the planetesimal belt to below the ABA-1 level over the system age. In contrast, ABA-2 is marginally consistent with the anticipated level of erosion, while ABA-3 and ABA-4 are more readily compatible with the system's age, only requiring planetesimal masses of a few times that of the Solar System's main asteroid belt. Nevertheless, we retain the ABA-1 model in this analysis for comparison purposes.

For reference, we also include an exozodi model that is similar to the Solar System zodi, even though its radial extent is non-physical around \acenA. This fiducial ``1-zodi" model is derived from the \citet{Kelsall1998} geometrical model for the Solar System's dust cloud, which was fitted to infrared zodiacal light observations by COBE/DIRBE. Here we use the radial surface density distribution approximation derived by \citet{Kennedy2015} for the \citet{Kelsall1998} model. Using the emissivities fitted by \citet{Kelsall1998}, we calculate the disk's corresponding surface brightness distribution at 15.5~$\mu$m, which is also shown in Figure~\ref{fig:zodi_surfbright}. We then use the same image synthesis pipeline as with our ABA exozodi models, the result of which closely matches the outcome of applying the \texttt{zodipic} model---an IDL implementation of the \citet{Kelsall1998} model \citep{Kuchner2012}---around \acenA \citep[see][]{Beichman2020}, and likewise inject this 1-zodi model into the MIRI datacubes.

While our ABA exozodi models are not strictly comparable to the Solar System's zodiacal cloud in terms of geometry, it is still useful to define a “zodi level” that quantifies the dust content of the disks relative to the Solar System. Two wavelength-independent metrics for this are the total disk luminosity and the surface density within the habitable zone (HZ). The luminosity-based zodi level is defined as the ratio of the exozodi's fractional luminosity to that of the Solar System's zodiacal cloud:
\begin{equation}
Z_L = \frac{L_\mathrm{d}}{L_\star} \;\Big/\; \frac{L_\mathrm{d,SS}}{L_\odot} \;\;,
\end{equation}
while the surface-density-based zodi level is given by the ratio of the disk surface density at the Earth-equivalent insolation distance (EEID), $r_0 = \sqrt{L_\star/L_\odot}\,\text{au}$---approximately $1.23\,\text{au}$ for \acenA{}---to that of the Solar System at $1\,\text{au}$ ($\Sigma_{0,\mathrm{SS}}$):
\begin{equation}
Z_\Sigma = \frac{\Sigma_0}{\Sigma_{0,\mathrm{SS}}} \;\;,
\end{equation}
with $\Sigma_{0,\mathrm{SS}} = 7.12 \times 10^{-8}$ \citep{Kennedy2015}.
As discussed by \citet{Kennedy2015}, $Z_\Sigma$ serves as a proxy for the exozodi's surface brightness in the HZ and is therefore useful for assessing its impact on direct imaging of Earth-like planets. Both definitions of the zodi level are provided in Table~\ref{tab:ExoZodi} for our set of models.

\subsubsection{Comparison with Previous Exozodi Searches}
The results of the injection-recovery analysis of our various exozodi models, presented in Paper II \citep{Aniket2025}, are summarized by the corresponding S/N values in Table~\ref{tab:ExoZodi}. \added{We find that the three brightest injected exozodi models (ABA 1--3) are reliably recovered by our method at $\text{S/N}\!\gtrsim\!20$. The measured S/N for the faintest model, ABA-4, is $\sim$6. However, this S/N level does not constitute a reliable detection as it is consistent with the range of S/N measured in the original image, when no disk model is injected \citep[see Paper II;][]{Aniket2025}. At the low flux level of ABA-4, the image is dominated by PSF subtraction artifacts from \acenB or the 4QPM transition boundaries.} \textit{In summary, these observations are sensitive to emission from an exozodiacal cloud that is coplanar with the binary orbit at a level of $Z_L\!\approx\!5$ or $Z_\Sigma\!\approx\!8$}. This represents an unprecedented sensitivity compared with previous observations and is facilitated by the system's proximity and the model disks' near-edge-on orientation, which our recovery method is tailored to \citep[see][]{Aniket2025}.

\begin{figure}[!tb]%
 \centering
 \includegraphics[width=0.45\textwidth]{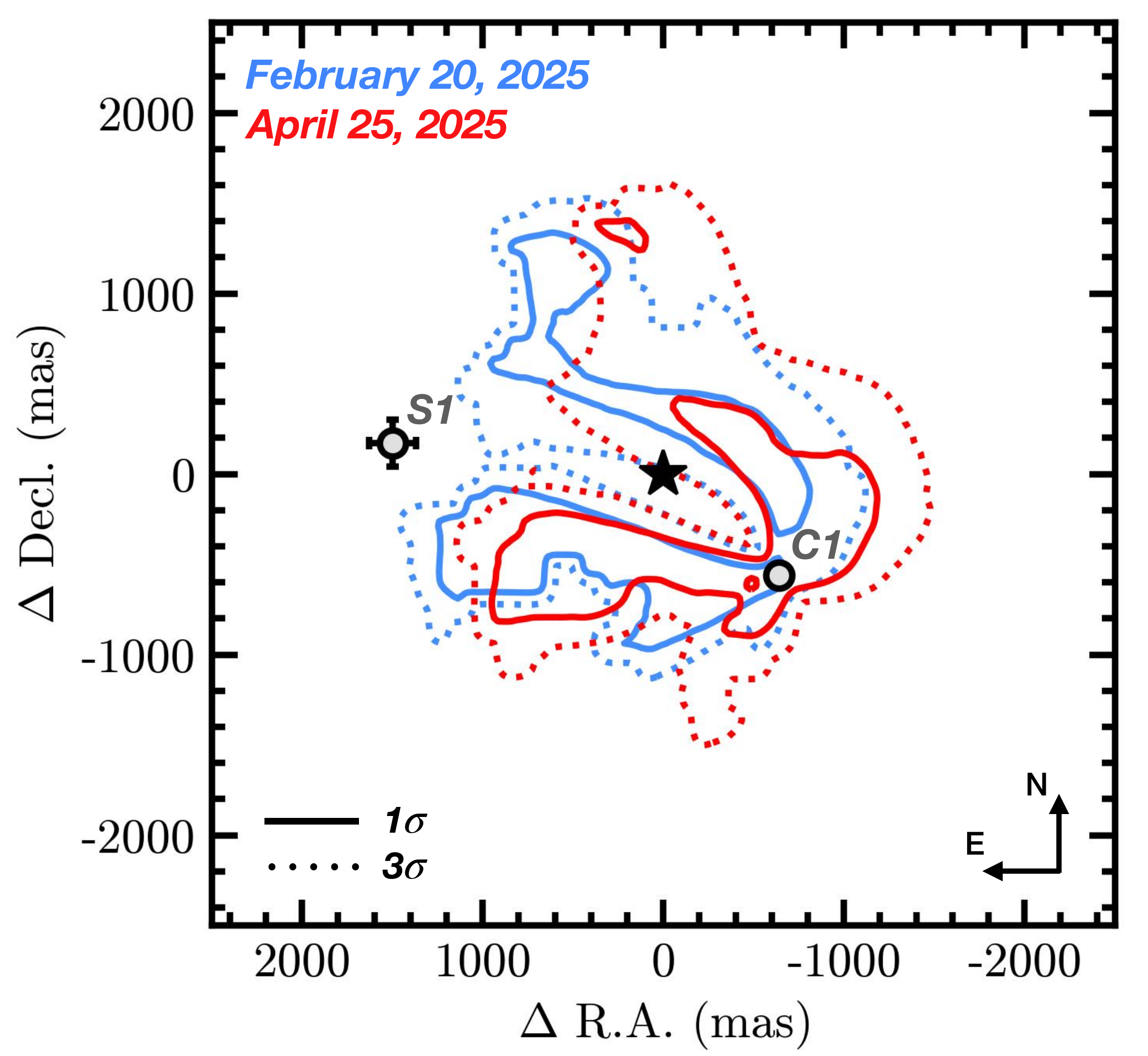} 
 \caption{1$\sigma$ (solid) and 3$\sigma$ (dotted) contours showing the sky-projected positions of the $S1$+$C1$ candidate consistent with a non-detection in the February (blue) and April (red)~2025 observation epochs.
 \label{fig:non-detect}}
\end{figure}

It is first worth acknowledging that previous photometric searches have not detected significant excess dust emission at any wavelength \citep{Wiegert2014,Yelverton2019}. While \citet{Wiegert2014} suggest a Spitzer/MIPS excess at 24~$\mu$m at $2.5\sigma$, even our brightest model disk, ABA-1, yields an excess at 24~$\mu$m of only around $1\sigma$ (see Table~\ref{tab:ExoZodi}).
Since the ABA-1 model would have easily been detected by our observations, we can confidently rule out the presence of a static (inclined) exozodi to have caused the reported feature. Excesses of our model disks in the far-IR for \textit{Herschel} PACS70 and PACS100 observations are of even lower significance, consistent with previous non-detections. This means that any circumstellar disk would have to be even more massive than ABA-1 to have shown up in previous mid- and far-IR photometric observations, indicating that our observations which could detect ABA-3 are at least 10 times more sensitive in terms of the belt's dust mass.

This improvement in sensitivity arises because photometric observations do not provide the most stringent limits on the presence of dust, due to calibration uncertainties that limit detectable excesses to typically more the 10\% of the stellar flux \citep{Beichman2005}, but with sensitivity approaching 2\% in recent studies with JWST \citep{Farihi2025}. By that metric it is clear that our resolved imaging approach is able to improve on that limit by about two orders of magnitude, since we were able to successfully suppress the stellar emission to recover the signal of the ABA-3 model which has an excess of $\sim 0.06$\% at $15.5$\,$\mu$m (see Table~\ref{tab:ExoZodi}). In principle, lower dust levels than simple photometry can be achieved using nulling interferometry to suppress the stellar emission. The largest and deepest survey of this kind was the HOSTS survey which used the LBT interferometer (LBTI) to search for exozodi emission in the habitable zone at $11$\,$\mu$m. While \acenA was not included in that survey, due to its Southern hemisphere location and its binarity, the survey results show that the best $1\sigma$ sensitivities achieved for (single) solar-type stars reached as low as $0.05$\% on the null depth, which corresponds to limit of $\sim 0.3$\% for the total flux required for a detection \citep{Ertel2018, Ertel2020}. That is, our imaging observations achieved a limit at least five times lower than is achievable with nulling interferometry. A similar conclusion is reached by comparing the 5--8 zodi levels of the ABA-3 model (see Table~\ref{tab:ExoZodi}) with the best reported zodi limits from the HOSTS survey of $\sim$70~zodis. This is similar to the sensitivity level of previous mid-infrared coronagraphic imaging of \acenA with VLT/VISIR, since \citet{Wagner2021} reported a resolved source ($C1$) that could be fitted with a $\sim$60~zodi ($3\sigma$) exozodi model. Such a disk would be in between our ABA-1 and ABA-2 models, and thus would have been easily detected by our observations. \added{The VLT/VISIR noise level is 5--10 times higher than JWST's.} We can thus rule out that $C1$ belongs to a static clump of exozodi material at this level. The comparisons with previous observations demonstrate the dramatic improvement in sensitivity achieved by the JWST measurements.

\begin{figure}[!tb]%
 \centering
 \includegraphics[width=0.47\textwidth]{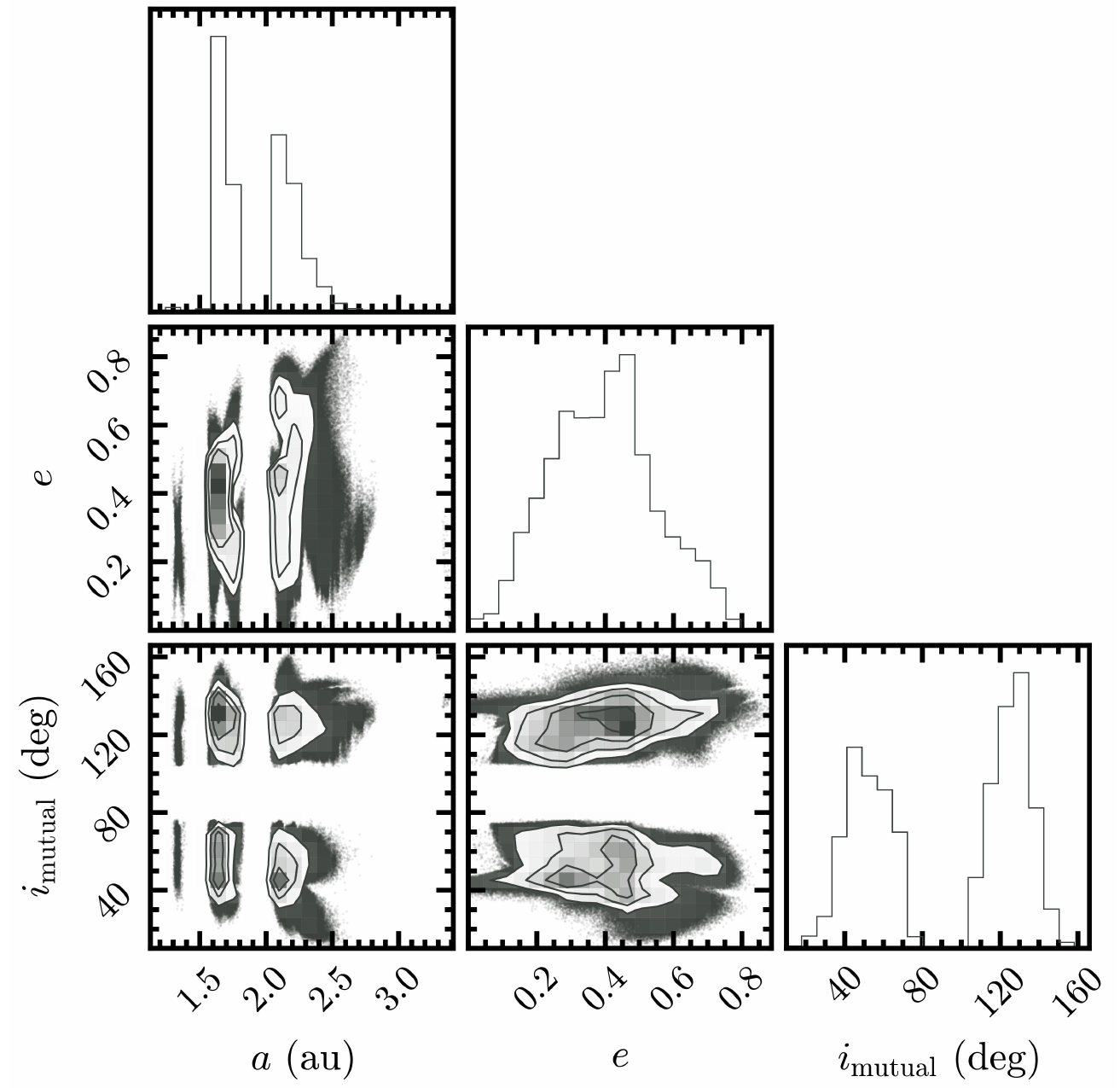} 
 \caption{Key parameters for stable planetary orbits fitting the $S1+C1$ astrometry and consistent with the February and April 2025 non-detections. Four orbital families are observed (prograde, retrograde, $a > 2$~au, $a < 2$~au). 
 \label{fig:orbits_stable_params}}
\end{figure}

\begin{deluxetable*}{ccccccc}
    \vspace{0.2in}
    \tablecaption{Key $S1+C1$ Orbital Parameters \label{tab:orbit}}
    \tablehead{
    Sightings Used 	&Orbit Type &$a$& $e$& $i_{\rm mutual}$\tablenotemark{a} & $i_{\rm sky}$\tablenotemark{b} &$T_{\rm eq}$\tablenotemark{c} \\
    &  &(au)&&($^\circ$) & ($^\circ$) &(K)}
    \startdata
    \hline
     $S1$, $C1$, \& ND\tablenotemark{d} & Prograde, $a<2$ au &   $1.66 \pm 0.06$ & $0.37 \pm 0.12$ & $54 \pm 11$ & $55 \pm 15$ or $124 \pm 13$ & $223 \pm 5$ \\
                                & No RV Constraint &            &           && \\
     $S1$, $C1$, \& ND &Prograde, $a<2$ au &  $1.64 \pm 0.07$ & $0.33 \pm 0.10$ & $58 \pm 11$ & $41 \pm 13$ or $136 \pm 9$ & $223 \pm 5$ \\
                                & $K_{RV}<6$ m/s&            &           &&& \\
     $S1$, $C1$, \& ND &Prograde, $a<2$ au &$1.58 \pm 0.08$ & $0.27 \pm 0.04$ & $70 \pm 4$ & $16 \pm 5$& $225 \pm 6$ \\
                                & $K_{RV}<3$ m/s&            &           &&& \\
    \hline
     $S1$, $C1$, \& ND & Prograde, $a>2$ au &$2.18 \pm 0.09$ & $0.43 \pm 0.18$ & $49 \pm 12$ & $78 \pm 26$& $197 \pm 6$ \\
                                & No RV Constraint &            &           &&& \\ 
     $S1$, $C1$, \& ND &Prograde, $a>2$ au &$2.14 \pm 0.07$ & $0.33 \pm 0.14$ & $51 \pm 11$ & $68 \pm 27$& $195 \pm 5$ \\
                                & $K_{RV}<6$ m/s&            &           &&& \\
     $S1$, $C1$, \& ND &Prograde, $a>2$ au &$2.09 \pm 0.02$ & $0.46 \pm 0.03$ & $64 \pm 6$ & $22 \pm 4$& $200 \pm 1$ \\
                                & $K_{RV}<3$ m/s&            &           &&& \\
    \hline
     $S1$, $C1$, \& ND &Retrograde, $a<2$ au&$1.68 \pm 0.06$ & $0.36 \pm 0.12$ & $126 \pm 10$ & $64 \pm 7$ or $132 \pm 19$& $221 \pm 6$ \\
                                & No RV Constraint &            &           &&& \\   
     $S1$, $C1$, \& ND &Retrograde, $a<2$ au&$1.67 \pm 0.08$ & $0.34 \pm 0.10$ & $123 \pm 11$ & $49 \pm 6$ or $144 \pm 14$& $221 \pm 6$ \\
                                & $K_{RV}<6$ m/s&            &           &&& \\
     $S1$, $C1$, \& ND &Retrograde, $a<2$ au&$1.65 \pm 0.08$ & $0.36 \pm 0.07$ & $115 \pm 5$ & $162 \pm 5$& $223 \pm 6$ \\
                                & $K_{RV}<3$ m/s&            &           &&& \\
    \hline
     $S1$, $C1$, \& ND &Retrograde, $a>2$ au&$2.23 \pm 0.14$ & $0.43 \pm 0.18$ & $126 \pm 10$ & $89 \pm 24$& $194 \pm 8$ \\
                                & No RV Constraint &            &           &&& \\  
     $S1$, $C1$, \& ND &Retrograde, $a>2$ au&$2.23 \pm 0.16$ & $0.32 \pm 0.14$ & $122 \pm 9$ & $88 \pm 29$& $192 \pm 8$ \\
                                & $K_{RV}<6$ m/s&            &           &&& \\
     $S1$, $C1$, \& ND &Retrograde, $a>2$ au&$2.09 \pm 0.02$ & $0.64 \pm 0.03$ & $115 \pm 5$ & $163 \pm 5$& $208 \pm 3$ \\
                                & $K_{RV}<3$ m/s&            &           &&& \\                            
    \hline
    \enddata
    \tablenotetext{a}{Inclination relative to the \acenAB orbital plane \citep[$i_{\rm AB} = 79.2430^\circ\pm 0.0089^\circ$, $\Omega_{\rm AB} = 205.073^\circ \pm 0.025^\circ$ from][]{Akeson2021}.}
    \tablenotetext{b}{Inclination relative to the plane of the sky. Bimodal distributions (about $i_{\rm sky} = 90^\circ$) are presented as two sets of values.}
    \tablenotetext{c}{Flux-averaged mean planet temperature for $A_B = 0.3$ (see \S\ref{sec:heating})}
    \tablenotetext{d}{ND denotes that orbits were checked for consistency with non-detections in the February and April 2025 epochs.}
    
    \tablecomments{Parameters are reported as mean $\pm$ standard deviation. $K_{RV}$ assumes a planet mass of 100~\mearth.}
\end{deluxetable*}

\section{Orbital Modeling of $S1$ + $C1$ \label{sec:orbits}}

With only a single JWST/MIRI sighting (and non-detections at two other epochs), it is challenging to uniquely constrain the orbit of \sone. To make progress, we consider the family of orbits that (a)~fit the relative astrometry of \sone and the VLT/NEAR 11.25~\mum\ candidate $C1$ \citep{Wagner2021}, which we treat as an earlier detection of the \sone object \added{(and in this context, referred to as the $S1+C1$ candidate)}; (b)~are dynamically stable in the presence of \acenB; and (c)~are consistent with the non-detection of $S1+C1$ in the February and April 2025 observation epochs. Additionally, we consider the consistency of the candidate's orbits with existing RV upper limits.

\begin{figure*}[!htb]%
 \centering
 \includegraphics[width=\textwidth]{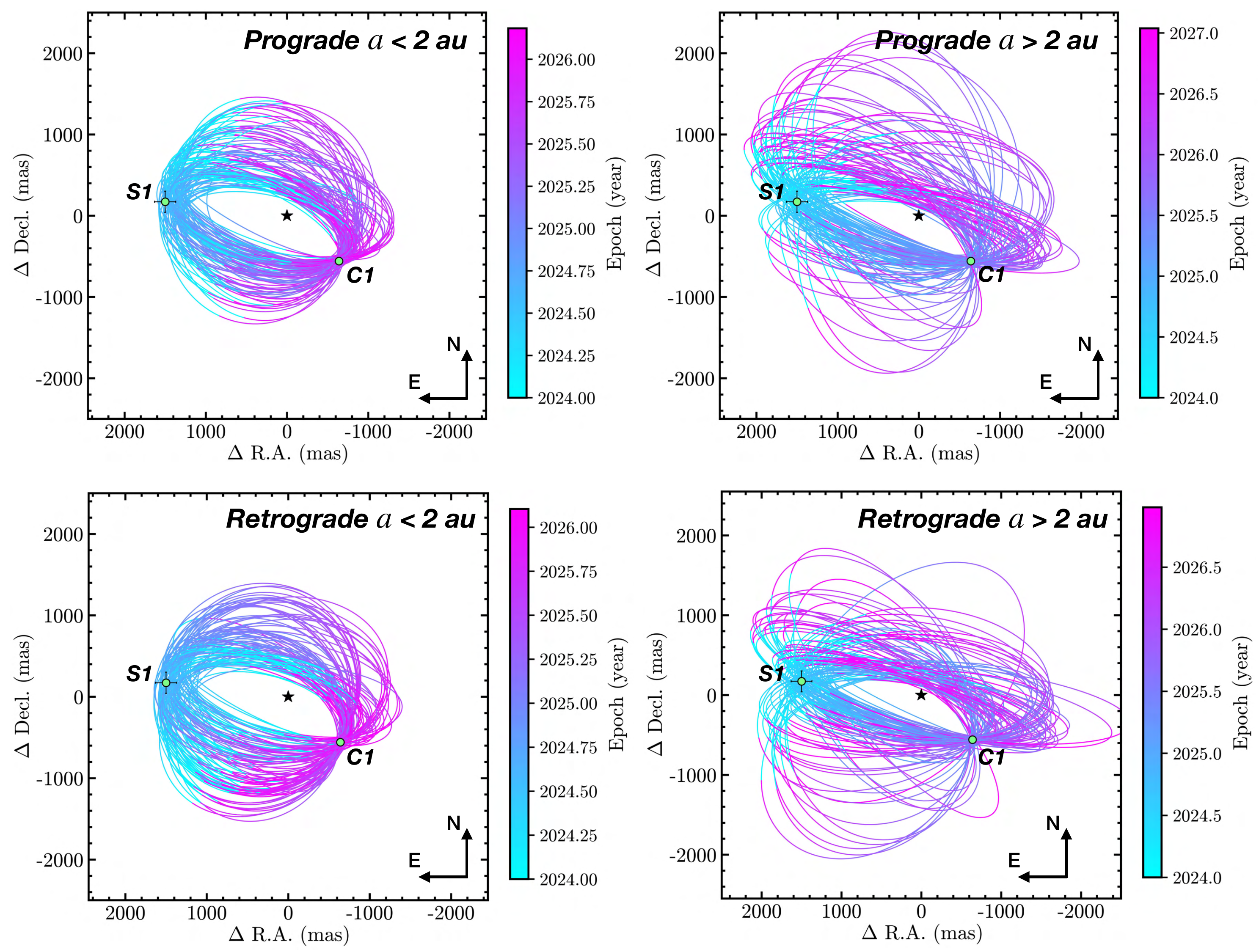} 
 \caption{100 randomly selected stable planetary orbits fitting the $S1+C1$ astrometry (marked as green points) and consistent with the February and April 2025 non-detections, for each orbital family. 
 \label{fig:orbits_stable}}
\end{figure*}
    
\begin{figure}[!tb]%
 \centering
 \includegraphics[width=0.47\textwidth]{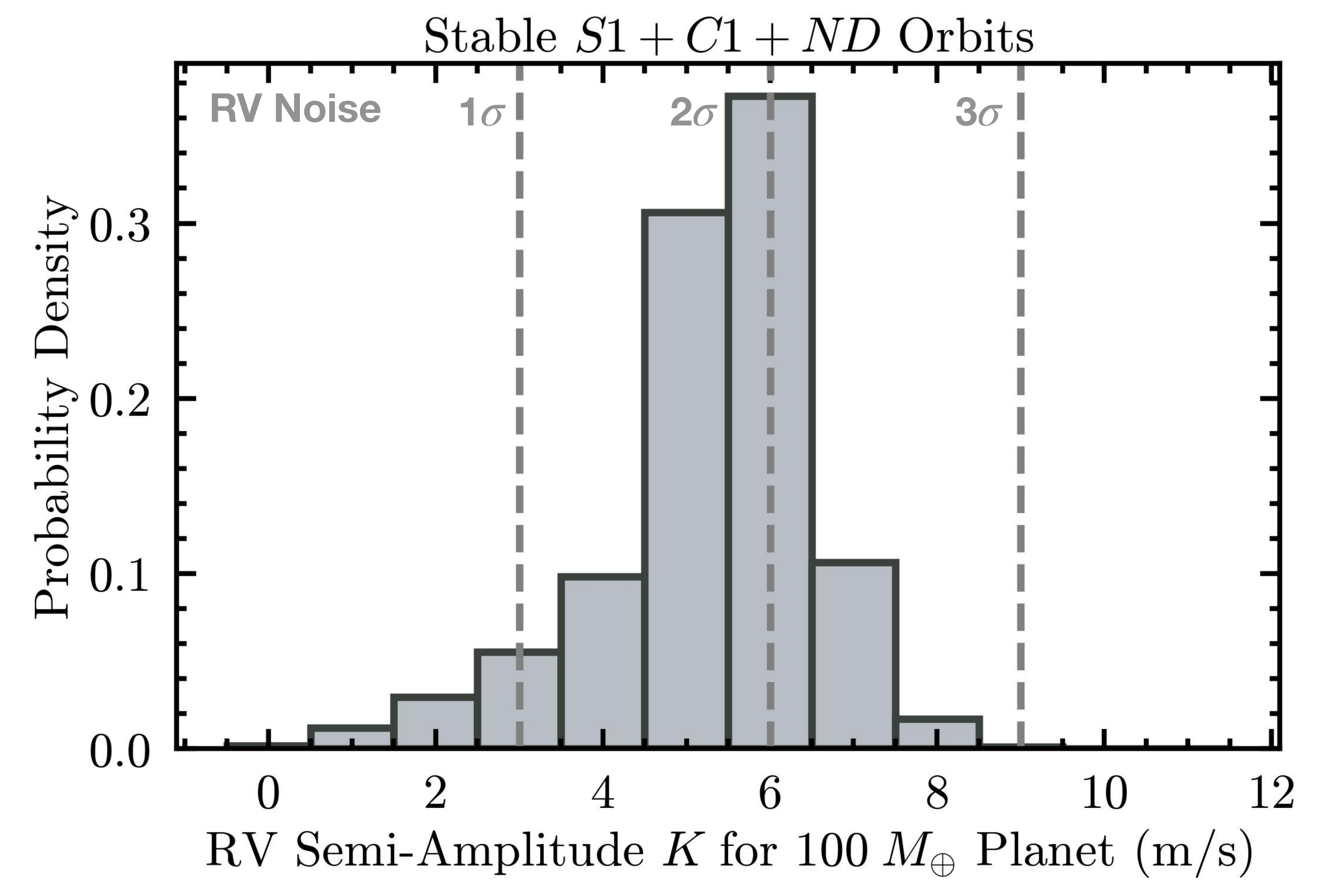} 
 \caption{The radial velocity (RV) semi-amplitude of a 100~\mearth\ planet in stable orbits fit to $S1+C1$ and consistent with non-detections in the February and April 2025 epochs. Note that $K_{RV}$ scales linearly with planet mass. The systematic RV noise floor of 3 m/s \citep[1$ \sigma$;][Kervella et al., in prep]{Zhao2018} is shown.
 \label{fig:rv_semi}}
\end{figure}

\subsection{Selection of Stable Orbits}
\label{sec:stable}
First, we randomly generate $10^7$ orbits matching the astrometry of \sone and $C1$ (Table~\ref{tab:S1props}) using the Orbits For The Impatient (OFTI) algorithm via the \texttt{orbitize!} package \citep{Blunt2017, Blunt2020}. We apply the default priors in the \texttt{orbitize!} code to the candidate planet's orbital elements and use a Gaussian prior for \acenA's mass and parallax \citep[$M_{\rm A} =1.0788 \pm 0.0029\;M_\odot$, $\pi = 750.81 \pm 0.38$ mas, from][]{Akeson2021}. Next, we evaluate the stability of the accepted orbits using the $N$-body simulation software \texttt{Rebound} \citep{Rein2012} over million year timescales using the \texttt{WHFAST} integrator \citep{Rein2019}. A given simulation is deemed unstable for very high planetary eccentricity $(e_p > 0.95)$ or large planetary distances from the host star $(d_p > 5\ {\rm au})$. Previous studies showed that orbits that meet either criterion are very likely to be unstable on billion-year timescales \citep{Quarles2016,Quarles2018b}, where more than $90\%$ of orbits stable on a million-year timescale were also stable for billion-year timescales. We find that 30\% of the orbits from the initial \texttt{orbitize!} sample are dynamically stable. 

\subsection{Incorporating Constraints From Non-Detections}
We investigate which of the above $S1+C1$ stable orbits are consistent with non-detections in the February and April 2025 observation epochs using the 2D sensitivity maps generated for both epochs in Paper II \citep{Aniket2025}. Specifically, we use the S/N~=~5 sensitivity map (rather than the 5$\sigma$ significance sensitivity map) to be stricter in eliminating orbits where $S1+C1$ would have been marginally recovered in our follow-up observations. The sensitivity maps provide the minimum point source flux detectable at a S/N~=~5 at different sky coordinates around \acenA. For each of the stable orbits above, we predict the sky position of the candidate planet in the February and April 2025 observations, and check the corresponding location in the sensitivity map to evaluate whether it would have been detected (for a flux of 3.5~mJy). Orbits where the candidate would have been recovered in either of the two epochs are eliminated. \textit{We find that 52\% of the stable orbits that fit the $S1+C1$ astrometry are also consistent with non-detections in both February and April 2025 (Figure~\ref{fig:non-detect}). There is, thus, an a priori significant chance that, if real, the planet candidate could have been missed in both follow-up observation epochs. }

The posteriors for the three key orbital elements (semi-major axis $a$, eccentricity $e$, and mutual inclination $i_{\rm mutual}$ with respect to the \acenAB binary orbit\footnote{The mutual inclination is calculated using Equation 19 in \citet{Xuan2020} and using the orbital parameters for \acenAB in \citet{Akeson2021}.}) of the stable orbits consistent with the non-detections (Figure~\ref{fig:orbits_stable_params}) show that there are four families of orbits\footnote{We do not consider the small fraction ($\sim$0.2\% of the total number) of $a < 1.5$~au orbits in Figure~\ref{fig:orbits_stable_params} as they do not agree with \sone's relative astrometry within 1$\sigma$ uncertainties.}. They correspond to the number of orbital periods that have elapsed between the VLT/NEAR observations in June 2019 and the JWST detection in August 2024 (either $\sim$1.5 or $\sim$2.5 periods, for $a \approx$~1.6~au and $a \approx$~2.1~au, respectively) and orbits in either the prograde ($i_{\rm mutual} \approx 50^\circ$) or retrograde ($i_{\rm mutual} \approx 130^\circ$) direction (Figure~\ref{fig:orbits_stable}). In addition to being significantly inclined, the $S1+C1$ planet candidate is in a moderately eccentric ($\approx0.4$) orbit. A summary of the mean orbital parameters for each family (no RV constraint case) is provided in Table~\ref{tab:orbit}. Note that all orbits presented are dynamically stable as evaluated previously (\S\ref{sec:stable}).

\subsection{Consistency with RV Limits}
To the family of stable orbits consistent with non-detections, we can add a constraint of $K_{RV}<3$~m~s$^{-1}$ \citep[$1\sigma$;][]{Butler2004, Zhao2018} on the radial velocity of \acenA, which is also observed in the HARPS RV residuals (Kervella et al., in preparation). This systematic noise floor constrains the minimum mass ($M_p\:\mathrm{sin}\:i$) of any planet around \acenA\ to be $<100$ \mearth\ (2$\sigma$) or $<150$ \mearth\ (3$\sigma$) within $\approx2$~au. Among the dynamically stable $S1+C1$ orbits consistent with the non-detections, assuming $M_p = 100$~\mearth, we find that 4\% of these orbits result in $K_{RV}\leq$~3~m/s, 50\% result in $K_{RV}\leq$~6~m/s, and 99.8\% of all orbits result in $K_{RV}\leq$~9~m/s (Figure~\ref{fig:rv_semi}, the maximum $K_{RV}$ is $\approx 11$~m/s). Table~\ref{tab:orbit} presents orbital parameters for each case. The semi-major axis and eccentricity remain largely unchanged after applying the RV constraints; however, the mutual and sky inclinations vary as the RV constraint becomes stricter (smaller reflex motion). In summary, the astrometric positions of $S1+C1$ can be fit by dynamically stable orbits consistent with both the non-detections in follow-up observations and existing RV limits. All dynamically stable $S1+C1$ orbits consistent with non-detections can be retrieved from\dataset[10.5281/zenodo.16280658]{10.5281/zenodo.16280658}.

\section{Photometric Modeling of $S1$ + $C1$ \label{sec:phot}}

In this section, we consider the available photometric data points for the \acenA planet candidate (JWST/MIRI 15.5~\mum\ and possibly, VLT/NEAR 11.25~\mum) to investigate its bulk physical properties. While the photometric data are sparse at the moment, there are some physical constraints that can be applied to aid modeling efforts. First, the effective temperature of the planet candidate is expected to be set by heating from \acenA. Second, the radius of a mature ($\sim$5 Gyr) gas giant planet cannot significantly exceed 1--1.2~\rj\ (allowing for some variations if the planet is rapidly rotating and viewed pole-on, for example) unless it is located in a very tight ``Hot Jupiter'' orbit, which is not the case as seen in the previous section on orbital modeling. Finally, the mass of the planet must be consistent with the limits set by the radial velocity measurements \citep[]{Zhao2018}. Subject to these constraints, we examine a range of plausible atmospheric models as well as thermal emission from a Saturn-like particle ring to explain the photometric data.

\subsection{Equilibrium Temperature \label{sec:heating}}

\begin{figure}[!tb]%
 \centering
 \includegraphics[width=0.47\textwidth]{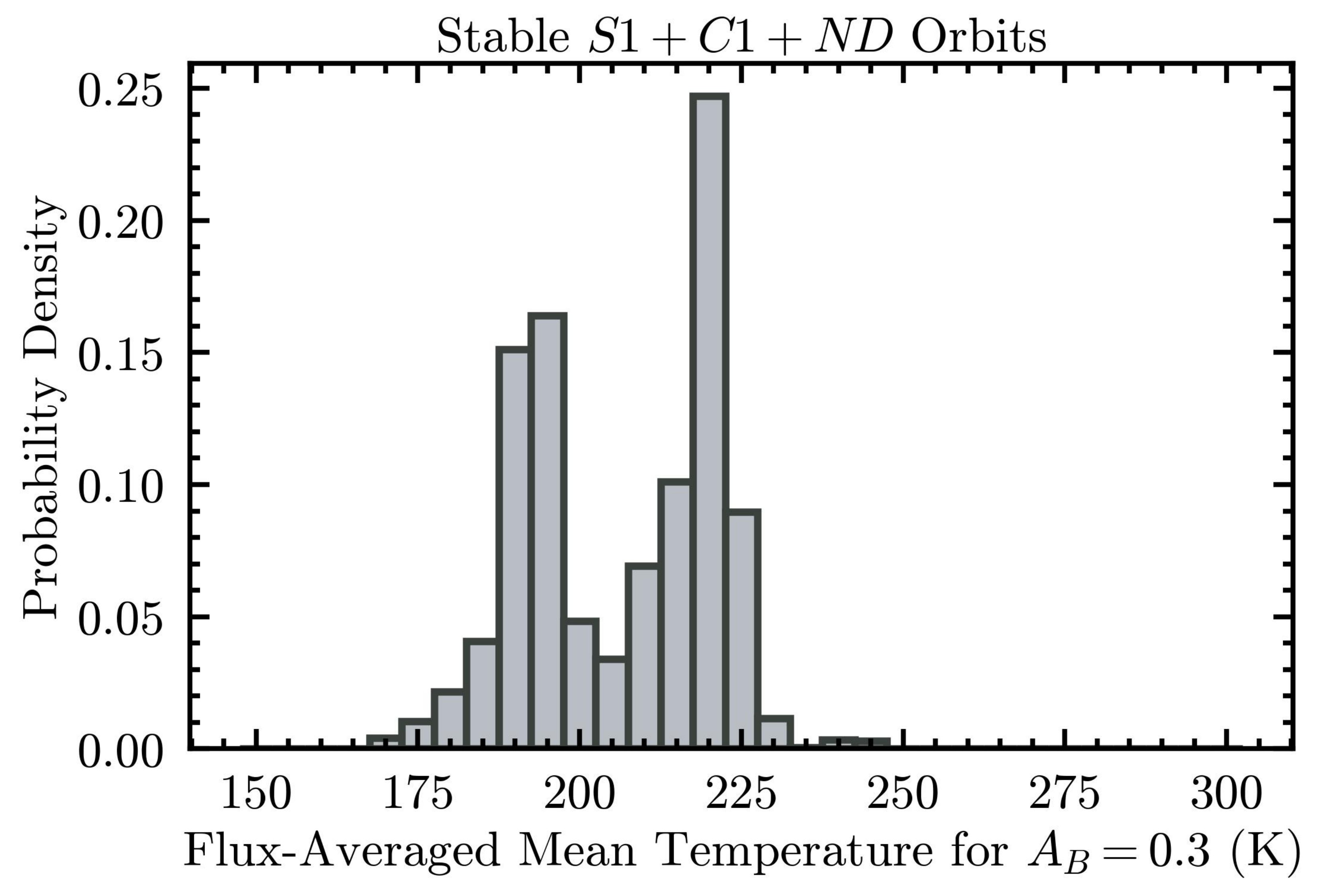} 
 \caption{Range of flux-averaged mean planet temperatures for $A_B=0.3$ corresponding to the orbits described in the preceding section. The lower temperatures in the bimodal distribution correspond to the $a > 2$~au orbits.\label{fig:planetTeff1}}
\end{figure}

We use the orbital information to infer the range of plausible effective temperatures for the planet candidate, heated by \acenA. The equilibrium temperature for a planet on an eccentric orbit depends on the instantaneous stellar input, the planet's thermal inertia, and radiative timescale. For a gas giant planet, any variations of temperature through the orbit are damped by the thermal inertia of the dense H/He atmosphere \citep{Quirrenbach2022}. In such a scenario, the correction to the equilibrium temperature to account for changes in the insolation averaged over an eccentric orbit are small, only a few percent, for eccentricities up to 0.5 \citep[][]{Johnson1976,Quirrenbach2022}. The flux-averaged temperature of a body heated by and orbiting \acenA in an eccentric orbit, $T_{\rm eq}$, at a distance $d$ is given by
\begin{equation}
 T_{\rm eq} = T_\star \cdot \left(\frac{1-A_B}{4f}\right)^{1/4} \cdot \sqrt{\frac{R_\star} {d}} \cdot (1-e^2)^{-1/8},\label{eqn:Teq}
\end{equation}
where $T_\star = 5766$ K \citep{Zhao2018}, $R_\star = 1.2175\:R_\odot$ \citep{Akeson2021}, $A_B$ is the Bond albedo, and full heat re-distribution ($f = 1$) is assumed. Adopting $A_B$ = 0.3 (intermediate between the values for most Hot Jupiters and those of the Solar System gas giants), we compute $T_{\rm eq}$ for all $S1+C1$ stable orbits consistent with the non-detections (\S\ref{sec:orbits}) and find a bimodal distribution with peaks at $\approx$195~K and $\approx$220~K (Figure~\ref{fig:planetTeff1}). The lower temperatures correspond to orbits in the $a > 2$~au families and the higher temperatures correspond to orbits in the $a < 2$~au families (Table~\ref{tab:orbit}).

\begin{figure*}[!ttb]%
 \centering
 \includegraphics[width=\textwidth]{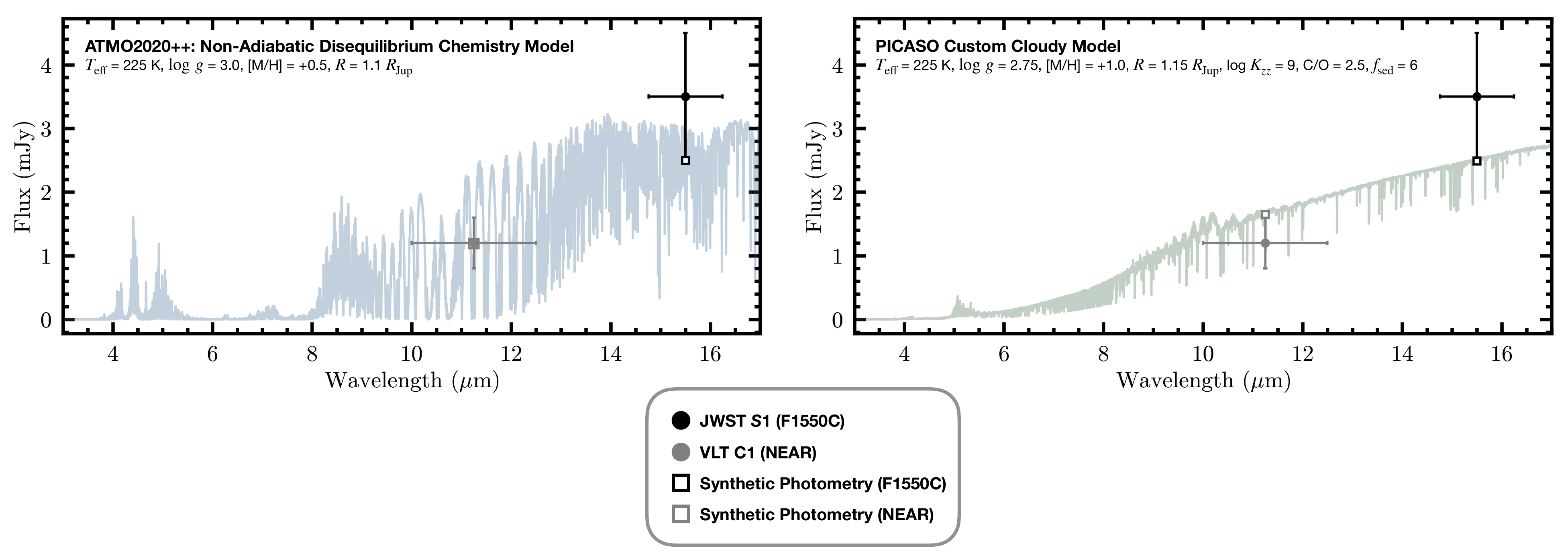} 
 \caption{$T_{\rm eff}$ = 225~K atmospheric models consistent with the \sone + $C1$ photometry (within 1$\sigma$) for a radius $<1.2$~\rj.
 \label{fig:planetmodel}}
\end{figure*}

The contributions of additional sources of heat for the planet candidate's temperature are negligible compared to stellar insolation. (1)~Residual heat of formation: \acenA is $\sim$5~Gyr old. For a similar internal radiation flux ($F$) as Jupiter or Saturn, $T_{\rm int}<$ 110~K \citep{Li2018}, the increase in temperature is negligible, due to $T \propto F^{1/4}$ and the planet candidate's higher expected $T_{\rm eq}$. (2)~Radiation from \acenB: \acenB is less luminous than \acenA by a factor of $\sim$3 \citep{Akeson2021} and at the time of JWST observations was $\sim$20~au away from \acenA. Thus, its contribution to heating is negligible. (3)~Tidal heating: the $S1+C1$ candidate is in an eccentric orbit. However, $\dot E_{\rm tide} \propto a^{-15/2}$ \citep{Peale1978} is negligible for $a \sim 2$~au. (4)~Heating from radioactivity: this is negligible for Neptune, which is both much colder than $S1+C1$ and likely has a larger fraction of radioactive isotopes.

\subsection{Planet Atmospheric Models \label{sec:AtmModels}}
The goal in this section is to obtain first estimates of the planet candidate's fundamental parameters ($T_{\rm eff}$, radius, and mass) using atmospheric model grids available in literature for cold planets and brown dwarfs. Given the numerous atmospheric parameter degeneracies involved in fitting two photometric data points, particularly in the observationally unexplored low-mass ($\lesssim$~200~\mearth), low temperature ($<$~300~K) planetary regime \citep[see for example,][]{Crotts2025}, we aim only to provide example scenarios that can explain the observed flux measurements. Detailed atmospheric modeling is appropriate for future studies when additional photometry and/or spectroscopy is available (see \S\ref{sec:future}).

We jointly fit the F1550C JWST/MIRI flux and the 11.25~$\mu$m\ VLT/NEAR flux (Table~\ref{tab:S1props}), assuming they are related (as indicated by the orbit fits in the previous section). The fitting procedure synthesizes model photometry in the F1550C bandpass ($\approx$15.15--15.85 $\mu$m, using the transmission curve from the SVO filter profile service\footnote{\url{http://svo2.cab.inta-csic.es/theory/fps/}}) and the VLT/NEAR bandpass ($\approx$10--12.5 $\mu$m, constant transmission assumed), and finds the minimum radius ($<$~1.2~\rj) that yields a model flux consistent with the measured photometry within 1$\sigma$. The effective temperature of the planet is set to 225 K for the atmospheric models fit below, matching that expected for $a < 2$~au orbits (Table~\ref{tab:orbit})\footnote{We were unable to fit the photometry with 200~K models for planet radii $<$~1.2~\rj.}. We also restrict the surface gravity (log $g$, in CGS units) of the models to 2.5--3.0~dex, chosen to yield a planet mass $\lesssim $150--200 $M_{\rm Earth}$ to be consistent with radial velocity limits ($M_p\:\mathrm{sin}\:i <150$~\mearth, 3$\sigma$; inclined orbits can raise the limit on the true planet mass).

ATMO2020++ \citep{Leggett2021, Meisner2023}: Using the ATMO2020 models with strong vertical mixing as a starting point, ATMO2020++ modifies the adiabatic ideal gas index $\gamma$ (and thus atmospheric temperature gradient) to account for the effect of processes responsible for producing a non-adiabatic cooling curve in giant planet and brown dwarf atmospheres. These processes include complex atmospheric dynamics (e.g., zones, spots, waves) due to rapid rotation, compositional changes due to condensation, upper atmosphere heating by cloud decks or breaking gravity waves, etc. Recent modeling with JWST data has shown that this grid provides an improved fit to Y dwarf spectra compared to the standard-adiabat models \citep{Leggett2023, Leggett2024, Luhman2024}. The default ATMO2020++ grid only extends to $T_{\rm eff} = 250$~K, so we generated custom models for $T_{\rm eff} = 225$~K, log~$g = 3.0$~dex, and [M/H] $= +0.5,\;+1.0$. We find that the $T_{\rm eff} = 225$~K, log~$g = 3.0$~dex, and [M/H] $= +0.5$ model agrees with the photometry for a radius of 1.1~\rj\ (Figure~\ref{fig:planetmodel}). For the above parameters, the planet candidate would have a mass $\approx$150 \mearth.

Sonora and PICASO models: The Sonora Flame Skimmer models (Mang et al., in prep) extend the cloud-free Sonora Elf Owl grid \citep{Mukherjee2024, Wogan2025} to colder effective temperatures, lower surface gravities, and a broader range of metallicities. These models incorporate rainout chemistry for H$_2$O, CH$_4$, and NH$_3$—even in cloud-free atmospheres—similar to the treatment in Sonora Bobcat. They also address the underestimation of CO$_2$ found in the Sonora Elf Owl models \citep{Mukherjee2024}, which has since been revised in \citet{Wogan2025}. In addition, we generated a custom grid of cloudy models using \texttt{PICASO} \citep{Batalha2019, Mukherjee2023}. This grid spans effective temperatures of $T_{\rm eff} = 200$ and 225 K, surface gravities of log $g$ = 2.75 and 3.0~dex (cgs), eddy diffusion coefficients $K_{\rm zz} = 10^2$ and $10^9$ cm$^2$ s$^{-1}$, metallicities of [M/H] = +0.5 and +1.0, and a C/O ratio of 2.5 (relative to solar). Cloudy models have $f_{\rm sed}$ = [4, 6, 8], with H$_2$O as the only condensing species. We find that the $T_{\rm eff} = 225$~K, log~$g = 2.75$~dex, [M/H] $= +1.0$, log $K_{zz}=9$, C/O = 2.5, $f_{\rm sed} = 6$ model agrees with the photometry for a radius of 1.15~\rj\ (Figure~\ref{fig:planetmodel}). For the above parameters, the planet would have a mass $\approx$90~\mearth.

Additional models applicable to cool giant planets: We also experimented with fitting the photometry using the Sonora Bobcat cloudless, chemical equilibrium model grid \citep{Marley2021}, the ATMO2020 solar metallicity, disequilibrium chemistry model grid \citep{Phillips2020}, the patchy water cloud models of \citet{Morley2014}, and a new grid of self-consistent models by \citet{Lacy2023} that incorporate both the effects of water clouds and disequilibrium chemistry. However, we did not find suitable solutions with these grids, as they all required a $T_{\rm eff} \geq 250$~K to fit the photometry for a radius $<1.2$~\rj. 

\begin{figure*}[!tb]%
 \centering
\includegraphics[width=\textwidth]{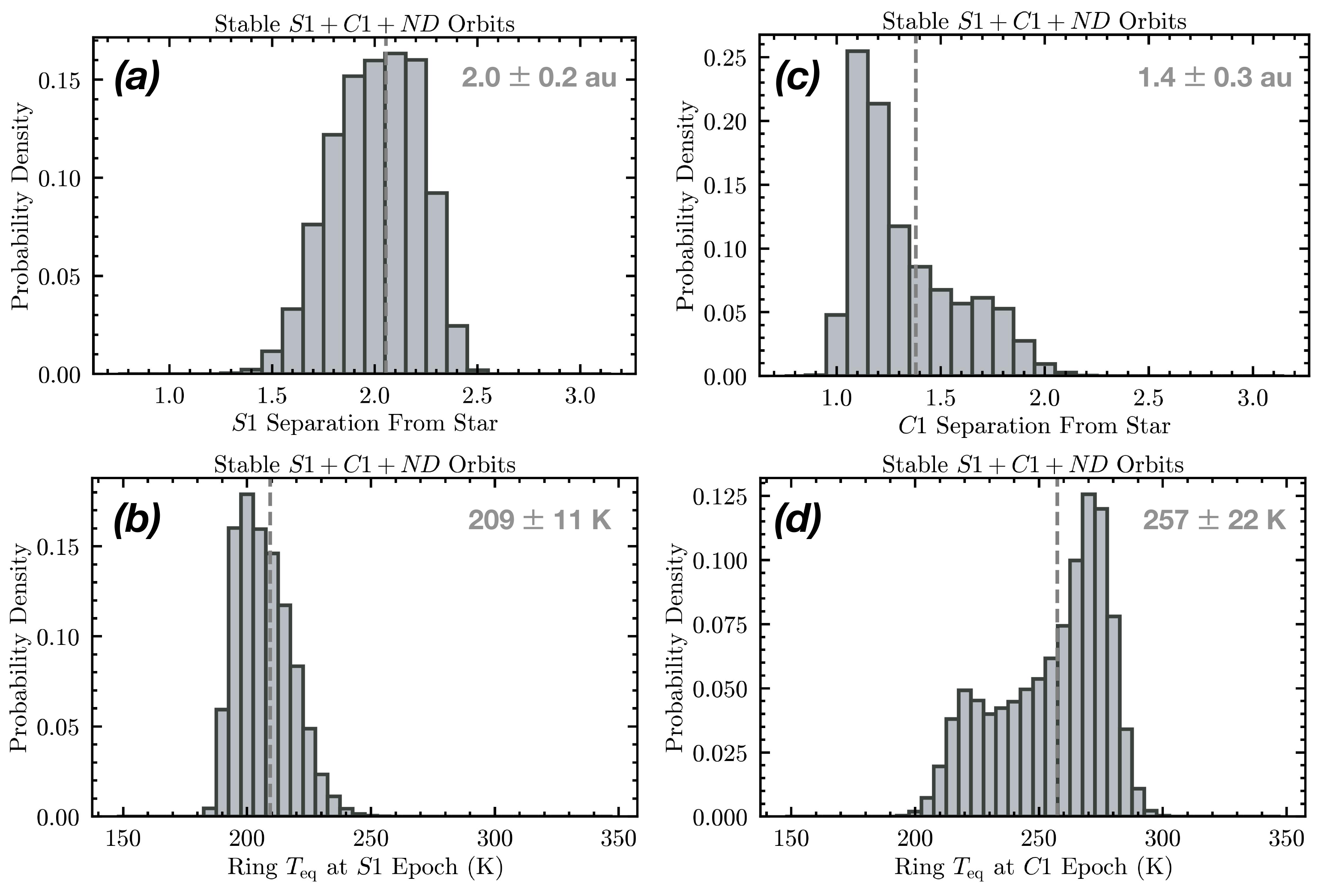} 
 \caption{Range of (true) star-planet separations (top) and instantaneous temperatures (bottom) for a planetary ring with $A_B = 0.1$, no thermal inertia, and $f = 1$, seen at the epochs of \sone (left) and $C1$ (right) based on the stable orbits consistent with non-detections in the prograde $a < 2$~au family as described in \S\ref{sec:orbits}. The mean value of each distribution is marked with a dashed line and noted with the standard deviation in the top right corner of each panel.
 \label{fig:ringTeff}}
\end{figure*}

\begin{figure}[!tb]%
 \centering
 \includegraphics[width=0.45\textwidth]{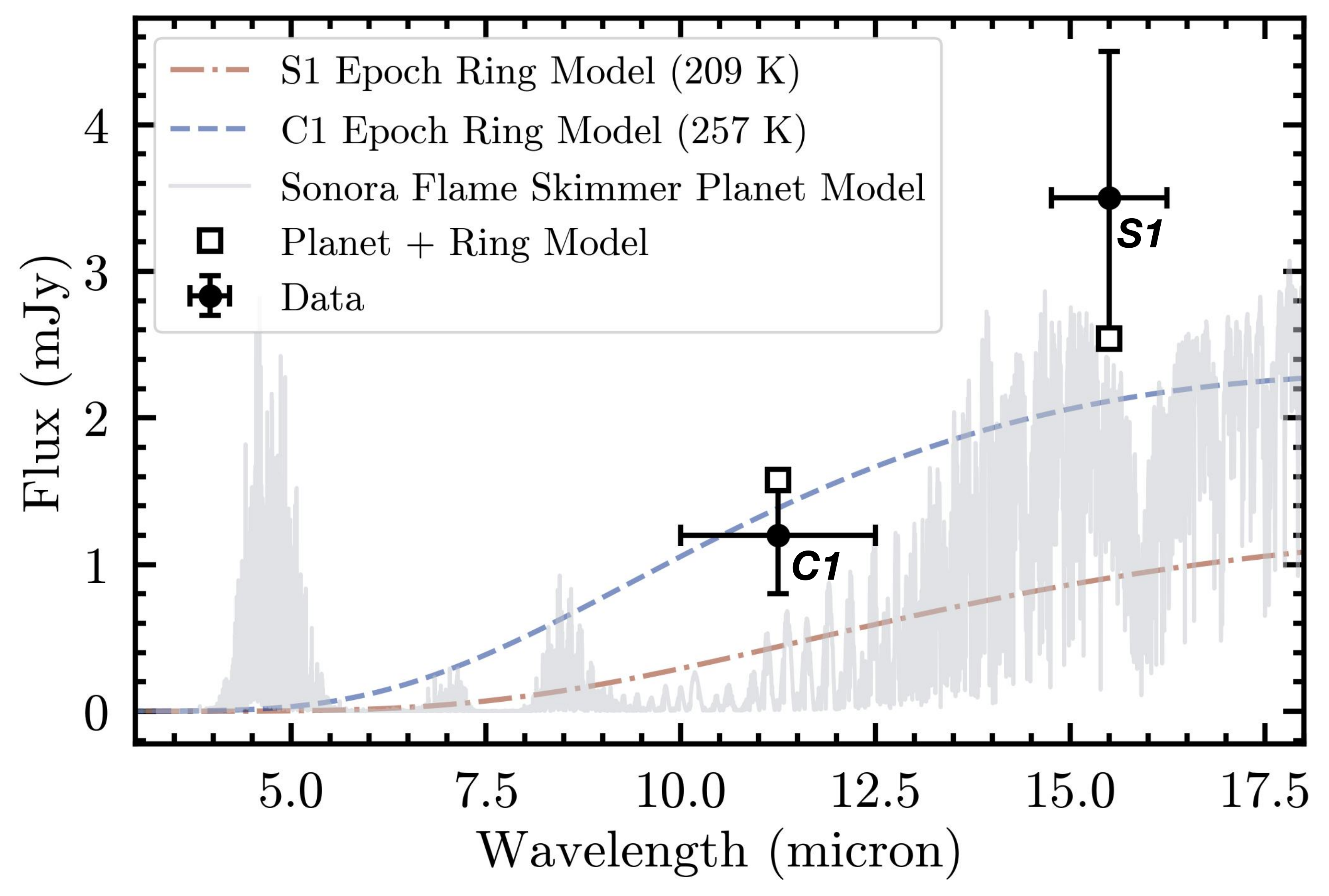} 
 \caption{Blackbody (BB) emission from a circumplanetary ring for a total cross-sectional area equivalent to the half the face-on cross-sectional area of Saturn's rings together with a fiducial planet atmospheric model ($T_{\rm eff} = 225$~K, log~$g$ = 3.0 dex, [M/H] = +1.0, C/O = 1.5, $R_p=$ 1~\rj). Each component contributing to the total model VLT/NEAR and F1550C flux (squares) is shown. There are two BB components as the ring temperature is different for each detection epoch, depending on the planet-star separation in Figure~\ref{fig:ringTeff}.
 \label{fig:planetring}}
\end{figure}

\subsection{Planet Ring System Models \label{sec:rings}}

The previous section presented example planet models which can reproduce the brightness of $S1+C1$, but required planet radii $\approx1.1$~\rj\ (driven by the observed F1550C brightness), more commonly observed for hotter planets, but plausible if a rapidly rotating planet is viewed closer to pole-on (the observed surface area can be higher). Alternate explanations for the F1550C brightness include (1)~a knot of exozodiacal emission; or (2) a smaller planet with a circumplanetary ring. Given the lack of exozodi detection reported in \S\ref{sec:zodi}, we do not consider the exozodi knot interpretation further, except to note that this would require the knot to dominate the exozodi emission, and for the knot to orbit the star with similar constraints to those reported for the planet scenario (\S\ref{sec:orbits}) and to have only been detected at one epoch of our observations. 

For an interpretation of the emission as circumplanetary material, a straight-forward model is to consider an optically thick ring. A ring is not expected to have significant thermal inertia (as opposed to a gas giant planet, as discussed in \S\ref{sec:heating}). Thus, the ring temperature at the \sone and $C1$ detection epochs will be the instantaneous equilibrium temperature calculated for the true planet-star separation at those epochs. For each stable $S1+C1$ orbit consistent with the non-detections in the prograde $a < 2$~au family, we calculate the planet-star separation and the corresponding equilibrium temperature, assuming $A_B = 0.1$ (similar to asteroids) and $f = 1$ (Figure~\ref{fig:ringTeff}). Orbits with $a < 2$~au are favored as they yield a higher planet $T_{\rm eq}$, which is required to better fit the F1550C brightness (the prograde and retrograde scenarios yield similar separation distributions and mean values). We find that an optically thick circumplanetary ring would be hotter at the $C1$ epoch ($T_{\rm eq} = 257 \pm 22$~K) than at the $S1$ epoch ($T_{\rm eq} = 209 \pm 11$~K).

We modeled the observed $S1+C1$ photometry using various 225~K planet atmospheric models (grids discussed in \S\ref{sec:AtmModels}) combined with a constant surface area (free parameter) blackbody ring with a temperature of 257~K for the VLT/NEAR 10--12.5~\mum\ flux and a temperature of $209$~K for the F1550C flux. The photometry agrees, within 1$\sigma$ uncertainties, with a Sonora Flame Skimmer clear, equilibrium, $T_{\rm eff} = 225$~K, log~$g$ = 3.0 dex, [M/H] = +1.0, and C/O = 1.5 model for a planet radius of 1~\rj\ (corresponding to $\approx120$~\mearth), together with a ring that has a cross-sectional area equivalent to a face-on disk of radius $\approx$ 64,000\,km or $\approx$~0.9~\rj\ (Figure~\ref{fig:planetring}). This is $\sim$ half the cross-sectional area of Saturn’s rings, which extend to 140,000 km (plus a more tenuous, more distant distribution). Planetary rings lie in their planet's Roche zone from 1.4--2.5~$R_{p}$, so this explanation seems plausible. \added{If the planet candidate is closer to the star at the $S1$ epoch or farther at the $C1$ epoch than the mean separation presently assumed at each epoch, then the agreement with the measured photometry improves.} 

\begin{figure}[!tb]%
 \centering
 \includegraphics[width=0.45\textwidth]{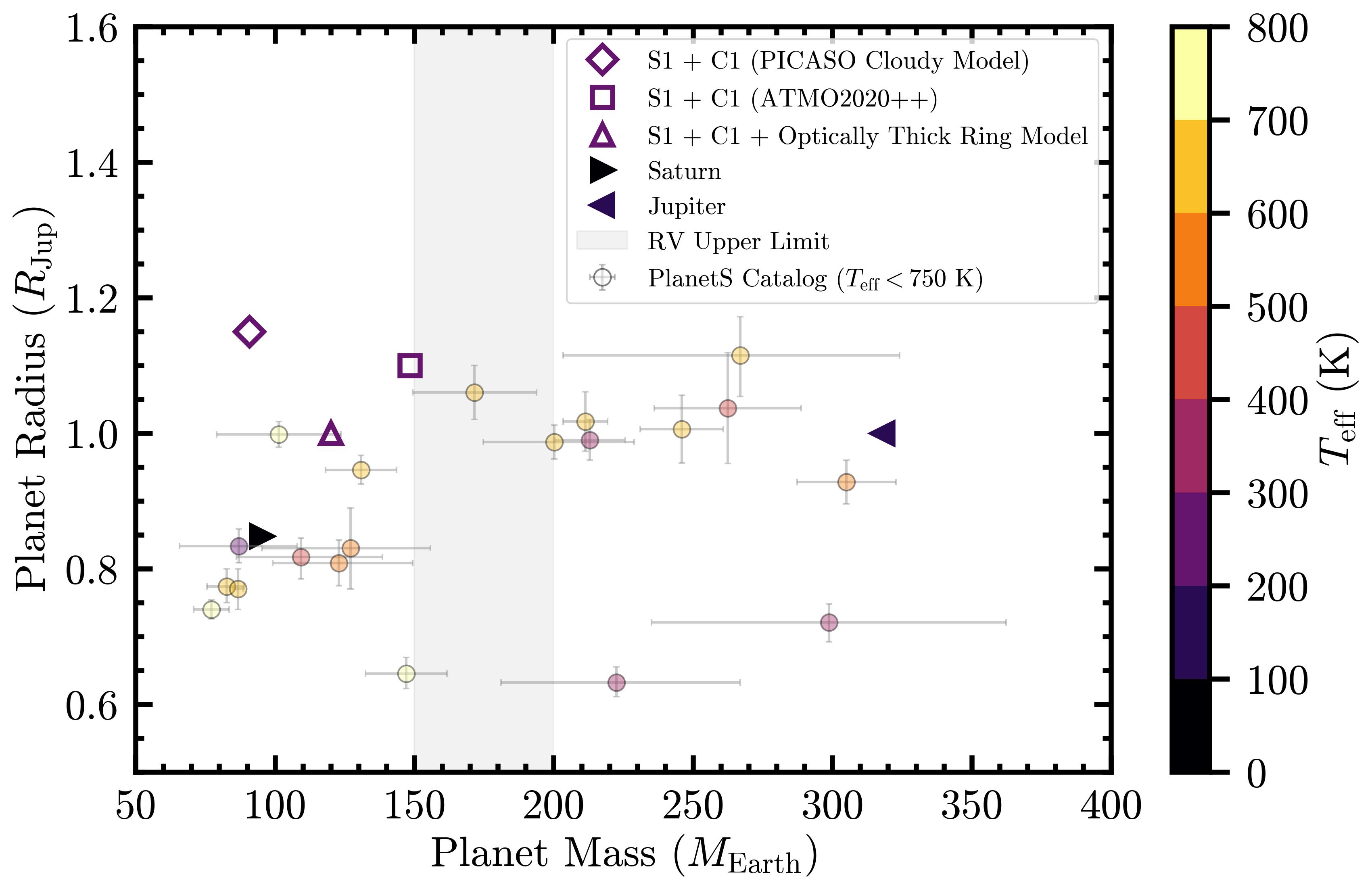} 
 \caption{Observed masses and radii derived from transit observations for planets cooler than 750~K \citep[from the PlanetS Catalog,][]{Muller2024} are shown along with the properties of the $S1+C1$ planet candidate inferred from the different models. The shaded area shows the approximate range of RV upper limits, depending on the exact inclination of the candidate's orbit. The positions of Saturn and Jupiter are included.
 \label{fig:planetMR}}
\end{figure}

We stress that the ring model discussed above is highly simplified. Geometrical effects make it challenging to develop a fully comprehensive and accurate optically thick ring model. The inclination of the ring with respect to the star affects the ring's temperature. The inclination of the ring to our line-of-sight together with shadowing of the ring by the planet affects the inferred size and visible emitting area. Additionally, a ring could both shade the planet from starlight, reducing planet temperature, and block planet light towards the observer. In the absence of strong constraints on the planet candidate's orbit and with only two photometric points, the problem is highly unconstrained. Overall, the key takeaway of the analysis presented above is that a circumplanetary ring around the $S1+C1$ planet candidate is a plausible hypothesis to explain the higher F1550C brightness for a smaller planet than inferred just using atmospheric models. A summary of the inferred mass and radius of the $S1+C1$ candidate from both atmospheric and ring models, as compared with the cold transiting planet population, is presented in Figure~\ref{fig:planetMR}.

For completeness, we consider an alternative model for circumplanetary material, where the cross-sectional area derived above comes from an optically thin dust distribution, such as one that might arise from the grinding down of a cloud of irregular satellites orbiting the planet \citep{Kennedy2011}. Such a cloud could extend out to roughly half the Hill radius of the planet ($R_\mathrm{H}\sim0.09$ au radius for a low eccentricity, 100~\mearth~planet at 2~au), which would correspond to a distribution with optical depth $\sim 5 \times 10^{-5}$. Small grains with realistic optical properties, in models in which the dust is optically thin, are heated above blackbody temperature. This increases the flux in the VLT/NEAR bandpass and makes it more challenging to simultaneously fit the $S1$ and $C1$ photometry, specifically because the ring is expected to have a higher temperature at the $C1$ detection epoch than the $S1$ detection epoch from planet-star separation calculations (Figure~\ref{fig:ringTeff}). This would require the dust distribution to be truncated to avoid the presence of $\mu$m-sized dust. This is in addition to the challenge of retaining the irregular satellites given their expected collisional erosion.

\begin{figure*}[!tb]%
 \centering
 \includegraphics[width=0.9 \textwidth]{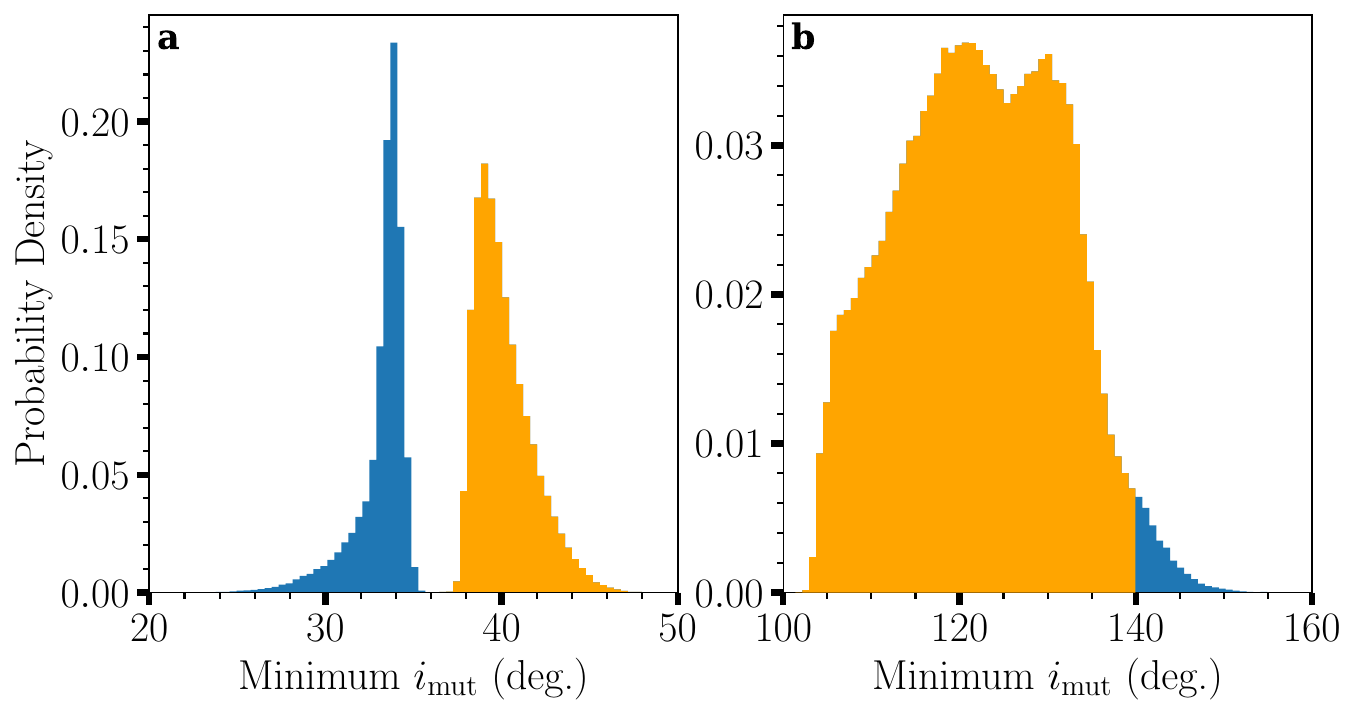} 
 \caption{Range of minimum mutual inclination $i_{\rm mut}$ experienced by the prograde (a) and retrograde (b) planet in the inner orbit ($a < 2\ {\rm au}$ family). Orbits colored yellow represent the $i_{\rm mut}$ range within the von Zeipel-Kozai-Lidov (vZKL) regime that would produce large amplitude oscillations, while blue denotes $i_{\rm mut}$ values that would have much lower oscillations.
 \label{fig:PlanetInclination}}
\end{figure*}

\begin{figure*}[tb]%
 \centering
 \includegraphics[width=\textwidth]{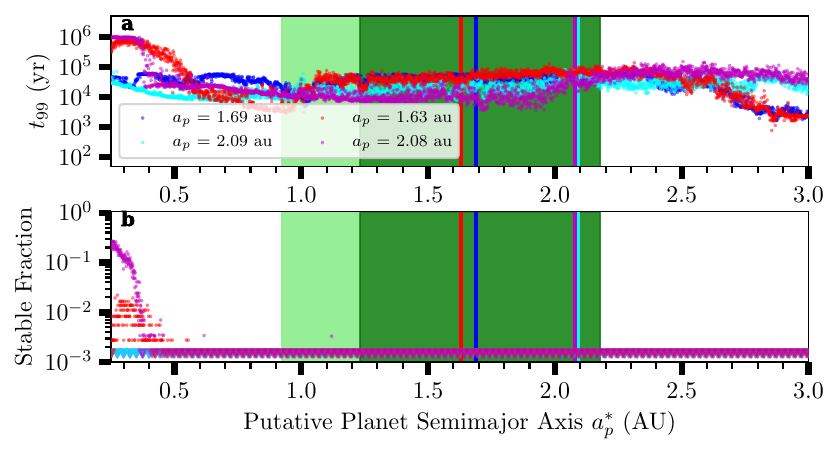} 
 \caption{The planet candidate is initialized using the mean values from Table~\ref{tab:orbit} with $K_{\rm RV}<3\ {\rm m/s}$, where the color-coded points denote the planet candidate with a prograde (blue/cyan) or retrograde (red/magenta) orbit. The top panel shows the lifetime $t_{99}$ when $99\%$ of the test particles are unstable, while the bottom panel measures the fraction of stable particles for a given initial semi-major axis. The triangles denote upper limits for the stable fraction due to the finite number of trials (360) per semimajor axis. The light green region denotes the optimistic HZ, while the dark green represents the more conservative HZ.
 \label{fig:PlanetStability}}
\end{figure*}

\section{Discussion \label{sec:discussion}}

We start by investigating possible mechanisms to explain the high eccentricity and inclination orbits inferred for the $S1+C1$ candidate in a close binary system like \acenAB. We then briefly discuss the prospects for other planets or an exozodiacal disk around \acenA in the presence of the $S1+C1$ candidate, and end with a discussion of implications of a gas giant in \acenAB system in relation to theories of planet formation in binary systems. 

\subsection{Planetary Companions}
\subsubsection{Secular Dynamics of S1 + C1 Planet Candidate}

Table \ref{tab:orbit} reveals that the best fitting orbits for $S1+C1$ consistent with the non-detections are significantly inclined with respect to the \acenAB binary orbital plane and eccentric, which naturally leads one to suspect that the planet candidate might undergo von Zeipel-Kozai-Lidov (vZKL) oscillations. \added{The von Zeipel-Kozai-Lidov mechanism is a secular gravitational effect in hierarchical triple systems whereby a distant companion's torque drives a periodic exchange between the inner orbit's eccentricity and inclination. This dynamical exchange can help explain why the candidate planet's best-fitting orbits exhibit moderate eccentricity, as it cycles through a wide range of eccentricities and mutual inclinations} \citep{vonZeipel1910,Kozai1962,Lidov1962,Ito2020}. In the test particle approximation \citep{Naoz2016} for vZKL oscillations, the less massive body ($S1+C1$) will have a minimum mutual inclination $i_{\rm mut}$ relative to the more massive binary, which is approximately $39.2^\circ$ for prograde and $141.8^\circ$ for retrograde orbits. Figure \ref{fig:PlanetInclination} shows the probability density of minimum inclination attained during our N-body stability simulations selecting on the stable orbits that fit $S1+C1$ and are consistent with non-detections. The minimum mutual inclination shows that a majority of orbits undergo vZKL oscillations, likely to be large amplitude, in the test particle approximation and consistent with the candidate's present configuration.

\subsubsection{Prospects For Other Planets Orbiting \acenA \label{sec:other}}

The Hill Radius, $R_H$, gives a measure of the relative gravitational influence of two bodies on a third and can be used to identify regions in semi-major axis for the stability of a third body. We use this to assess the possibility that another planet might exist in the Habitable Zone (HZ) of \acenA given the presence of a planet with the properties of \sone + $C1$. The Hill Radius, $R_H$ is given by:

\begin{equation}
R_{\mathrm {H} }\approx a(1-e){\sqrt[{3}]{\frac {m_{p}}{3(m_{*}+m_{p})}}} \label{eq:RH}.
\end{equation}

Given the best-fit semi-major axis, $a$, and eccentricity, $e$, for each of the potential orbital families (Table~\ref{tab:orbit}), using the host star mass $(m_\star=1.08$ M$_\odot)$, and assuming a candidate planet mass $(m_p\sim 100$ M$_\oplus)$, the Hill Radius $R_H$ for semi-major axis, $a$=1.62--2.16 au and \added{eccentricity $\sim$0.4}, ranges from 0.044--0.060 au. \citet{Quarles2018a} argue that for stability in a coplanar system, a buffer zone of $\approx 7.5\, R_H$ is required to establish stable orbits in a two planet system. Thus, it is unlikely for there to be any other planets between $a(1 - e) - 7.5 R_\mathrm{H}$ and $a(1 + e) + 7.5 R_\mathrm{H}$, i.e., \added{from 0.6~au to 3.5~au, depending on the value of $a$}. \textit{Thus, assuming the $S1+C1$ candidate is real, there are probably no other planets within or exterior to \acenA's HZ.}

These arguments are bolstered with numerical simulations which show, with the parameters of $S1+C1$, there are no stable orbits exterior to ${\sim}0.5$ au (Figure~\ref{fig:PlanetStability}). These simulations use the N-body simulation package \texttt{Rebound} with the \texttt{IAS15} integrator, where the massive bodies (binary + $S1+C1$) begin with the mean orbital parameters from the $K_{\rm RV} < 3\ {\rm m/s}$ cases in Table~\ref{tab:orbit} for $S1+C1$ and stellar parameters from \citet{Akeson2021}. The putative second planet begins as a test particle on a slightly eccentric orbit $(e^*_p = 0.05)$ that is apsidally aligned and coplanar with $S1+C1$, where we vary the putative second planet semi-major axis $a^*_p$ from $0.25-3\ {\rm au}$ with steps of $0.001\ {\rm au}$ and initialize the mean anomaly of the body from $0^\circ-359^\circ$ in steps of $1^\circ$. The simulations are evolved for 1 Myr, where an individual simulation is stopped depending on the state of the test planet, which can either be ejected (distance $r^*_p$ of the test planet exceeding $5\ {\rm au}$), have high eccentricity $(e^*_p > 0.95)$, or collide with either $S1+C1$ or the host star.

From these simulations, we calculate the lifetime $t_{99}$ when $99\%$ of the test particles are unstable (in Figure~\ref{fig:PlanetStability}a) and the fraction of test particles that are stable for a given semi-major axis (in Figure~\ref{fig:PlanetStability}b). For $a^*_p\gtrsim 0.5\ {\rm au}$, virtually all test particles are unstable within $10^5$~yr, while for $a^*_p \gtrsim 0.4\ {\rm au}$ very few remain in orbit for 1~Myr. The prospects for stability for the test particles increase for $a^*_p \lesssim 0.4\ {\rm au}$, but only when considering a retrograde-orbiting planet candidate. In summary, Figure~\ref{fig:PlanetStability}b strongly suggests that the region exterior to $\sim$0.4~au will be inhospitable to any other planets in the presence of $S1+C1$. The dynamics of the \acenAB system were already known to be inhospitable to planets outside of $\sim$3~au \citep{Quarles2018b}.

\subsubsection{Planets in Binary Systems}

\begin{deluxetable*}{cccccccc}
    \tablecaption{Known S-type Planetary Systems with Similar Orbital Architecture as \acenAB + $S1+C1$ \label{tab:orbit-compare}}
    \tablehead{
    Planetary System	& $a_{A-B}$ &$e_{A-B}$& $a_{A-Ab}$ &$e_{A-Ab}$ & $i_{\rm mutual}$ & References \\
    & (au) &&(au)& & ($^\circ$)&}
    \startdata     
    \hline
     \acenAB + $S1+C1$ Candidate& $\approx$23 &$\approx$0.5& $\approx$1.6 or $\approx$2.1 &$\approx$0.4 & $\approx$50 or $\approx$130 & 1, 2 \\
     HD 196885 AB + HD 196885 Ab & $\approx$21 &$\approx$0.4 & $\approx$2.6& $\approx$0.5 & $\approx$25 & 3 \\
     $\gamma$~Cep~AB + $\gamma$~Cep~Ab  & $\approx$19 & $\approx$0.4 & $\approx$2.1 & $\approx$0.1 & $\approx$114 & 4 \\
    \hline
    \enddata
    \tablecomments{The orbital parameters are generally well-constrained in all cases but are quoted without uncertainties for the purposes of an approximate, order-of-magnitude comparison.}
    \tablerefs{(1)~\citet{Akeson2021}; (2)~This work; (3)~\citet{Chauvin2023}; (4)~\citet{Huang2022}.}
\end{deluxetable*}

Planets in multiple star systems are not rare. As of this writing, the NASA Exoplanet Archive \citep{Christiansen2025} lists over 500 such systems, although their number decreases with the number of stars (only 71 in triple systems such as \acenAB + Proxima Cen, i.e., \acenABC, but stellar companions that are as intrinsically-faint as Proxima Cen may not have been identified yet), including cases like Kepler-132 (KOI-284) where planets have been found in circumstellar orbits about both stars \citep{Lissauer2014}. There is strong observational evidence and robust physical arguments suggesting that for systems with $a<100 $ au, the formation of planets larger than sub-Neptunes in stable configurations is suppressed in multiple systems \citep{Kraus2016,Moe2021, Dupuy2022, Sullivan2024}. \citet[their Figure 3]{Moe2021} describe a suppression factor of 0.4 for a binary system with \acenAB's semi-major axis of 23 au. Yet although there is observational evidence for the suppression of planet formation in binary systems, there are numerous analyses of multiple star systems which show islands of stability close to either or both of the stars in multiple systems \citep{Quarles2016,Quarles2018b}. Two systems in particular, HD 196885~AB + HD 196885~Ab and $\gamma$~Cep AB + $\gamma$~Cep Ab, are notable for their similarity in S-type orbital architectures to the candidate \acenAB + $S1+C1$ system (Table~\ref{tab:orbit-compare}). In each case, the stellar system is a close, eccentric binary and hosts a moderately eccentric planet that is inclined with respect to the stellar binary orbital plane (prograde or retrograde). Thus, the existence of an exoplanet with the properties of $S1+C1$ in the \acenAB system is not impossible.

\subsection{Exozodiacal disks}
\subsubsection{Exozodiacal Disks in Binary Systems} \label{sec:exozodi_in_binaries}
While the formation of circumstellar sub-Neptunes and larger planets appears to be significantly suppressed in close binaries like \acenAB, the fate of the smaller bodies that constitute terrestrial planets and debris disks is more nuanced. On the one hand, there has so far been no clear detection of circumstellar debris disks in binaries of separations $<\!100\,\text{au}$ by means of infrared excess \citep{Trilling2007,Yelverton2019}. But recent observational findings suggest that the presence of circumstellar super-Earths in such binaries is relatively less suppressed than that of sub-Neptunes \citep{Sullivan2024}. This implies that the early stages of planet formation, that is, planetesimal formation and accretion, remain relatively effective. 

\begin{figure*}[!tb]%
 \centering
 \includegraphics[width=\textwidth]{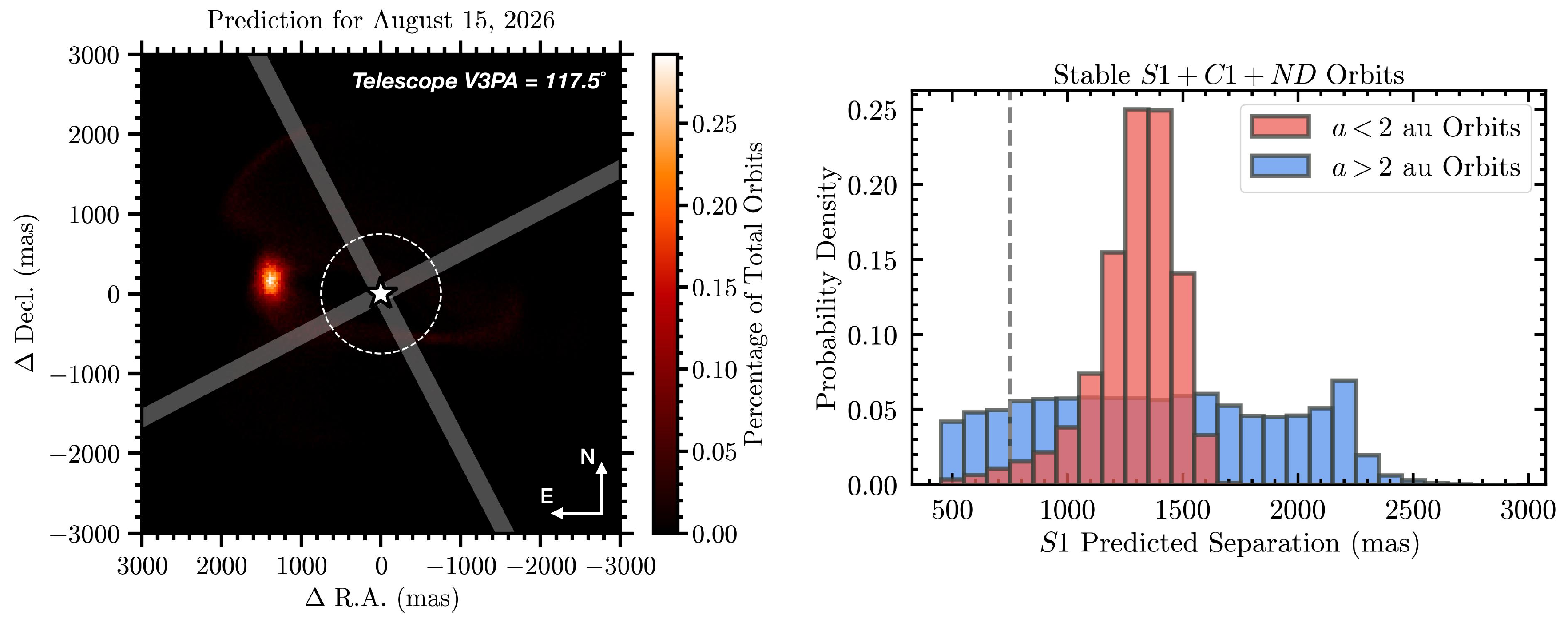} 
 \caption{A prediction for the location of \sone based on the family of dynamically stable orbits for $S1+C1$ consistent with the non-detections in 2025 suggests that \sone will be well-positioned for recovery in August 2026. \emph{Left:} predicted position in sky coordinates. The approximate orientation of the 4QPM transition boundaries at the selected example observation date is shown as a shaded region. The dashed circle marks 0\farcs75 separation. \emph{Right:} Histogram of the predicted separation of \sone in August~2026 for the two orbital semi-major axis families. A dashed line marks 0\farcs75 separation. The $a < 2$~au orbits constitute a greater fraction of the total number of orbits and are also favored based on photometric modeling in \S\ref{sec:phot}. \label{fig:PredictOrbit}}
\end{figure*}

It is thus reasonable to assume that debris disks can form and exist around these objects, with the caveat that they must lie within the stable region around either stellar component, stretching only a small fraction of their separation, e.g., $\sim\!12\%$ in the case of \acenAB \citep{Thebault2021,Cuello2024}. For binaries with similar separations, this suggests that only asteroid belt analogues within a few AU of each star are dynamically viable as circumstellar debris disks. Any dust produced from them would be relatively warm ($\sim\!100$--$300\,\text{K}$), exozodiacal dust, and would emit predominantly in the mid-infrared---where it is outshone by the star---potentially explaining the lack of photometric detections.

Finally, recent observational studies indicate that the orbits of planets orbiting one of the stars in binary (an S-type exoplanet) are roughly aligned with the binary orbit, particularly for separations below $\sim100\,\text{au}$. Astrometric monitoring of \textit{Kepler} planet hosts in binaries has shown that mutual inclinations are typically small, likely within 0–30° \citep{Dupuy2022,Lester2023}. These findings suggest that long-lived debris disks might exist and might be aligned with binary orbital plane. 

\subsubsection{Can an Exozodi Disk and an $S1+C1$-like Planet Coexist Around \acenA?} 

The prospects of finding a stable debris disk orbiting a star in a binary system must also be considered in the context of the presence of planets. Figure~\ref{fig:PlanetStability} shows effect of a planet on the possibility of stable orbits either for other planets or for particles in a disk. This shows that even if a planetesimal belt was able to form in this system it would not be able to survive in the face of dynamical perturbations from both \acenB, which prevents orbits surviving beyond $\sim 2.8$\,au \citep{Quarles2016}, and a planet like the $S1+C1$ candidate which causes an unstable region that extends to within $\sim$0.4~au of \acenA. While a planetesimal belt could survive interior to the planet, the current JWST/MIRI observations are not sensitive to any possible exozodi so close to the star.

\subsection{Future Opportunities with \acenA}
\label{sec:future}

The most pressing task for further work is to capture a second sighting of \sone with JWST. Figure~\ref{fig:PredictOrbit} identifies an excellent opportunity in August 2026 to recover \sone based on the family of stable $S1+C1$ orbits consistent with non-detections described in $\S$\ref{sec:orbits}. Around this date, the separation exceeds $\sim$1\arcsec\ and the predicted location is clear of the 4QPM boundaries. There is an urgency to this given the rapid approach of \acenA to the known background star denoted KS5 \citep{Kervella2016}. Between mid-2027 and mid-2028, the two will be within 3\arcsec\ of one another. \added{We note here that if \sone is unrelated to $C1$, then the orbits are much less constrained and there is significant uncertainty in its position at any given observation date.} Additionally, there are numerous opportunities to follow-up the detection of the candidate exoplanet for further characterization with upcoming and future facilities:

\begin{itemize}
\item The clear photospheric model has higher flux between 4--5 \mum\ (Figure~\ref{fig:planetring}) compared to the non-adiabatic and cloudy models (Figure~\ref{fig:planetmodel}). NIRCam coronagraphy could serve as a powerful diagnostic tool for the different models presented.
 
\item The coronagraphic instrument on the Nancy Roman Space Telescope has a mask specifically designed to work in the presence of a binary star system \citep{Bendek2021} and could be used to detect reflected visible light from a gas giant around 1--2~au. 

\item The METIS instrument on the European Extremely Large Telescope (EELT) should be capable of spectroscopic observations of the \sone candidate \citep{Birkby2024} and could even look for radial velocity shifts in the motion of the planet due to the presence of an exomoon.

\item Direct mass measurements should be possible with additional RV monitoring and with differential astrometry at millimeter wavelengths with ALMA \citep{Akeson2021}, or at visible wavelengths with the proposed Toliman \citep{Tuthill2018} or SHERA (J.~Christiansen, private comm.) space telescopes. 

\item Finally, the proposed Habitable Worlds Observatory (HWO) could, if equipped with appropriate binary star rejection capabilities, search for terrestrial-sized planets which might be found within the \acenA system despite the pessimistic concerns about the stability of orbits exterior to 0.4~au ($\S$\ref{sec:other}).

\end{itemize}

\section{Conclusions\label{sec:conclusions}}

We conducted JWST/MIRI F1550C coronagraphic imaging observations of the nearest solar-type star, $\alpha$~Centauri~A, over three epochs between August 2024 and April 2025 to directly resolve \acenA's habitable zone and perform a deep search for planets and exozodiacal disk emission. The key results from our program are summarized below.

\textbf{Detection of a candidate gas giant exoplanet in orbit around our nearest Sun-like star, \acenA.} We detected a point source ($S1$) in the August 2024 epoch of JWST/MIRI 15.5~\mum\ coronagraphic imaging. Detailed analysis, including various tests, presented in Paper~II \citep{Aniket2025} show that the source is unlikely to be a detector or speckle artifact. We definitively show that \sone is neither a foreground nor a background object. However, with only a single sighting by JWST, the candidate cannot be unambiguously confirmed as a bona fide planet.

\textbf{Deep upper limits on an exozodiacal disk around \acenA.} These observations have set stringent upper bounds on the presence of extended ``exozodiacal" dust disk in the habitable zone of \acenA. A limit of $<$ 5--8$\times$ the dust level within our own zodiacal cloud (for a disk coplanar with the \acenAB orbit) is a factor of $\gtrsim$5--10 more sensitive than those set by either photometric or interferometric methods toward more distant stars. Simulations show that for a planet with the candidate's properties, it is unlikely that a debris disk could remain stable and survive the planet's dynamical influence unless located within 0.4~au of the star.

\textbf{Orbital properties of the \acenA planet candidate.} By linking the sighting of JWST/MIRI \sone to another candidate, $C1$, detected by the VLT/NEAR experiment in 2019, we found a set of dynamically stable orbits. 52\% of the stable orbits were consistent with a non-detection of the planet candidate in the February and April 2025 epochs, indicating that it was likely missed in both follow-up observations due to orbital motion. The $S1+C1$ candidate is in a highly inclined ($\approx$50$^\circ$ or $\approx$130$^\circ$ with respect to the \acenAB binary orbital plane) and eccentric ($\sim0.4$) orbit, not unlike other S-type planets in close binary systems (e.g., HD~196885~Ab and $\gamma$~Cep~Ab), and is expected to undergo large amplitude von~Zeipel-Kozai-Lidov (vZKL) oscillations. 

\textbf{Physical properties of the \acenA planet candidate.}
$S1+C1$'s effective temperature is set by heating from \acenA and is expected to be $\sim$225 K based on the candidate's orbital properties. We found plausible atmospheric model solutions to the $S1+C1$ photometry for a planet radius between $\approx$1.1--1.15~\rj\ and mass between 90--150~\mearth\ (consistent with RV limits). Alternatively, we showed that a simplified optically thick ring with a cross-section equivalent to half of Saturn's ring could increase the mid-infrared flux of a smaller ($\sim1$~\rj) planet to explain the estimated photometry. 

\textbf{Importance of a confirmed planet around \acenA.} A confirmation of the $S1$ candidate as a gas giant planet orbiting our closest solar-type star, \acenA, would present an exciting new opportunity for exoplanet research. Such an object would be the nearest (1.33 pc), coldest ($\sim$225~K), oldest ($\sim$5 Gyr), shortest period ($\sim$2--3~years), and lowest mass ($\lesssim$~200~\mearth) planet imaged in orbit around a solar-type star, to date. Its extremely cold temperature would make it more analogous to our own gas giant planets and an important target for atmospheric characterization studies. Its very existence would challenge our understanding of the formation and subsequent dynamical evolution of planets in complex hierarchical systems. Future observations will confirm or reject its existence and then refine its mass and orbital properties, while multi-filter photometric and, eventually, spectroscopic observations will probe its physical nature. 

\begin{acknowledgments}
% STScI planning team and ALMA team acknowledgments
The STScI support staff provided invaluable assistance in the planning and execution of this program. In particular, we thank George Chapman and the FGS team for their dedicated work in finding and vetting guide stars for this program and Wilson Joy Skipper and the short- and long-range planning teams for their contributions to this challenging observational program. The STScI's Director's Office provided strong support for this program, from its initial selection as a high-risk, high-reward project, granting time to conduct test observations needed to validate the target acquisition strategy, to the execution of the follow-up DDT programs. We gratefully acknowledge the ALMA Director's Discretionary Time program. In particular we wish to thank Richard Simon, Bill Dent, Brian Mason, Erica Keller and Ilsang Yoon for their assistance in completing these critical observations in a timely manner. \added{Dan Sirbu provided constructive comments on an earlier version of this manuscript. We thank the referee for a prompt report and helpful comments that improved this manuscript.}

% Data use acknowledgments
The specific observations analyzed can be accessed via
\dataset[doi:10.17909/v8nv-vx17]{http://dx.doi.org/10.17909/v8nv-vx17} for the August 2024 observations,\dataset[doi:10.17909/cb0x-rn85]{http://dx.doi.org/10.17909/cb0x-rn85} for the February 2025 observations, and\dataset[doi:10.17909/3z9q-9f65]{http://dx.doi.org/10.17909/3z9q-9f65} for the April 2025 observations. STScI is operated by the Association of Universities for Research in Astronomy, Inc., under NASA contract NAS5-26555. Support to MAST for these data is provided by the NASA Office of Space Science via grant NAG5-7584 and by other grants and contracts. This research has made use of NASA's Astrophysics Data System. \added{Portions of this research were conducted with the advanced computing resources provided by Texas A\&M High Performance Research Computing that was supported in part by the National Science Foundation (NSF) under grant \#2232895.} This paper makes use of the following ALMA data: ADS/JAO.ALMA\#2022.A.00017.S. ALMA is a partnership of ESO (representing its member states), NSF (USA) and NINS (Japan), together with NRC (Canada), MOST and ASIAA (Taiwan), and KASI (Republic of Korea), in cooperation with the Republic of Chile. The Joint ALMA Observatory is operated by ESO, AUI/NRAO and NAOJ. The National Radio Astronomy Observatory is a facility of the National Science Foundation operated under cooperative agreement by Associated Universities, Inc. This work has made use of data from the European Space Agency (ESA) mission {\it Gaia} (\url{https://www.cosmos.esa.int/gaia}), processed by the {\it Gaia} Data Processing and Analysis Consortium (DPAC,
\url{https://www.cosmos.esa.int/web/gaia/dpac/consortium}). Funding for the DPAC has been provided by national institutions, in particular the institutions participating in the {\it Gaia} Multilateral Agreement. This publication makes use of VOSA, developed under the Spanish Virtual Observatory (https://svo.cab.inta-csic.es) project funded by MCIN/AEI/10.13039/501100011033/ through grant PID2020-112949GB-I00. VOSA has been partially updated by using funding from the European Union's Horizon 2020 Research and Innovation Programme, under Grant Agreement \#776403 (EXOPLANETS-A). This research has made use of the NASA Exoplanet Archive, which is operated by the California Institute of Technology, under contract with the National Aeronautics and Space Administration under the Exoplanet Exploration Program.

% Grant acknowledgments
This material is based on work supported by the National Science Foundation Graduate Research Fellowship under Grant No.~2139433. Part of this work was carried out at the Jet Propulsion Laboratory, California Institute of Technology, under a contract with the National Aeronautics and Space Administration (80NM0018D0004). Program PID\#1618, \#6797, and \#9252 are supported through contract JWST-GO-01618.001, JWST-GO-06797.001, and JWST-GO-09252.001, respectively. \added{Research done at NASA's Ames Research Center was supported by the NASA Astrophysics Division's Internal Scientist Funding Model (ISFM) program.} This project is co-funded by the European Union (ERC, ESCAPE, project No 101044152). Views and opinions expressed are however those of the author(s) only and do not necessarily reflect those of the European Union or the European Research Council Executive Agency. Neither the European Union nor the granting authority can be held responsible for them. 

\end{acknowledgments}

\begin{contribution}
C.~A.~Beichman and A.~Sanghi led the writing and submission of this manuscript. C.~A.~Beichman developed the observational strategy and sequences in conjunction with D.~Hines, J.~Aguilar, and M.~Ressler. R.~Akeson and E.~Fomalont executed and reduced the ALMA data. P.~Kervella developed the detailed astrometric solution used for \acenAB. A.~Sanghi led the post-processing of the MIRI observations with the assistance of D.~Mawet, W.~Balmer, L.~Pueyo, A.~Boccaletti, and J.~Llop-Sayson. Analysis of the possible orbits of the \acenA\ candidate was conducted by A.~Sanghi, K.~Wagner, B.~Quarles, and J.~Lissauer. Photometric modeling of the \acenA\ candidate was carried out by A.~Sanghi with the assistance of M.~Zilinskas and R.~Hu. Custom atmospheric models were generated by J.~Mang and P.~Tremblin. Dust emission models for the zodiacal cloud and an exoplanet ring system were developed by M.~Sommer, M.~Wyatt, and C.~A.~Beichman. Analysis of the extended emission was carried out by N.~Godoy and E.~Choquet. Other authors assisted with the preparation of the original JWST proposal and the manuscript.
\end{contribution}

\facilities{JWST(MIRI), ALMA, Gaia}

\software{
\texttt{astropy} \citep{astropy2022},
\texttt{emcee} \citep{Foreman-Mackey2013},
\texttt{jwst} \citep{jwst2022},
\texttt{multinest} \citep{feroz2009},
\texttt{NIRCoS} \citep{Kammerer2022},
 \texttt{pyNRC} \citep{Leisenring2024},
\texttt{pysynphot} \citep{stsci2013}, 
\texttt{spaceKLIP} \citep{Kammerer2022},
\texttt{matplotlib
} \citep{Hunter2007}, \texttt{numpy
} \citep{Harris2020}, \texttt{scipy
} \citep{Virtanen2020},
\texttt{STPSF} \citep{perrin14},
\texttt{webbpsf\_ext} \citep{Leisenring2024}.}

\appendix
\restartappendixnumbering

\section{Details of the Observation Strategy \label{app:obs-details}}

\begin{deluxetable}{cccccc}
    \centering
    \tabletypesize{\scriptsize}
    \tablecaption{Position of \acenA at JWST Observation Epochs\label{tab:AcenAPosns}}
    \tablehead{&  & \colhead{R.A.} & \colhead{R.A.\tablenotemark{a}} &\colhead{Decl.}&\colhead{Decl.\tablenotemark{b}}\\
    \colhead{Date}&\colhead{JD}	&\colhead{(deg)}&\colhead{(sec)}&\colhead{(deg)}&\colhead{(\arcsec)}}
    \startdata
    8/10/2024	&	2460750.363 &	 219.84748601	&	(23.3967) &			$-$60.83161739	&(53.8226) \\
    2/20/2025	&	2460727.234	&	219.8472157	&	(23.3318)	&	$-$60.8317325	&	(54.2370)\\
    4/25/2025	&	2460790.995	&	219.8464966	&	(23.1592)	&	$-$60.83177278	&	(54.3820)\\
    4/25/2025	&	2460791.2344&	 219.8464935	&	(23.1584) &	$-$60.8317725	&	(54.3813) \\
    \enddata
    \tablecomments{$^a$Relative to $\alpha=14^h$ 39$^m$. $^b$Relative to $\delta=-60^\circ$ 49$^\prime$. Positions incorporate proper motion and parallax as seen from vantage point of JWST.}
\end{deluxetable}

\begin{deluxetable*}{ccccccccc}
    \centering
    \tabletypesize{\scriptsize}
    \tablecaption{Gaia Astrometry for \emus and Offset Stars\label{tab:Gaia}}
    \tablehead{\colhead{Star}& \colhead{Gaia ID} & \colhead{R.A.} & \colhead{Decl.} &\colhead{$\mu$R.A.}&\colhead{$\mu$Decl.}&\colhead{Parallax}&\colhead{Total Uncertainty\tablenotemark{\scriptsize a}}&\colhead{F$_\nu$(F1000W)\tablenotemark{\scriptsize b}} \\
    \colhead{}&\colhead{}	&\colhead{(deg, 2016.0)}&\colhead{(deg, 2016.0)}&\colhead{(mas yr$^{-1}$)}&\colhead{(mas yr$^{-1}$)}&\colhead{(mas)}&\colhead{ (mas)}&\colhead{($\mu$Jy)}}
    \startdata
    \emus & 	5859405805013401984&184.3887799	&$-$67.960909&	$-230.60\pm0.19$	&$-26.21\pm0.26$&	$9.99\pm0.20$&	2.68&\nodata\\
    \emus TA\tablenotemark{\scriptsize c} (G9)&	5859405804986931200&184.3941452	&$-$67.953630&$-7.11\pm0.02$&	$0.43\pm0.02$&	$0.14\pm0.02$&	0.21&13450\\
    \acen TA\tablenotemark{\scriptsize d} (G0)&	5877725249280411392&219.8776270	&$-$60.828383&	$-1.79\pm0.11$&	$-1.01\pm0.11$&	$0.32\pm0.09$&	1.28&1350\\
    \acen TA\tablenotemark{\scriptsize e}  (G5)&	 5877725146201190144&219.8800796 &$-$60.8445635&$-2.71 \pm	0.17$ &	$-2.63\pm0.28$&$0.001\pm0.230$&0.23&580\\
    \enddata
    \tablenotetext{a}{Combined uncertainty from parallax and proper motion between 2016.0 and 2024.3.}
    \vspace{-0.1in}
    \tablenotetext{b}{Flux density measured in June 2023 test images.}
    \vspace{-0.1in}
    \tablenotetext{c}{Offset star used for target acquisition (TA) of \emus in all three epochs.}
    \vspace{-0.1in}
    \tablenotetext{d}{Offset star used for target acquisition (TA) of \acenA in August 2024 and February 2025.}
    \vspace{-0.1in}
    \tablenotetext{e}{Offset star used for target acquisition (TA) of \acenA in April 2025.}
\end{deluxetable*}

\begin{deluxetable*}{clccccc}
    \tabletypesize{\scriptsize}
    \tablecaption{Log of Successful JWST/MIRI Observations of $\alpha$ Cen A\label{tab:MIRIObs}}
    \tablehead{\colhead{PID-Obs.~\#} & \colhead{Target} & \colhead{\# Dither} &\colhead{Science Time}&\colhead{Start Time}&\colhead{Observation Mid-Point}&\colhead{MIRI (X,Y) Offsets\tablenotemark{\scriptsize a}}\\
    \colhead{}& \colhead{}& \colhead{} & \colhead{(hr)}&\colhead{(UTC)}& \colhead{(UTC)}&(arcsec)}
    \startdata
    1618-11	&	$\alpha$ Cen Snapshot (F1000W)\tablenotemark{\scriptsize b}& 4& 0.19&06/19/2023 10:00&\nodata&\nodata\\
    1618-50	&	$\epsilon$ Mus Snapshot (F1000W)\tablenotemark{\scriptsize b}& 4& 0.19&06/19/2023 10:00&\nodata&\nodata\\
    1618-61 &$\epsilon$ Mus Snapshot (F1550C) &1&0.04&07/07/2024 22:31&\nodata&\nodata\\
    1618-62 &$\alpha$ Cen Snapshot (F1550C)&1&0.04&07/08/2024 00:29&\nodata&\nodata\\
    1618-63 &$\epsilon$ Mus Background (F1550C)&1&0.04&07/08/2024 02:21&\nodata&\nodata\\
    1618-64 &$\alpha$ Cen Background (F1550C)&1&0.04&07/08/2024 02:52&\nodata&\nodata\\ \hline
    1618-52&$\epsilon$ Mus Visit 1 (V3=135.0$^\circ$)&9&7.19&08/11/2024 13:00	&	08/11/2024 17:04
    &($-47.7588$, $-5.2736$)\\
    1618-53&$\epsilon$ Mus Visit 1 Background &1&0.80&08/11/2024 21:16	&	\nodata
    &\nodata\\
    1618-56&Alpha Cen Roll 2 (V3=112.7$^\circ$) &1&2.50&08/12/2024 04:56&	8/12/2024 06:32
    &($32.7863$, $42.8539$)\\
    1618-57&Alpha Cen Visit 2 Background &1&2.50& 08/12/2024 08:09& \nodata
    &\nodata\\
    1618-65&$\epsilon$ Mus at \acenB\ (V3=135$^\circ$) &1&0.80&08/12/2024 17:11	&	08/12/2024 18:48
     &($-39.372$, $-8.104$) \\ \hline
    6797-01&$\epsilon$ Mus Visit 1 (V3=133.0$^\circ$)&9&7.19&02/20/2025 01:00	&	02/20/2025 04:57
    &($46.9405$, $-9.7562$)\\
    6797-02&$\epsilon$ Mus Visit 1 Background &1&0.80&02/20/2025 09:04	&	\nodata
    &\nodata\\
    6797-03&$\epsilon$ Mus at \acenB\ &1&0.80&02/20/2025 10:13	&	02/20/2025 11:42
    &($38.4623$, $-7.294$)\\
     6797-06&$\alpha$ Cen Roll 2 (V3=294.5$^\circ$)&1&2.50&02/20/2025 19:51	&	02/20/2025 21:20
    &($-38.7111$, $-38.6047$)
    \\
    6797-07&$\alpha$ Cen Roll 2 Background&1&2.50&02/20/2025 23:10	&	\nodata
     & \nodata\\
     6797-08&$\epsilon$ Mus Visit 2 (V3=133.0$^\circ$)&9&7.19&02/21/2025 02:17	&	02/21/2025 06:15&($46.9399$, $-9.7563$)\\
    6797-09&$\epsilon$ Mus Visit 2 Background &1&0.80&02/21/2025 10:22	&	\nodata
    & \nodata\\
    6797-10&$\epsilon$ Mus at \acenB\ (V3=133.0$^\circ$)&1&2.50&02/21/2025 11:30	&	02/21/2025 13:00 &($38.629$, $6.781$)\\ \hline
    9252-01&$\epsilon$ Mus Visit 1 (V3=38.0$^\circ$)&9&7.19&04/25/2025 01:00	&	04/25/2025 04:57
    &($10.96213$, $-46.62994$)\\
    9252-02&$\epsilon$ Mus Visit 1 Background &1&0.80&04/25/2025 09:04	&	\nodata
    &\nodata\\
    9252-03&$\epsilon$ Mus at \acenB\ &1&0.80&04/25/2025 10:13	&	04/25/2025 11:42
    & ($8.01166$, $-38.20160$)\\
    9252-04&$\alpha$ Cen Roll 1 (V3=346$^\circ$)&1&2.50&04/25/2025 13:35	&	04/25/2025 15:04
    &($-50.83941$, $-54.82857$)
    \\
    9252-05&$\alpha$ Cen Roll 1 Background&1&2.50&04/25/2025 16:53&	\nodata
     & \nodata\\
     9252-06&$\alpha$ Cen Roll 2 (V3=356$^\circ$)&1&2.50&04/25/2025 19:51	&	04/25/2025 21:20
    &($-59.593412$, $-45.16806$)
    \\
    9252-07&$\alpha$ Cen Roll 2 Background&1&2.50&04/25/2025 23:10	&	\nodata
     & \nodata\\
     9252-08&$\epsilon$ Mus Visit 2 (V3=38.0$^\circ$)&9&7.19&04/26/2025 02:17	&	04/26/2025 06:15
    &($10.96213$, $-46.62994$)\\
    9252-09&$\epsilon$ Mus Visit 2 Background &1&0.80&04/26/2025 10:22	&	\nodata
    & \nodata\\
    9252-10&$\epsilon$ Mus at \acenB\ (V3=38.0$^\circ$)&1&2.50&04/26/2025 11:30	&	04/26/2025 13:00
    &($9.51997$, $-37.81736$)
    \\
    \enddata
    \tablenotetext{a}{Offsets from the Gaia star to the target star were calculated based on the epoch positions using the STScI software \texttt{pysiaf}.}
    \tablenotetext{b}{The F1000W broadband image was obtained with FASTR1 using 60 groups.}
    \tablecomments{All F1550C coronagraphic data was obtained with FASTR1 using 30 groups with 400 integrations for \emus (behind the 4QPM) and 1250 integrations for \acenA (behind the 4QPM) and \emus at the off-axis position of \acenB. Observations 1618-1 to 1618-8 were obtained on 07/26/2023 and 07/27/2023 but failed due to either a guide star issues or incorrect offsets from the Gaia stars. Observations 1618-54 and 1618-58 in August 2024 and 6797-04 in February 2025 failed due to guide star issues.}
    \end{deluxetable*}

\subsection{Reference Star Selection \label{sec:refstar}}

For stars as bright as \acenA\ ([F1550C] $=-$1.51~mag), the selection of reference stars is limited. The IRAS Low Resolution Spectrometer (LRS) Catalog \citep{LRS} was used to identify potential reference stars: $F_\nu(12\;\mu$m)~$>$~50~Jy within 20$^\circ$ of \acenA, clean Rayleigh-Jeans photospheric emission, constant ratio ($<$10\%) of LRS brightness ($F_{\rm \alpha\;Cen\;A}/F_{\rm star}$) across the F1550C band, a low probability of variability during the 300 day IRAS mission (\texttt{VAR}~$<15$\%), and no bright companions within 100\arcsec. These criteria resulted in the selection of $\epsilon$ Mus ([F1550C]~=~$-1.3$~mag), a long period variable star located 17$^\circ$ away on the sky with a $K$ band variability $\lesssim$0.5~mag \citep{Murakami2007,Tabur2009}. The ratio of the LRS spectra of the (unresolved) \acenAB system to these stars is constant across the F1550C bandpass to $<$ 1\% which means that the effects of wavelength mismatches in the reference star subtraction will be negligible.

\subsection{Target Acquisition}
\subsubsection{Astrometry of \acen and \emus}

\begin{figure}[!tb]
    \centering
    \includegraphics[width=0.48\textwidth]{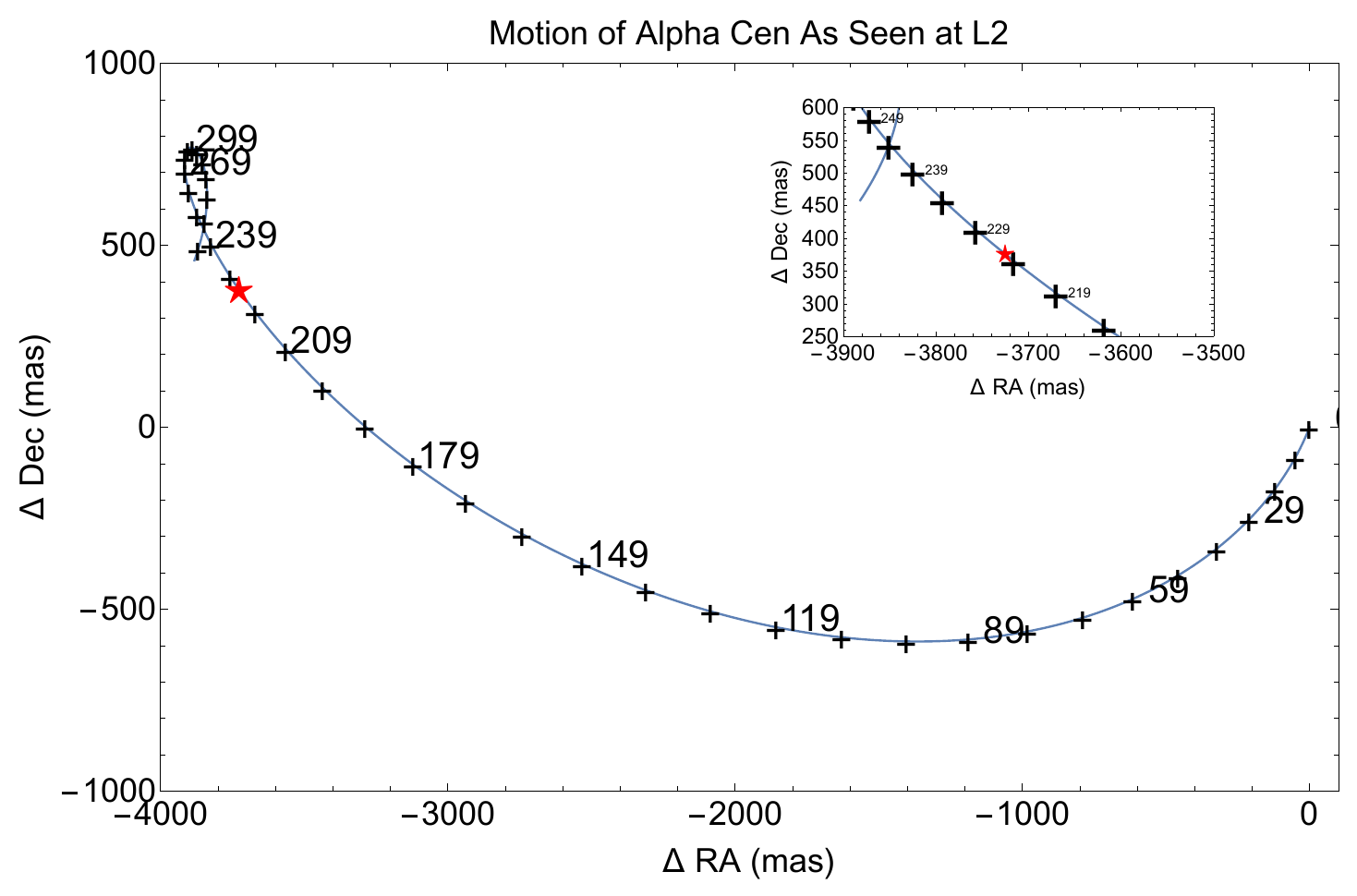}
    \caption{The change in the position of \acenA\ relative to 2024-Jan-1 ($\Delta\alpha$ $=0$, $\Delta\delta$ $=0$) computed using proper motion and parallax as seen from JWST's L2 vantage point. Markers denote 10-day intervals and days of the year. As a reference, the red star denotes the date of the August 2024 JWST observations (first epoch). The inset plot zooms-in near the August 2024 observation date with markers spaced in 5-day intervals. Details of the astrometry for \acen are given in Appendix~\ref{sec:system} and in \citet{Akeson2021}. \label{fig:ALMA}}
    \end{figure}

\begin{figure*}[!tb]
    \centering
    \includegraphics[width=\textwidth]{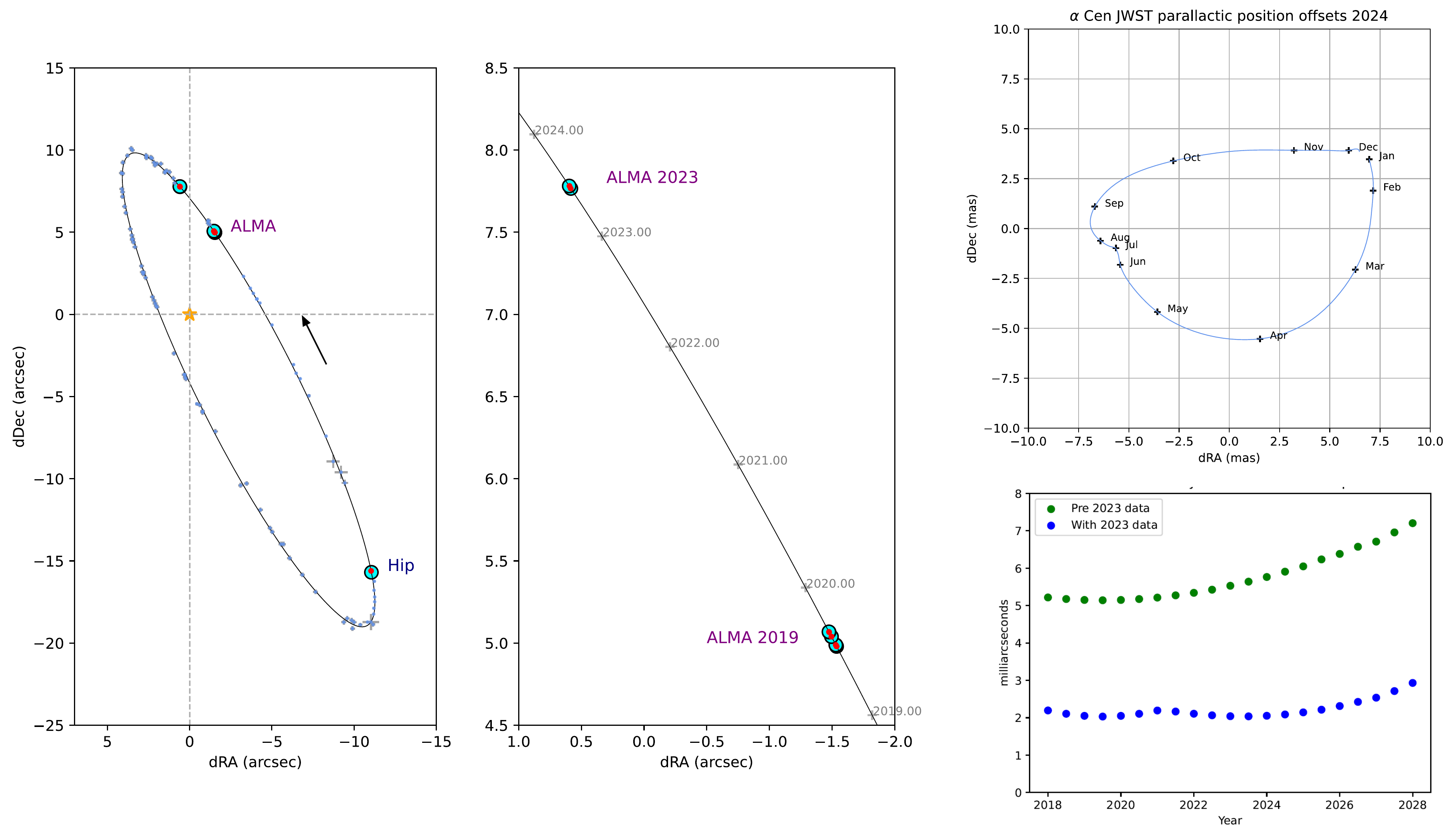}
    \caption{\emph{Left:} Orbit of \acenB\ relative to \acenA showing one epoch of Hipparcos and two epochs of ALMA data. \emph{Center:} zoom-in showing the ALMA data. \emph{Top right:} The difference between the parallax effect as seen from Earth and from JWST at L2. \emph{Bottom right:} The uncertainty in position of \acen\ as a function of year before and after the addition of the new 2023 ALMA observations.\label{fig:ALMA1}}
\end{figure*}

The \acenAB system has a parallax of 750 mas and an annual proper motion of ($-$3640, $+$700) mas yr${^{-1}}$, which corresponds to a mean motion of $\sim$10 mas/day. The description of the procedures leading to the detailed ephemeris used for the observations are provided in \citet{Akeson2021}. As described in Appendix~\ref{sec:system}, the ephemeris is based on a combination of absolute astrometry from Hipparcos and ALMA. Radial velocity observations of both \acenA and \acenB help determine the motions of \acenAB in their 80 year orbit. The ephemeris was calculated on an hour-by-hour basis, including the effects of parallax as observed from the vantage point of JWST's L2 orbit (Figure~\ref{fig:ALMA}). The location of JWST at L2, an additional 1.5 million km from Earth, increases the parallactic effect by 1\% or 7.5 mas, but in a non-intuitive manner due to JWST's motion at L2 (Figure~\ref{fig:ALMA1}). This effect is not negligible compared to the $\theta_\mathrm{LD} = 8.5$\,mas angular diameter of \acenA \citep{Kervella2017} and the centering accuracy requirement ($\sim$10 mas) for best performance behind the MIRI coronagraphic mask \citep{Boccaletti2022}.

The precise location of JWST at the epoch of these observations was obtained from the JPL Horizons website\footnote{\url{https://ssd.jpl.nasa.gov/horizons/app.html/}}. The combination of visible data and two epochs of ALMA data \citep[2018/2019 from][]{Akeson2021} and the 2023 ALMA DDT observations (described in Appendix~\ref{sec:ALMA1}) for \acenAB yields a precision of $\sim$2 mas in the predicted position of \acenA (Figure~\ref{fig:ALMA1} and Table~\ref{tab:AcenAPosns}). Taking into account the astrometric precision of the Gaia stars and of \acen\ itself, we estimate that the overall astrometric precision of the blind offset between the offset stars and the two targets, \emus and \acenA, will be $\sim$2.5 mas ($1\sigma$), to which must be added the $\sim$5--7~mas (1$\sigma$, one axis) offsetting precision of JWST itself\footnote{\url{https://jwst-docs.stsci.edu/jwst-observatory-characteristics/jwst-pointing-performance}}. Astrometry for \emus, its associated Gaia offset star, and \acenA's associated Gaia offset star was obtained from the Gaia DR3 catalog (Table~\ref{tab:Gaia}). The effects of the proper motion and parallax values were taken into account but were relatively minor compared to those for \acenA.

In planning the observational sequences in APT, we specified the exact V3 rotation angles (with a precision of 0.001$^\circ$) for each observation and used that information to derive the shift from the Gaia offset star to either \acenA\ or \emus in instrument $(x,y)$ coordinates. For \acenA, these calculations used the position of \acenA \citep[including proper motion, parallax and other smaller effects;][]{Akeson2021} at the expected midpoint of the observation based on a detailed timeline of each observation (Table~\ref{tab:MIRIObs}). The small amount of smearing during the 2.8 hr duration of each \acenA\ observation ($<$2 mas) was deemed acceptable compared to the complexity of designating \acenA\ as a moving target. The conversion of the offset in $(\Delta\alpha,\:\Delta\delta)$ to instrument $(x,y)$ was calculated for the exact epoch of observation and desired V3 angle using a model of the MIRI focal plane using STScI's \texttt{pysiaf} routine\footnote{\url{https://github.com/spacetelescope/pysiaf}}.

\subsubsection{Offset Star Selection and Validation}

The offset stars used for Target Acquisition were drawn from the Gaia DR3 catalog and had to have a mid-IR brightness suitable for easy measurement in a short TA observation. A preliminary search of DR3 revealed 92 targets within 60\arcsec\ of \acenA and 24 within 30\arcsec\ of \emus. Cuts in magnitude ($G$~$<$~16 mag) and the requirement that each star have a quoted parallax and proper motion measurement reduced the number to a handful for each source. However, the proximity of both \acen and \emus to the Galactic Plane ($b=-0.67^\circ$ and $-5.30^\circ$, respectively) means that the effects of extinction can make predictions of mid-IR brightness highly uncertain. For this reason, we scheduled test observations of both \acen\ and \emus in the MIRI TA filter (F1000W) without any associated coronagraphic observations. These were executed in June 2023. Figure~\ref{fig:TAobs} shows images of \acen\ and \emus. Simulations using \texttt{STPSF} were developed to assess the influence of the bright target stars (\acenA, \acenB, and \emus) to ensure that diffraction effects would not affect the detectability of the much fainter Gaia stars in the TA procedure. The star denoted G0 was suitable for both of the V3 roll angles selected for August 2024 and February 2025 (Table~\ref{tab:Gaia}). The star denoted G5 was used for April 2025. A single star (G9) was suitable for use with \emus at its V3 angle of observation in all three epochs. 

\begin{figure}[t]
    \centering
    \includegraphics[width=0.47\textwidth]{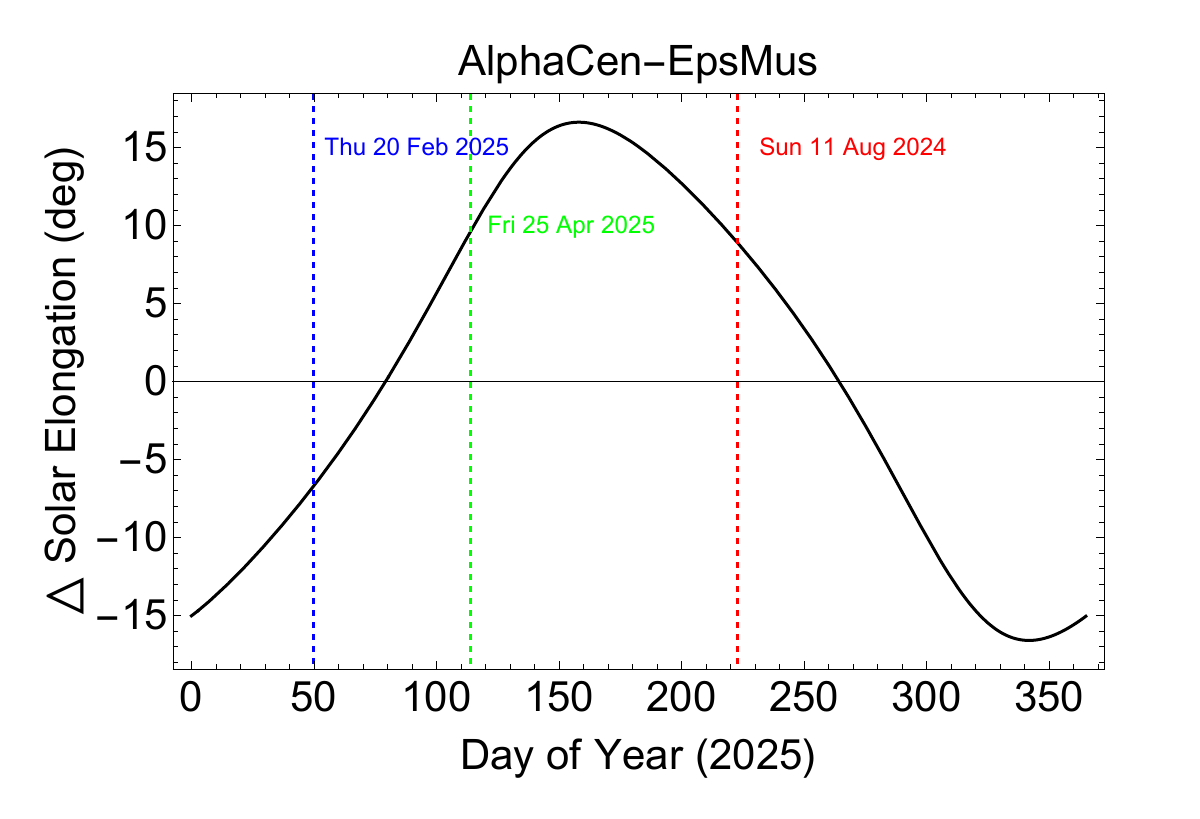}
    \caption{Change in the solar elongation angle between \acen and \emus as a function of the day of year. The red, blue, and green dashed vertical lines correspond to the observation dates in August 2024, February 2025, and April 2025 (respectively). \label{fig:solar}}
\end{figure}

\subsection{Minimizing Effects of Telescope Slews}
The ability to detect faint companions is dominated by the stability of the wavefront of the JWST telescope and the accurate placement of the star behind the coronagraphic mask. Pre-launch expectations were that there would be slow-varying wavefront error (WFE) drifts $\sim$10~nm depending on changes in solar elongation (and thus in the telescope's thermal balance) and due to stresses from internal structures such as the ``frill" surrounding the primary mirror \citep{Perrin2018}. On-orbit performance proved to be significantly better than pre-launch estimates, particularly as the telescope assembly continued to stabilize thermally, with drifts as low as $\sim$2--5~nm as measured using semi-daily WFE monitoring observations \citep{Rigby2023,Lajoie2023}. One factor in selecting the requested observing window was minimizing the change in solar elongation angle between \acen\ and \emus. The final scheduled dates in August 2024, February 2025, and April 2025 resulted in changes in the absolute value of solar angle $\lesssim10^\circ$ (Figure~\ref{fig:solar}). To mitigate further the effects of wavefront drift, we bracketed observations of \acen with observations of \emus. This choice provided redundancy against any failure during guide star and target acquisition by the telescope and thus would still leave one valid PSF reference star observation. Indeed, this proved to be important for the August 2024 observations where one \emus reference observation failed. 

\subsection{Mitigating the Effects of \acenB \label{sec:AcenB}}
We initially considered strategies either using \acenB as a reference star for \acenA or placing \acenB along the quadrant boundaries of the 4QPM to suppress its starlight. Both strategies had significant disadvantages. In the former case, the brighter star \acenA would be unocculted on the detector in the reference images. In the latter case, very few opportunities were available to schedule observations in the desired configuration. Furthermore, small errors in the position of \acenB along the boundaries would lead to substantial stellar leakage and changes in the PSF. Our pre-launch simulations and current data analysis demonstrated that the influence of \acenB\ at 7\arcsec--9\arcsec\ separation is sufficiently large in the 1\arcsec--2\arcsec\ region around \acenA that it is necessary to subtract the image obtained by placing \emus, unocculted but through the F1550C mask, at the same offset of \acenB relative to \acenA to provide an off-axis PSF reference. Analysis with the test observations obtained in July 2024 showed that the use of \texttt{STPSF} models was inadequate to remove speckles from \acenB\ at the required level. 

\subsection{Mitigation of MIRI Background: The ``Glow Sticks"}

Following practice recommended by STScI to mitigate the effects of the ``Glow Stick" phenomenon \citep[excess telescope background scattered off of telescope structure;][]{Boccaletti2022, Carter2023}, we included MIRI background observations in exactly the same detector and instrument setup as the on-target observations for both \emus and \acen. These background observations were placed in positions which appeared relatively blank in Spitzer or WISE images. Each background field was observed twice for each object (Table~\ref{tab:MIRIObs}) with 5\arcsec\ shifts in the center position to help mitigate the effects of sources in the fields.

\section{Astrometry of the $\alpha$ Cen System}\label{sec:system}
\setcounter{table}{0}
Determining the astrometric properties of the \acen\ system is complex due to the proximity of \acen\ to Earth and the orbits of the two stars around their common center of mass. High precision visible light observations are scarce due to the brightness of the two stars relative to much fainter reference stars. \citet{Akeson2021} made millimeter-wavelength observations of \acenAB\ using the ALMA array. At these wavelengths, the two stars are bright enough to yield high SNR data with milli-arcsec precision in $\sim$ an hour of observing time with absolute positions on the ICRF reference frame. The ALMA observations are described in $\S$\ref{sec:ALMA1} and the astrometric information used for the JWST observations in $\S$\ref{sec:ALMA2}.

\begin{deluxetable*}{ccccc}[!tb]
    \tabletypesize{\small}
    \tablecaption{New ALMA Astrometry\label{tab:ALMA}}
    \tablehead{\colhead{Date} & \colhead{Start Time} & \colhead{Star} & \colhead{R.A.} & \colhead{Decl.} \\ 
    \colhead {} & \colhead{(UT hours)} &\colhead{} & \colhead{} & \colhead{}}
    \startdata
     01 June 2023 & 02:35 & A & 219.864944115$^\circ$
    ($\pm$3 mas) & $-$60.848591790$^\circ$ ($\pm$3 mas) \\
     & & B & 219.865261605$^\circ$ ($\pm$4 mas)
     & $-$60.846446520$^\circ$ ($\pm$4 mas) \\
     & & A$-$B & $-$0\farcs5935 $\pm$ 0\farcs0004 &  $-$7\farcs6756 $\pm$ 0\farcs0005 \\
    17 June 2023 & 01:55 & A &219.850279410$^\circ$ ($\pm$1 mas)
     & $-$60.831926083$^\circ$ ($\pm$1 mas) \\
    & & B &219.850610280$^\circ$
    ($\pm$1 mas) & $-$60.829773243$^\circ$ ($\pm$1 mas) \\
     & & A$-$B & $-$0\farcs5806 $\pm$ 0\farcs0005 &  $-$7\farcs7502 $\pm$ 0\farcs0005  \\
    26 June 2023 & 00:59 & A & 219.850180635$^\circ$
    ($\pm$0.8 mas) & $-$60.831902697$^\circ$ ($\pm$1.1 mas) \\
    & & B & 219.850518885$^\circ$
     ($\pm$1.2 mas) & $-$60.829745590$^\circ$ ($\pm$1.2 mas) \\ 
     & & A$-$B & $-$0\farcs5715 $\pm$ 0\farcs0020 &  $-$7\farcs7234 $\pm$ 0\farcs0020\\
    \enddata
\end{deluxetable*}

\subsection{New ALMA Astrometry of \acenAB \label{sec:ALMA1}}

The positions of \acenA\ and \acenB\ were observed with ALMA between 2018 Oct 14 and 2019 Aug 26 using about 40 25-m antennas as described in detail in \citet{Akeson2021} and summarized here. The average observing frequency was 343.5 GHz, but the ALMA configurations varied and produced a resolution of 140, 33, 28, 62, and 62 mas for the five observing blocks. Each block was about 80 min long and consisted of about 120 scans of which 15\% included calibration sources. The pointing center of each experiment was located near the expected barycenter of the A and B stars to minimize errors caused by small antenna pointing offsets. The \acenAB positions were determined using the phase referencing technique with the nearby quasar, J1452-6502, with an ICRF position accuracy $<$0.5 mas. This uncertainty produced the main limit to the absolute radio position of the AB system. However, the separation of the A and B stars do not depend on the quasar position accuracy and some of the atmospheric position jitter between the A and B stars also canceled. Further details of the reduction, imaging and multi-calibrator checks to the astrometric accuracy are given in \citet{Akeson2021}.

\begin{deluxetable*}{cc|cc|cc|cc}[!htb]
    \tablecaption{Astrometry of \acenA and \acenB (2024--2027) \label{tab:ALMA2}}
    \tabletypesize{\scriptsize}
    \tablehead{\colhead{Julian Year} & \colhead{Time} & \colhead{R.A. (\acenA)} & \colhead{Decl.~(\acenA)} & \colhead{R.A. (\acenB)} & \colhead{Decl. (\acenB)} &  \colhead{$\rho$(A$\rightarrow$B)} & \colhead{$\theta$(A$\rightarrow$B)}\\
    & \colhead{} & \colhead{(J2000, deg)}& \colhead{(J2000, deg)}& \colhead{(J2000, deg)}& \colhead{(J2000, deg)}&(arcsec)&\colhead{(deg, North $=0$)}}
    \startdata
     2024.0000000& 2024-01-01T12:00:00.000&  219.84969992 & $-$60.83174299& 219.85019709& $-$60.82949888 &  8.1258  & 6.1630\\
     2024.0027379 & 2024-01-02T12:00:00.000 & 219.84969758& $-$60.83174528 &219.85019558& $-$60.82950072 & 8.1275 &  6.1721\\
    2024.0054757 & 2024-01-03T12:00:00.000 & 219.84969512 & $-$60.83174758 &219.85019396& $-$60.82950258&  8.1293  & 6.1812\\
    2024.0082136 & 2024-01-04T12:00:00.000 & 219.84969256 & $-$60.83174989 &219.85019224 & $-$60.82950444&  8.1310  & 6.1903\\
    2024.0109514 & 2024-01-05T12:00:00.000 & 219.84968989 & $-$60.83175221 &219.85019041& $-$60.82950632 &  8.1328 & 6.1994\\
    \enddata
    \tablecomments{The coordinates are apparent coordinates from the location of JWST. The astrometry is available at\dataset[10.5281/zenodo.16280658]{10.5281/zenodo.16280658}.}
\end{deluxetable*}

New observations were obtained through an accepted ALMA DDT proposal on 2023 June 1, 17 and 26 (DDT proposal 2022.A.00017.S). The data were taken in Band 7 (343.5 GHz) with the correlator configured for maximum broadband sensitivity. The 2023 positions are listed in Table~\ref{tab:ALMA} and were derived from pipeline-processed data using additional analysis to determine the internal position uncertainties. Due to the large proper motion of \acen, roughly 10 mas/day on average, the phase center tracking has a significant impact on the measured positions. For the 2023 June 1 observations, the \acenA position was tracked with time, while for the 2023 June 17 and 26 observations, the phase center was located near \acenA, but was not tracked with time. 

The main improvement from the previous ALMA observations \citep{Akeson2021} is the measurement of internal position errors. These were estimated by splitting each 50-min experiment into three independent parts and then determining the mean stellar position and the error from the scatter
among the three parts. The slight stellar motion during an hour observation (0.4 mas) is much less than that caused by the typical temporal ``atmospheric" variations. For the 2023 June 17 and 26 observations, where the phase center was located near \acenA, but was not tracked with time, the \acenA\ and \acenB\ 1-sigma position errors are about 1.5 mas. For the 2023 June 1 observation, during which the \acenA position was phased tracked, the location of the absolute frame of the images is uncertain, resulting in a larger absolute position error estimate (Table~\ref{tab:ALMA}).

The absolute star positions in the DDT observations are tied to the phase calibrator (J1408-5712) whose absolute position is measured by global VLBI observations. Its absolute position has an uncertainty of about 1.5 mas which is, unfortunately, relatively large for an ICRF source since it has not been observed very often. The calibrator is located only 5.4$^\circ$ away from \acen\ and is the closest of the brighter available calibrators. If the J1408-5712 position error and the \acenA internal position errors are combined, then the ICRF absolute position errors should be no larger than 2 mas in R.A. and in Decl.

The separation of \acenA and \acenB was also obtained for these observations, again by splitting up each 50-min experiment into three parts. The estimated separation error is less than 1 mas for the 2023 June 17 and 26 observations because much of the error that affect the absolute position of A and B cancels when calculating the stellar position difference. The accuracy is mostly signal-to-noise limited. Since the A--B separation for the 2023 June 1 observation does not depend on the somewhat uncertain definition of the absolute coordinate grid, its accuracy is only a bit larger than the two later observations.

\subsection{Astrometry of \acenA, \acenB and \acenAB \label{sec:ALMA2}}

The analysis method for determining the astrometric properties of the \acenAB\ system using the combined visible and ALMA data is described in \citet{Akeson2021}. The updated orbit is visualized in Figure~\ref{fig:ALMA1}. Accurate knowledge of the position of \acenA depends on accounting for the location of JWST at L2. The differences in the parallactic motion between the two observing sites, Earth and L2 is $\sim\pm$5 mas (Figure~\ref{fig:ALMA1}). Finally, we note that the addition of the ALMA DDT observations reduced the uncertainty in the positions of \acenA\ from $\sim$5--6 mas to $\sim$2 mas through 2028 (Figure~\ref{fig:ALMA1}). A detailed analysis of the orbit of \acenAB, including all ALMA epochs and new HARPS radial velocity data will be presented in a forthcoming paper (Kervella et al., in prep).

Table~\ref{tab:ALMA2} lists the position of \acenA\ and \acenB\ as seen from JWST's location (which is obtained from the HORIZONS database\footnote{\url{https://ssd.jpl.nasa.gov/horizons/app.html/}}) and lists the relative positions of \acenA\ and \acenB, which were used in positioning the reference star \emus at the position of \acenB\ to mitigate the speckles from the unocculted star. The first few entries of the full ephemeris are displayed and the complete table is available in electronic form at\dataset[10.5281/zenodo.16280658]{10.5281/zenodo.16280658}.

\end{document}